\documentclass[12pt]{article}

\usepackage{amssymb}
\usepackage{graphicx}
\usepackage{amsmath}
\usepackage{verbatim}
\usepackage{setspace}
\usepackage{ulem}
\usepackage{textpos}
\usepackage{changepage}
\usepackage{url}
\usepackage{amsfonts}
\usepackage{rotating}
\usepackage{subcaption}
\usepackage{natbib}

\usepackage{amsthm}

\newtheorem{assumption}{Assumption}
\newtheorem{proposition}{Proposition}
\newtheorem{remark}{Remark}

\newtheorem{example}{Example}

\def\bs{\boldsymbol}

    \DeclareMathOperator*{\argmax}{\arg\!\max}

\graphicspath{{../figure/}}
\begin{document}

{
\singlespacing
\title{\textbf{Identification and Estimation of Production Function with Unobserved Heterogeneity}\thanks{The first and second authors acknowledge financial support from the Social Sciences and Humanities Research Council while the third author acknowledges financial support from JSPS Grant-in-Aid for Young Scientists (B) No.16K21004 and Grant-in-Aid for Scientific Research (C) No.19K01584. All remaining errors are our own. This paper represents the views of the authors and not necessarily those of the Cabinet Office, the Government of Japan or the institutions they represent.}
\author{Hiroyuki Kasahara\thanks{Address for correspondence: Hiroyuki Kasahara, Vancouver School of Economics, University of British Columbia, 6000 Iona Dr., Vancouver, BC, V6T 1L4 Canada.}\\
Vancouver School of Economics\\
University of British Columbia\\
hkasahar@mail.ubc.ca \and Paul Schrimpf\\
Vancouver School of Economics\\
University of British Columbia\\
schrimpf@mail.ubc.ca
\and Michio Suzuki\\
Tohoku University\\
Cabinet Office, Government of Japan\\
michio.suzuki.a3@tohoku.ac.jp
}}
\maketitle
}
\begin{center}
\end{center}
\begin{abstract}This paper examines the nonparametric identifiability of production functions, considering firm heterogeneity beyond Hicks-neutral technology terms. We propose a finite mixture model to account for unobserved heterogeneity in production technology and productivity growth processes. Our analysis demonstrates that the production function for each latent type can be nonparametrically identified using four periods of panel data, relying on assumptions similar to those employed in existing literature on production function and panel data identification. By analyzing Japanese plant-level panel data, we uncover significant disparities in estimated input elasticities and productivity growth processes among latent types within narrowly defined industries. We further show that neglecting unobserved heterogeneity in input elasticities may lead to substantial and systematic bias in the estimation of productivity growth.
\end{abstract}


\section{Introduction}
Estimating a firm's production function and productivity is  a critical topic in empirical economics.\footnote{Understanding how the input is related to the output is a fundamental issue in empirical industrial organization \citep[see, for example,][]{Ackerberg07} while a measure of total factor productivity is necessary to examine the effect of trade policy on productivity and to analyze the role of resource allocation on aggregate productivity \citep[e.g.,][]{Pavcnik00,KASAHARA2008,KASAHARA2013,Hsieh09}. Production function estimation is also important for markup estimation \citep{hall1988relation, de2012markups,Raval2023}.} Despite its importance, the standard production function estimation procedures impose an implausible assumption that production functions are common across firms except for separable Hicks-neutral productivity terms \citep{Olley1996, Levinsohn03, Wooldridge09, Ackerberg15, gandhi20}. In the presence of identification issues due to the simultaneity problem \citep{Marschak44}, the literature on identifying production functions that incorporate unobserved heterogeneity beyond the Hicks-neutral productivity term is scarce  but a rapidly growing research area \citep{Li2017, doraszelski2018measuring, balat19, ZHANG2019,  Demirer20, chen21, Raval2023}. 

This paper establishes the nonparametric identification of production functions from panel data when production functions and productivity growth processes are heterogeneous across firms in unobserved time-varying ways. 
We consider a finite mixture specification in which there are $J$ distinct time-varying production technologies, and each firm belongs to one of the $J$ latent types. Econometricians do not observe the latent type of firms. Without making any functional form assumption on production technology and productivity growth processes, we establish nonparametric identification of $J$ distinct production functions, productivity processes, and a population proportion of each type under assumptions similar to those used in the existing production function and panel data identification literature. We also address potential measurement errors in labor inputs because of unobserved working hours and labor quality.

Building on our nonparametric identification result and considering computational ease, we propose an estimation procedure for the production function with random coefficients. Under the assumption of Gaussian error terms, we develop a penalized maximum likelihood estimator for a finite mixture model of random coefficient production functions, where the form of the likelihood function is motivated by our identification argument. The EM algorithm is employed to simplify the computational complexity of maximizing the log-likelihood function of the mixture model.


As an empirical application, we investigate the extent of production technology heterogeneity across plants using panel data from Japanese manufacturing plants between 1986 and 2010. We show that the ratios of intermediate cost to sales and intermediate cost to variable cost—averaging from 1986 to 2010 at the plant level—are substantially different across plants within narrowly defined industries. For example, the 90th-10th percentile differences in the intermediate cost shares in variable costs for concrete products and electric audio equipment are large, at 0.27 and 0.67, respectively.\footnote{\cite{ZHANG2019} finds considerable heterogeneity in the labor share in sales across firms in the Chinese steel industry.} Differences in input levels cannot explain these heterogeneities; that is, a considerable cross-plant variation in the ratio of intermediate cost to sales or variable costs remains after controlling for observable inputs, presenting evidence for persistent and substantial heterogeneity in production technologies across plants.

By employing a finite mixture of random coefficients production functions, we also find substantial differences in estimated input elasticities across latent types within narrowly defined industries. To understand the consequences of neglecting unobserved heterogeneity in input elasticities on productivity growth measurement, we adopt a specification with unobserved heterogeneity as the true model and calculate the bias in productivity growth measurement when using a misspecified production function model that omits unobserved heterogeneity. Our findings indicate that ignoring unobserved heterogeneity in input elasticities can result in substantial and systematic bias in estimated productivity growth, contingent on the heterogeneous parameter estimates and the direction of productivity changes. Additionally, our analysis reveals a significantly stronger correlation between estimated productivity and investment among high capital-intensive latent type firms compared to low capital-intensive type firms, implying that unobserved disparities in input elasticities are vital in plant-level investment decisions.

As first discussed by \cite{Marschak44}, ordinary least squares estimation of production functions is subject to simultaneity bias when firms make input decisions based on their productivity level \citep{Griliches98}. To address the simultaneity issue, \cite{Olley1996} and \cite{Levinsohn03} develop control function approaches, which have been widely applied in empirical studies \citep[see also][]{Wooldridge09, Ackerberg15}. Despite their popularity, the control function approach has faced potential identification issues as highlighted in the literature.  \cite{bond05} and \cite{Ackerberg15} discuss identification issues due to collinearity under two flexible inputs (i.e., material and labor) in Cobb-Douglas specification. Furthermore, \citet[][hereafter GNR]{gandhi20} contend that if the firm's decision follows a Markovian strategy,  the moment restriction utilized in the control function approach fails to provide sufficient restriction to identify flexible input elasticities due to a lack of instrumental power.

GNR exploit the first-order condition for flexible input under profit maximization and establish the identification of production functions without making any functional form assumptions. However, their result presumes that production technology is identical across plants, except for the Hicks-neutral productivity term. This paper extends the nonparametric identification approach of GNR to accommodate settings where production technologies exhibit unobserved heterogeneity across plants.

Several papers  employ the first-order condition as a restriction to identify heterogenous elasticities of flexible inputs under functional form assumptions \citep{VanBiesebroeck03, Li2017, doraszelski2018measuring, balat19, ZHANG2019}.\footnote{As \cite{Solow57} first illustrates, the flexible input elasticities are identified with their input revenue share under the Cobb-Douglas production functions.} \cite{doraszelski2018measuring} develop a framework to identify plant-level time-varying labor-augmenting productivity in addition to Hicks-neutral productivity under the constant elasticity of substitution (CES) production function, allowing for two-dimensional heterogeneity.  \cite{ZHANG2019} proposes an estimation method based on the CES production function that accounts for heterogeneity in capital, labor, and material-augmenting efficiency across firms.  \cite{Li2017} use the flexible input cost ratio to construct a control variable for latent technology to identify flexible inputs' elasticities while imposing the timing assumption suggested by \cite{Ackerberg15} to identify the labor and capital coefficients under the Cobb-Douglas specification. \cite{balat19} also consider the Cobb-Douglas production function with heterogeneity in the efficiency of using skilled and unskilled labor. \cite{Demirer20} extends the framework of \cite{doraszelski2018measuring} by relaxing the parametric assumption of the CES production function but assumes that the labor-augmenting technology is the only additional source of individual-level heterogeneity other than the Hicks neutral productivity.  \cite{Raval2023} demonstrates the importance of accommodating non-neutral productivity differences across firms when estimating markups using flexible inputs. \cite{dewitte2022} illustrate the significance of accounting for unobserved heterogeneity in productivity growth processes when analyzing export premia and the contributions of exporting firms to aggregate productivity.

These papers identify firm-specific input elasticities, factor-augmenting technologies, or productivity growth processes but impose parametric assumptions or limit the sources of heterogeneity. Our paper complements these studies by establishing nonparametric identification of heterogenous production functions and productivity growth processes without imposing any functional form assumptions or limiting sources of unobserved heterogeneity. 

\cite{Cheng21} extend the k-means clustering approach of \cite{Bonhomme15ecma} to multi-dimensional clustering in random coefficient production functions in a nonlinear GMM framework, building upon the dynamic panel approach  \citep{Arellanobond91, Blundell1998, Blundell2000}.  \cite{Cheng21} consider an asymptotic setting when  the time dimension $T\rightarrow \infty$ while our identification is based on $T$ being fixed. Our identification result with fixed $T$ is useful in empirical applications where  firm-level panel datasets have limited time dimensions.

Our paper also contributes to the literature on identifying dynamic panel data models with unobserved heterogeneity by relaxing the existing identification conditions. Specifically, Proposition 3 demonstrates that the mixing probabilities and the type-specific time-varying probability distributions across latent types can be identified from panel data with four periods under the Markov assumption and other regularity conditions. This result improves  upon the findings of  \cite{Kasahara09}, who established the identification of dynamic panel data models under the Markov assumption but imposed the stationarity  and required panel data with six periods.\footnote{\cite{HigginsJochmans21} point out that the type-specific distribution is identified  only up to an arbitrary ordering of the latent types that differs across different points  in \cite{Kasahara09}. Our  argument for identifying a common order of the latent types is based on that of \cite{HigginsJochmans21}. } \cite{Hu12} consider a non-stationary case and establish the identification of a continuous mixture dynamic panel data model using a panel dataset with five time periods. However, their result is limited, as they only establish the identification of type-specific distributions for the third to fifth periods, leaving the identification for the first two periods unresolved. In contrast, we identify the type-specific distributions across all four periods from panel data of length four. 

A key condition for our identification analysis is that the observed variable must follow a first-order Markov process within the subpopulation specified by latent type.
Proposition \ref{P-1} demonstrates that this Markov assumption is satisfied under our structural model assumptions, including the Markovian investment strategy (Assumption \ref{A-3}(b)) and the monotonicity of flexible input demands for productivity and wage shocks (Assumption \ref{A-4}(b)), both of which are standard assumptions in the production function literature \citep[e.g.,][]{Olley1996,Levinsohn03}. Another identifying condition is a rank condition in Assumption \ref{A-P-2}  which requires that the changes in the value of the observed vector, $\bs z_{t}$, must induce sufficiently different changes in the value of the type specific conditional density function of $\bs z_t$ given the past value $\bs z_{t-1}$ across latent types. As illustrated in the Cobb-Douglas example in Appendix \ref{appendix-6}, this condition is satisfied when input elasticities are sufficiently different across latent types.

The remainder of this paper is organized as follows. Section 2 presents evidence of heterogeneity in production technologies across plants, using panel data from Japanese manufacturing plants. Section 3 introduces the setup for our production function models and discusses the assumptions. Section 4 provides the main identification results, while Section 5 develops an estimator for the production function using a finite mixture model. In Section 6, we present empirical results based on the Japanese manufacturing plants. Section 7 concludes the paper.

\section{Evidence for production technology heterogeneity}\label{sec:evidence-hetero}


In order to underscore the significance of accounting for unobserved heterogeneity in production functions, we first present  a set of stylized facts that clearly indicate the presence of heterogeneity beyond Hicks-neutral technology components in production functions. Our analysis employs panel data derived from Japanese manufacturing facilities, spanning the years 1986 to 2010. A comprehensive discussion of the dataset can be found in Section \ref{data}.

For illustration, consider a plant with the Cobb-Douglas production technology:
\[
\log Y_{it} =\beta_0+  \beta^i_{m} \log M_{it}+  \beta^i_{\ell} \log L_{it} +  \beta^i_{k} \log K_{it} +\omega_{it},\smallskip
\]
where $Y_{it}$, $M_{it}$, $L_{it}$, and $K_{it}$ denote the output, intermediate input, labor, and capital of plant $i$ in year $t$, respectively, while $\omega_{it}$ represents the total factor productivity (TFP) that follows a first-order Markov process. The superscript $i$ in $\beta^i_m$, $\beta_\ell^i$, and $\beta^i_k$ signifies the variation in output elasticities of inputs across different plants.

We assume that firms consider their output and input prices as given, and that the intermediate input and labor are flexibly selected after $\omega_{it}$ has been fully observed. Consequently, a plant's profit maximization implies the following relationships:
\begin{equation}\label{input-share}
\beta^i_m = \frac{P_{M,t}M_{it}}{P_{Y,t} Y_{it}}\quad\text{and}\quad \frac{\beta^i_m}{\beta_m^i +\beta_\ell^i}= \frac{P_{M,t}M_{it}}{P_{M,t}M_{it}+W_{t}L_{it}},
\end{equation}
where $P_{Y,t}$, $P_{M,t}$, and $W_{t}$ represent the prices of output, intermediate input, and labor, respectively.

In most existing empirical studies, production functions are estimated under the assumption that the coefficients $\beta^i_m$, $\beta^i_\ell$, and $\beta^i_k$ do not vary across plants. This assumption can be tested in light of (\ref{input-share}) by examining whether the intermediate input share, $ \frac{P_{M,t}M_{it}}{P_{Y,t} Y_{it}}$, and the ratio of intermediate cost to variable cost (i.e., the sum of intermediate and labor costs), $\frac{P_{M,t}M_{it}}{P_{M,t}M_{it}+W_{t}L_{it}}$, remain constant across plants. To investigate this, we calculate the plant-level averages over the maximum 25-year period, during which a plant remained in the market between 1986 and 2010, as follows:
\[
\overline{ \left(\frac{P_{M,t}M_{it}}{P_{Y,t} Y_{it}}\right)}_i=\frac{1}{25}\sum_{t=1986}^{2010} \frac{P_{M,t}M_{it}}{P_{Y,t} Y_{it}}\quad\text{and}\quad \overline{\left(\frac{P_{M,t}M_{it}}{P_{M,t}M_{it}+W_{t}L_{it}}\right)}_i=\frac{1}{25}\sum_{t=1986}^{2010}\frac{P_{M,t}M_{it}}{P_{M,t}M_{it}+W_{t}L_{it}}.
\]
Subsequently, we analyze the extent of variation across plants within a narrowly defined industry.

\begin{figure}[tb]
     \caption{Histogram of $\overline{\left(\frac{P_{M,t}M_{it}}{P_{Y,t}Y_{it}}\right)}_{i}$}
     \begin{minipage}{.45\linewidth}
     \includegraphics[width=\linewidth]{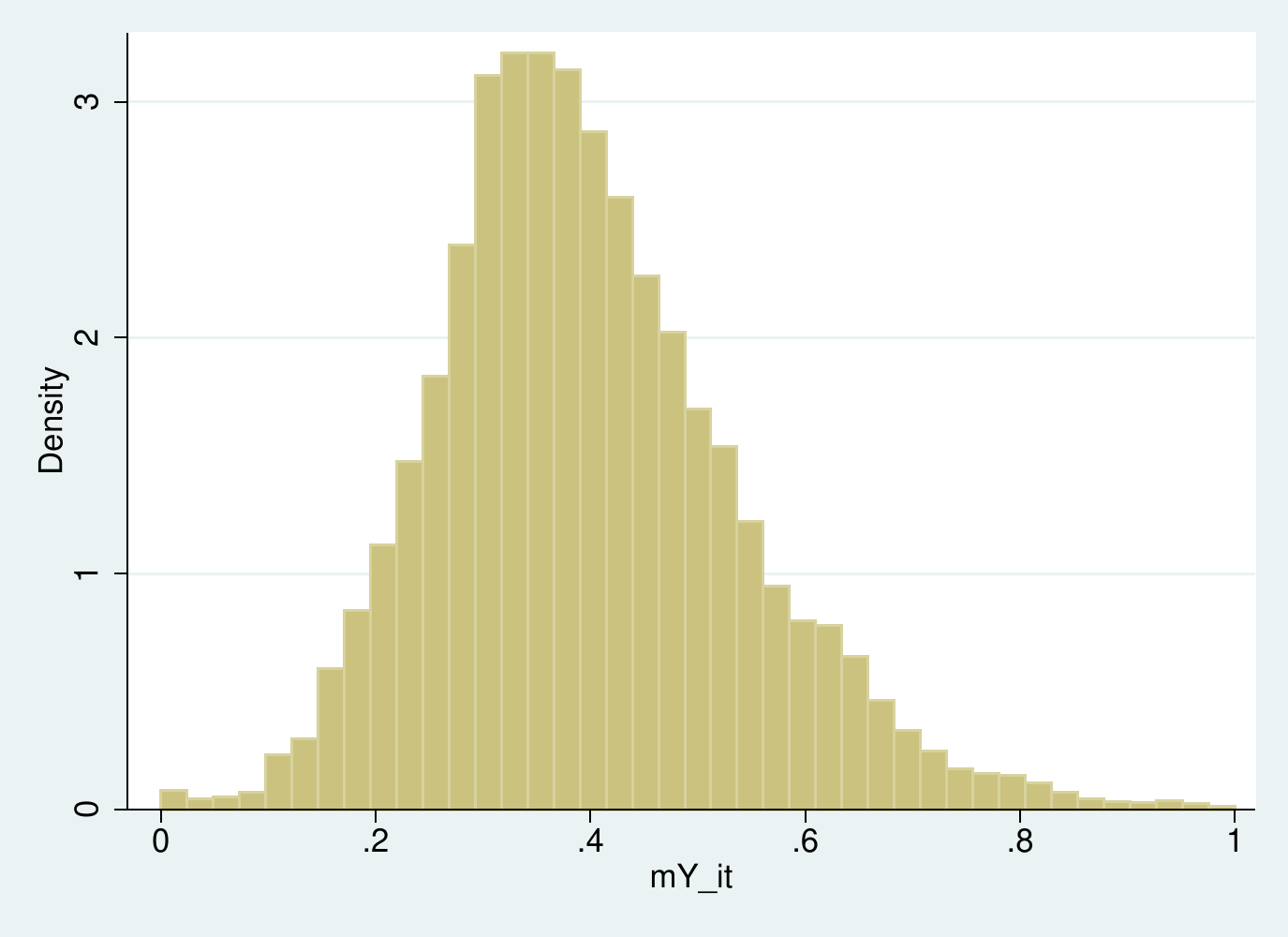}
     \subcaption{Concrete Product}\label{fig:hist-concrete}
     \end{minipage}
     \begin{minipage}{.45\linewidth}
     \includegraphics[width=\linewidth]{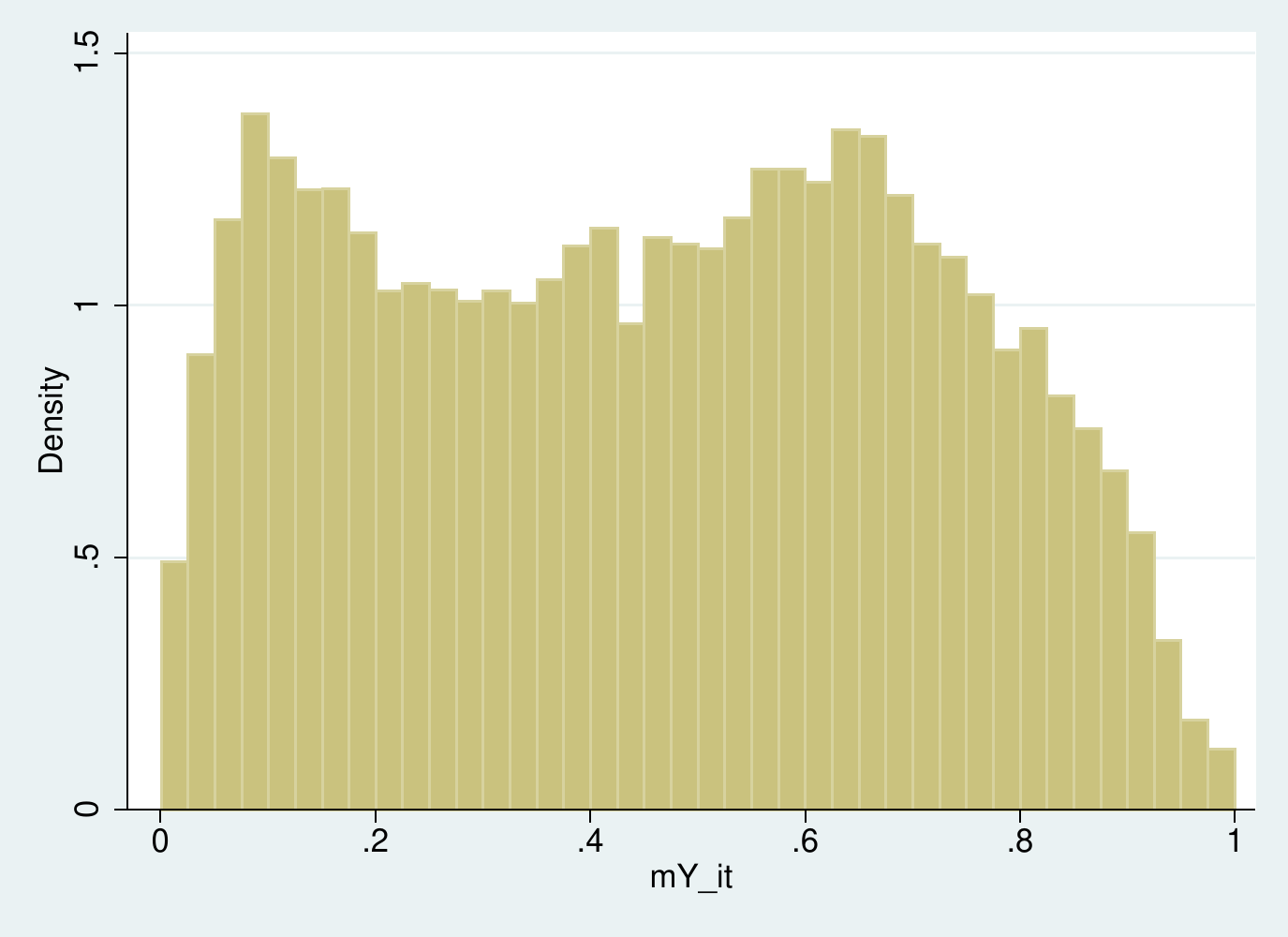}
     \subcaption{Electric Audio}\label{fig:hist-audio}
     \end{minipage}
\end{figure}

Figure \ref{fig:hist-concrete} and Figure \ref{fig:hist-audio} display histograms illustrating plant-level averages of intermediate input shares, $\overline{ \left(\frac{P_{M,t}M_{it}}{P_{Y,t} Y_{it}}\right)}_i$, for all plants within the concrete products and electric audio equipment industries, respectively. Both figures exhibit substantial variation in intermediate shares. The disparity between the 90th and 10th percentiles reaches up to 0.28 for concrete products, an industry typically regarded as having homogeneous technology.


\begin{figure}[tb]
     \caption{Histogram of $\overline{\left(\frac{P_{M,t}M_{it}}{P_{M,t}M_{it}+W_{t}L_{it}}\right)}_i$}
     \begin{minipage}{.45\linewidth}
     \includegraphics[width=\linewidth]{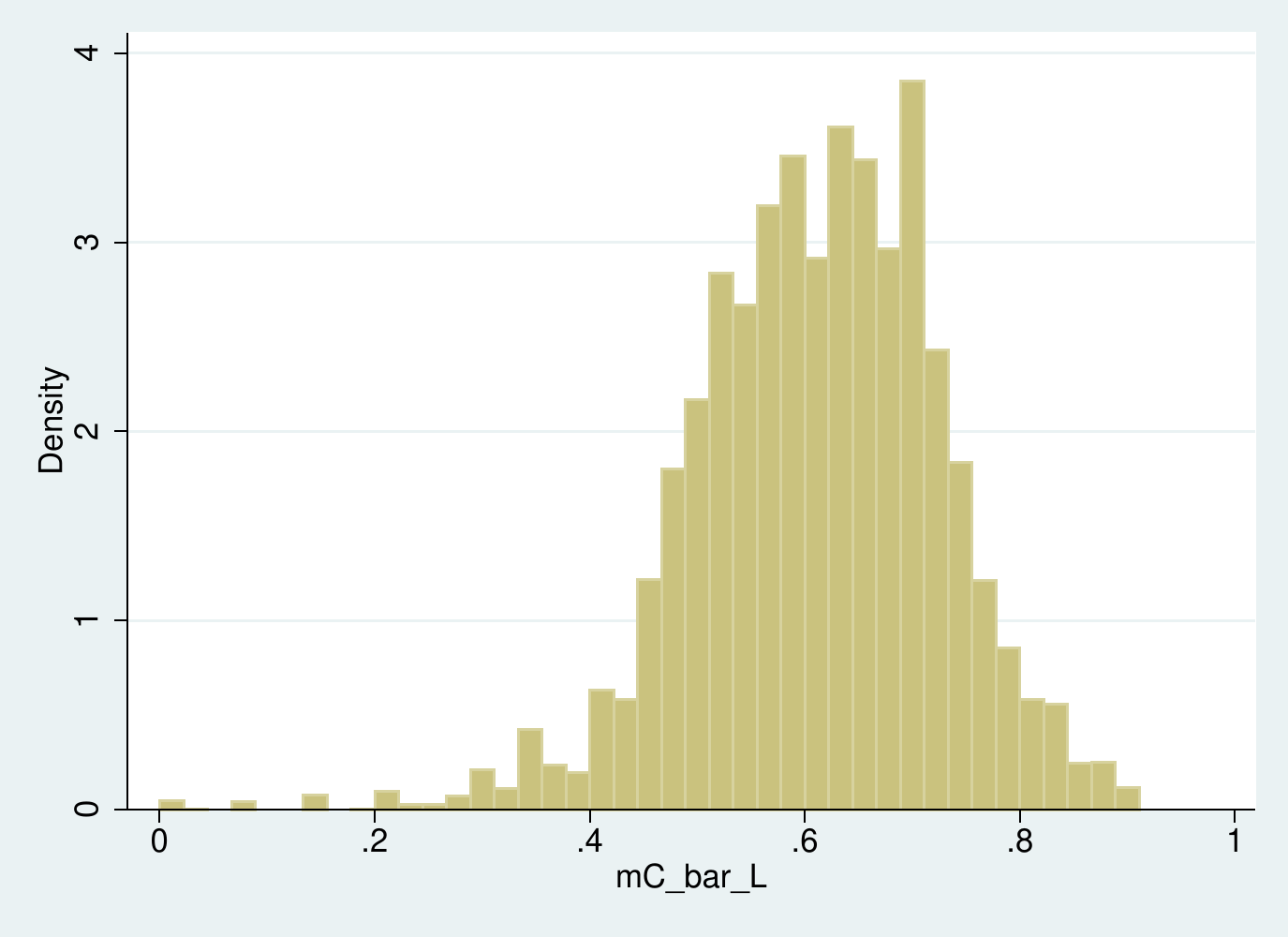}
     \subcaption{Concrete Product}\label{fig:hist-concrete_cost}
     \end{minipage}
     \begin{minipage}{.45\linewidth}
     \includegraphics[width=\linewidth]{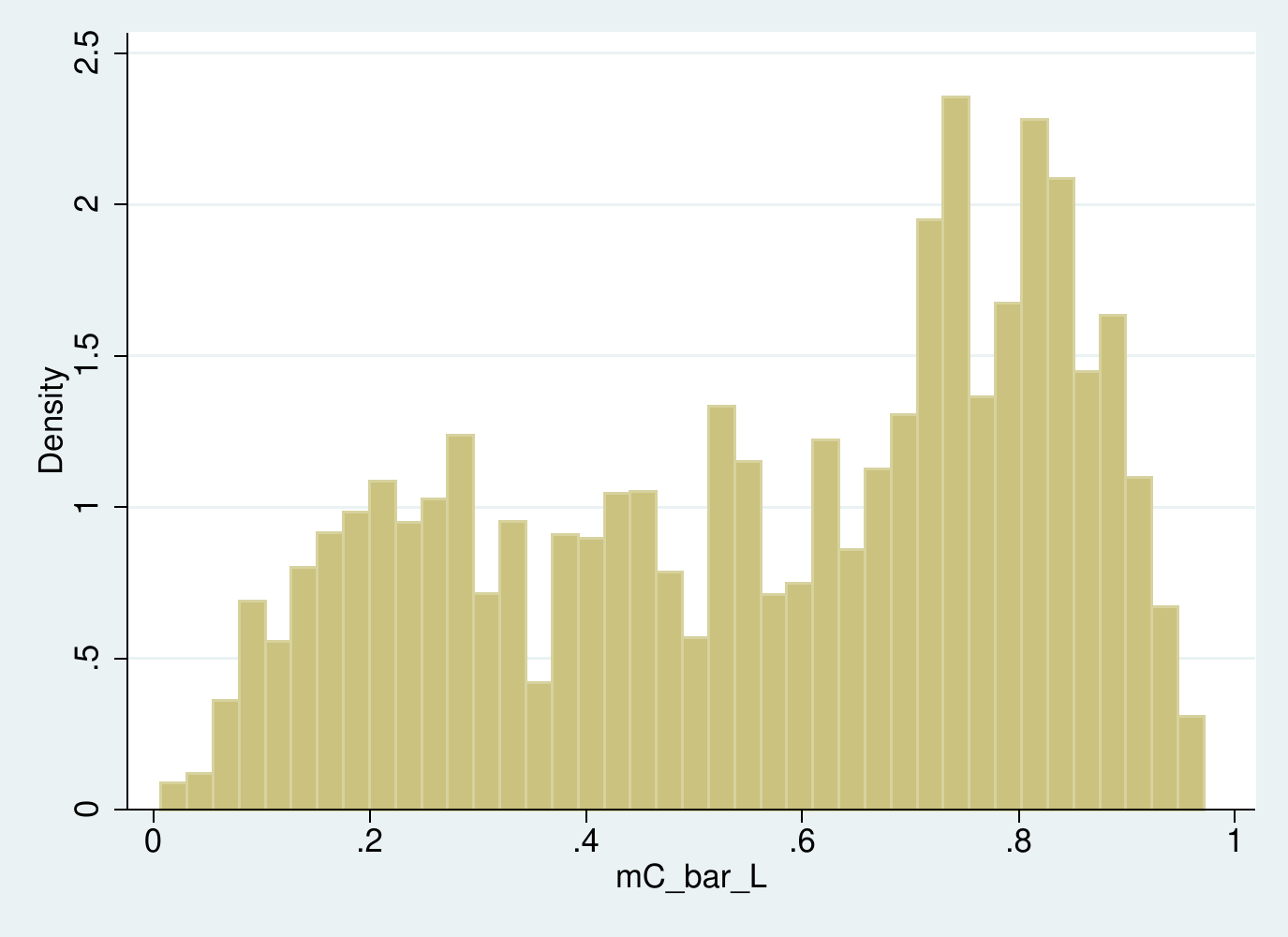}
     \subcaption{Electric Audio Equipment}\label{fig:hist-audio_cost}
     \end{minipage}
\end{figure}

The variation in intermediate shares between plants might be indicative of differences in markups; however, the ratio of intermediate costs to total variable costs is less susceptible to markup discrepancies. Figures \ref{fig:hist-concrete_cost} and \ref{fig:hist-audio_cost} depict histograms of plant-level averages of the ratio of intermediate costs to total variable costs, $\overline{\left(\frac{P_{M,t}M_{it}}{P_{M,t}M_{it}+W_{t}L_{it}}\right)}_i$, for the concrete products and electric audio equipment industries, respectively. The significant variation in intermediate cost shares implies that heterogeneous markups are not the primary explanation for the observed variation in intermediate input shares presented in Figures \ref{fig:hist-concrete} and \ref{fig:hist-audio}.

By comparing the degree of dispersion in input shares within the 2-digit industry classification with that within the 3-digit or 4-digit industry classification, we can examine the extent to which classifying industries at a more refined level helps control for heterogeneity in production technology.

\begin{figure}[tb]
     \caption{Histogram of $\overline{\left(\frac{P_{M,t}M_{it}}{P_{Y,t}Y_{it}}\right)}_{i}$}
     \begin{minipage}{.30\linewidth}
     \includegraphics[width=\linewidth]{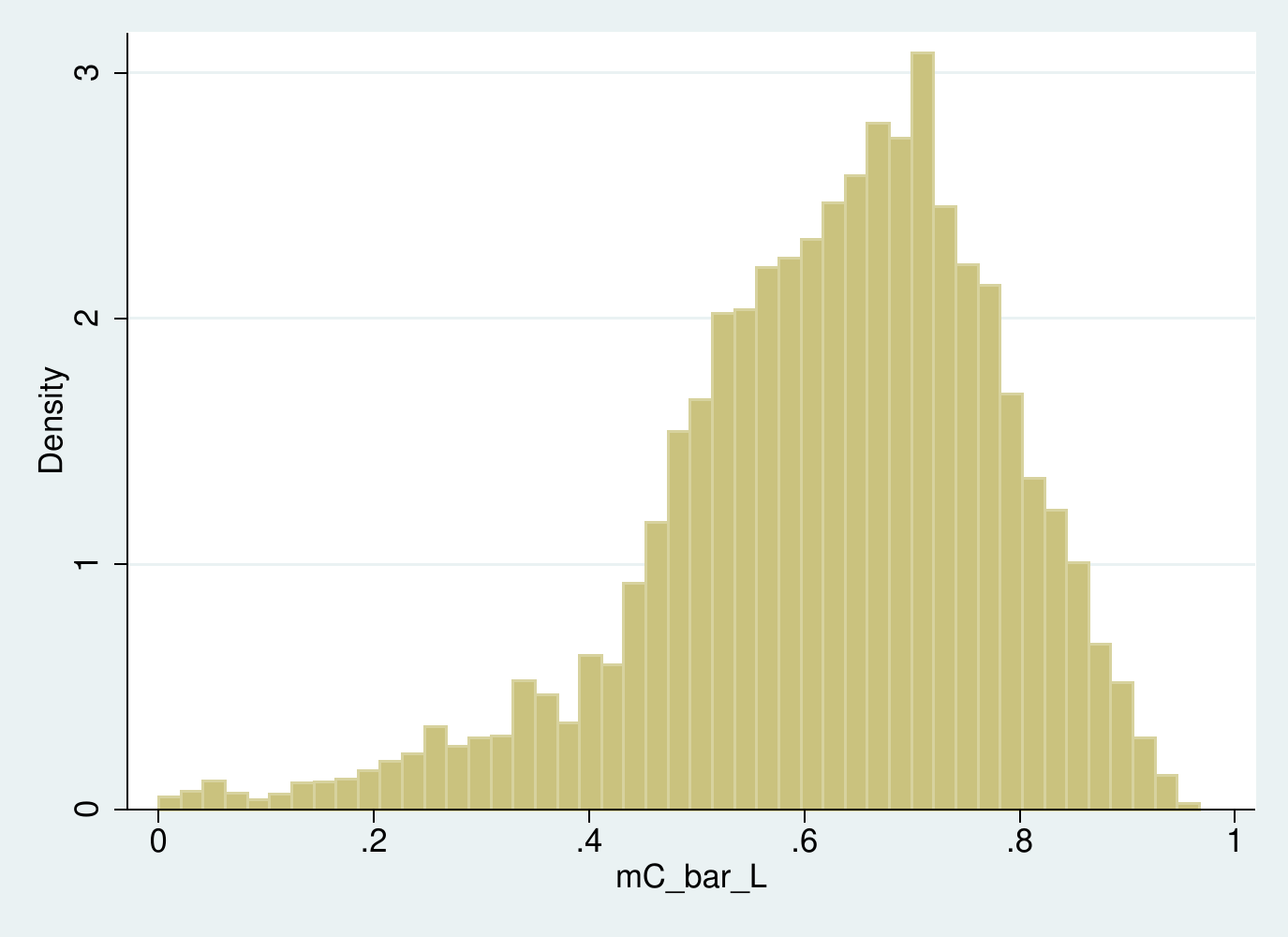}
     \subcaption{2-digit  Ceramics and Clay}\label{fig:hist-concrete-2dgt}
     \end{minipage}
     \begin{minipage}{.30\linewidth}
     \includegraphics[width=\linewidth]{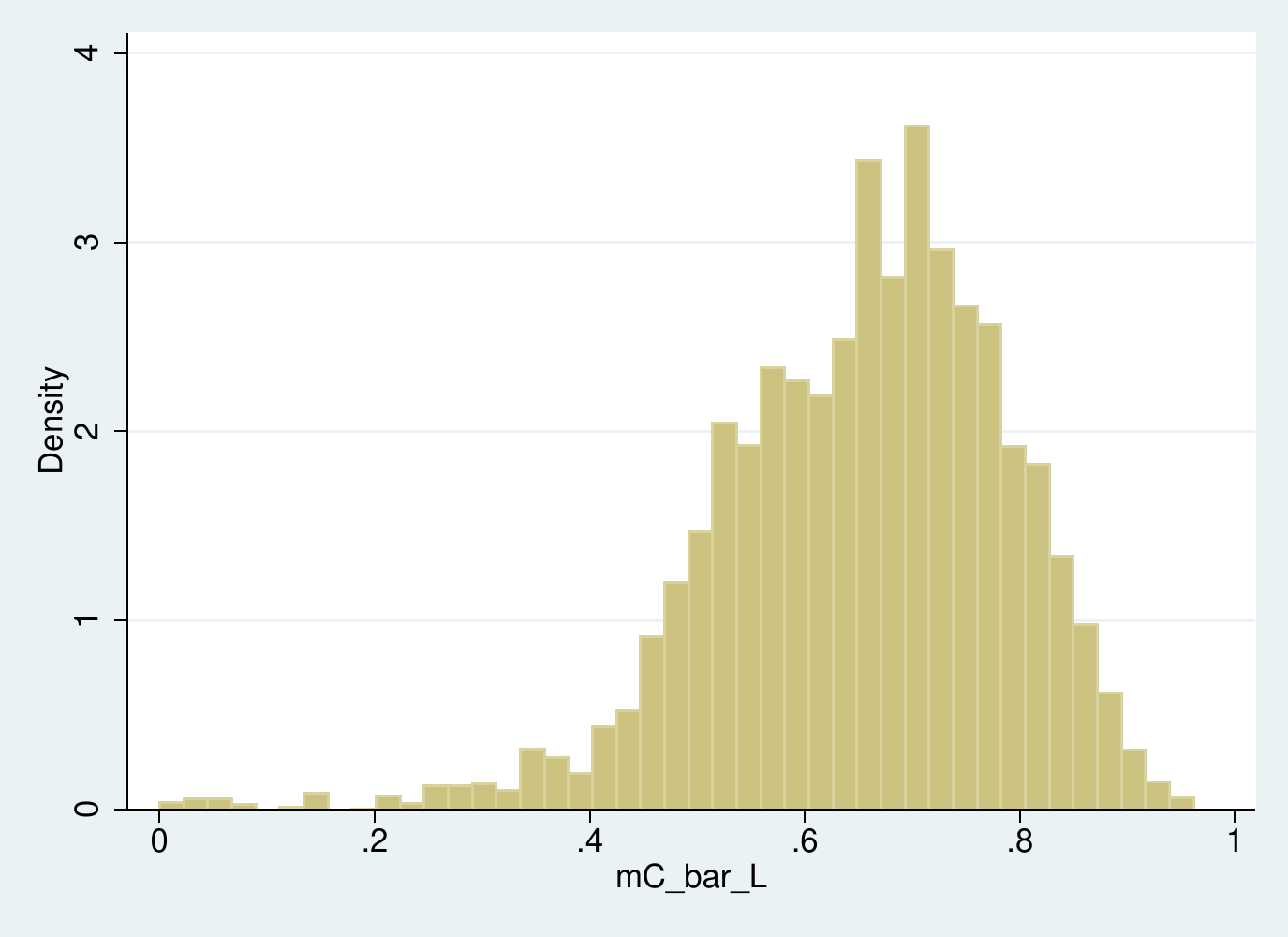}
     \subcaption{3-digit Cement  Product}\label{fig:hist-concrete-3dgt}
     \end{minipage}
     \begin{minipage}{.30\linewidth}
     \includegraphics[width=\linewidth]{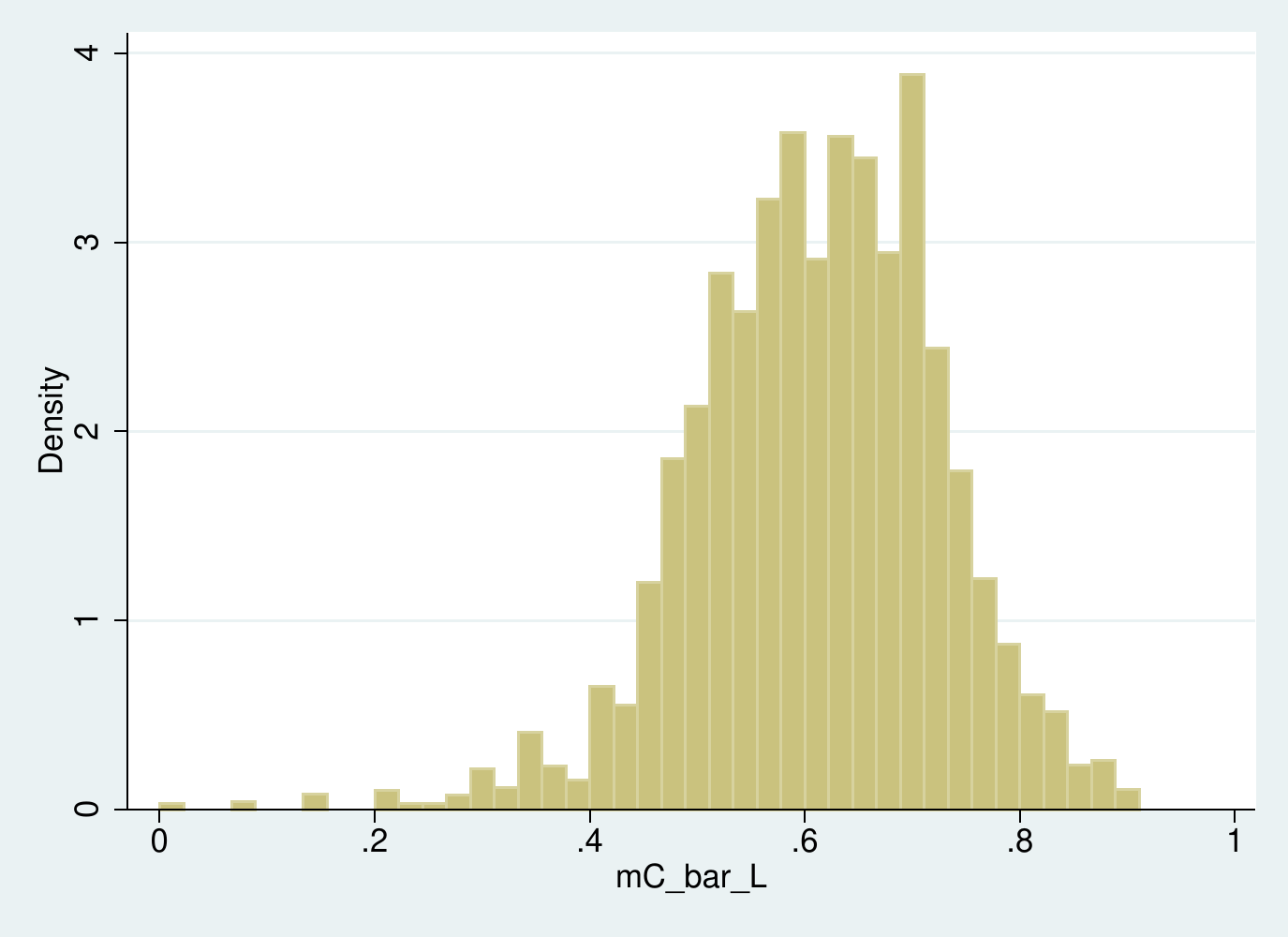}
     \subcaption{4-digit  Concrete Product}\label{fig:hist-concrete-4dgt}
     \end{minipage}
\end{figure}
\begin{figure}[tb]
     \caption{Histogram of $\overline{\left(\frac{P_{M,t}M_{it}}{P_{Y,t}Y_{it}}\right)}_{i}$}
     \begin{minipage}{.30\linewidth}
     \includegraphics[width=\linewidth]{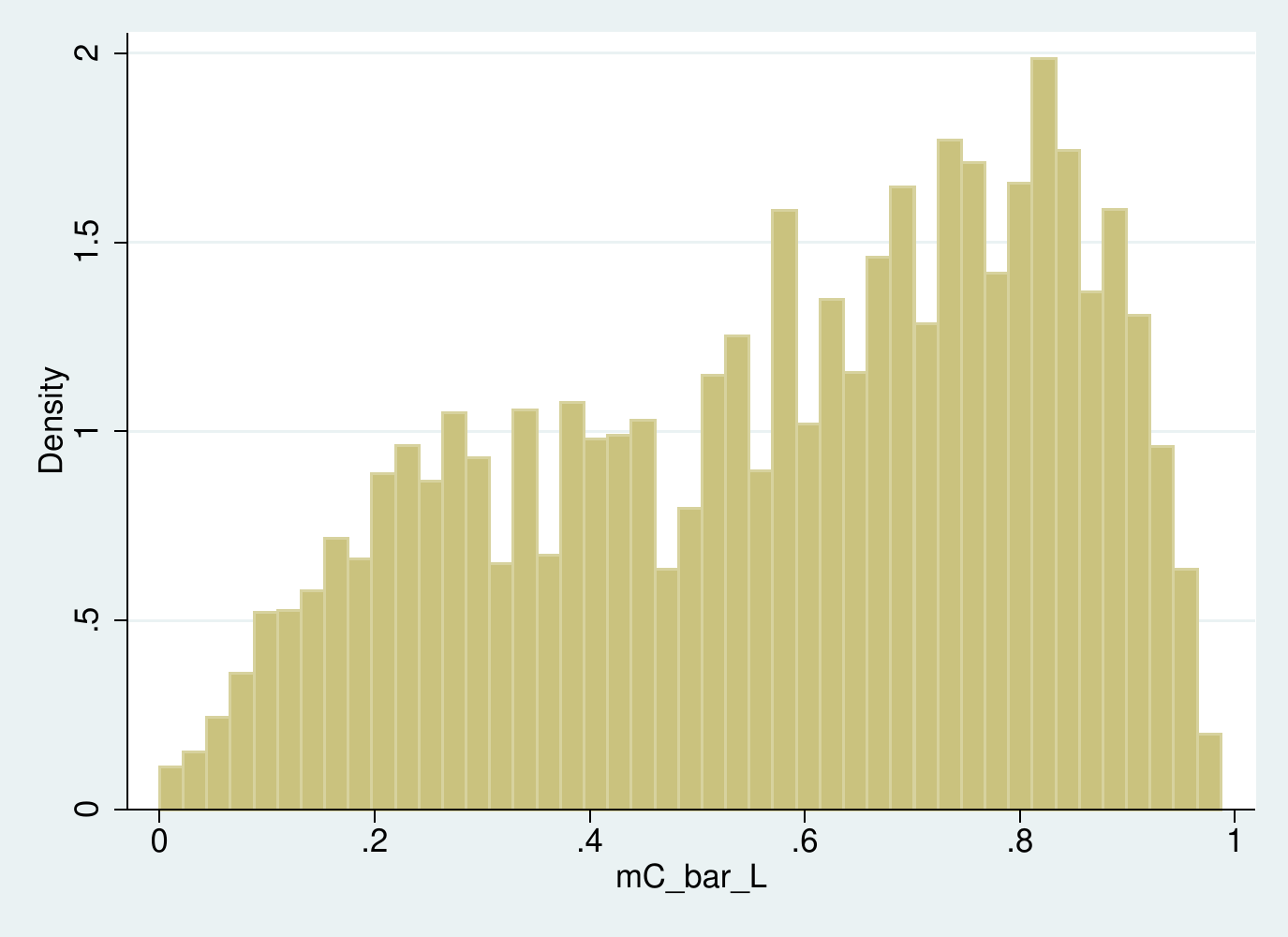}
     \subcaption{2-digit  Electric Parts, Device, Circuit}\label{fig:hist-audio-2dgt}
     \end{minipage}
     \begin{minipage}{.30\linewidth}
     \includegraphics[width=\linewidth]{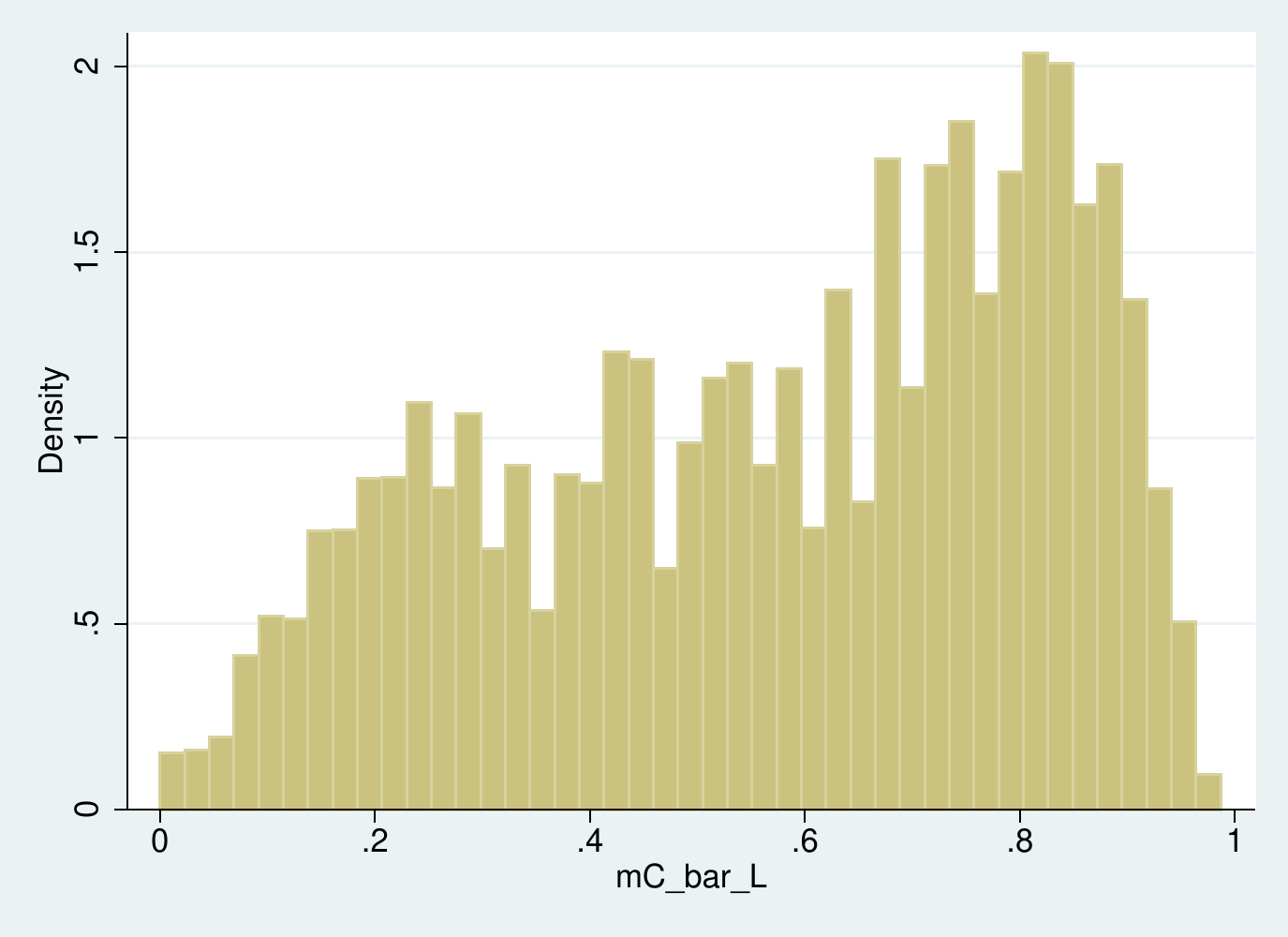}
     \subcaption{3-digit Electric Device}\label{fig:hist-audio-3dgt}
     \end{minipage}
     \begin{minipage}{.30\linewidth}
     \includegraphics[width=\linewidth]{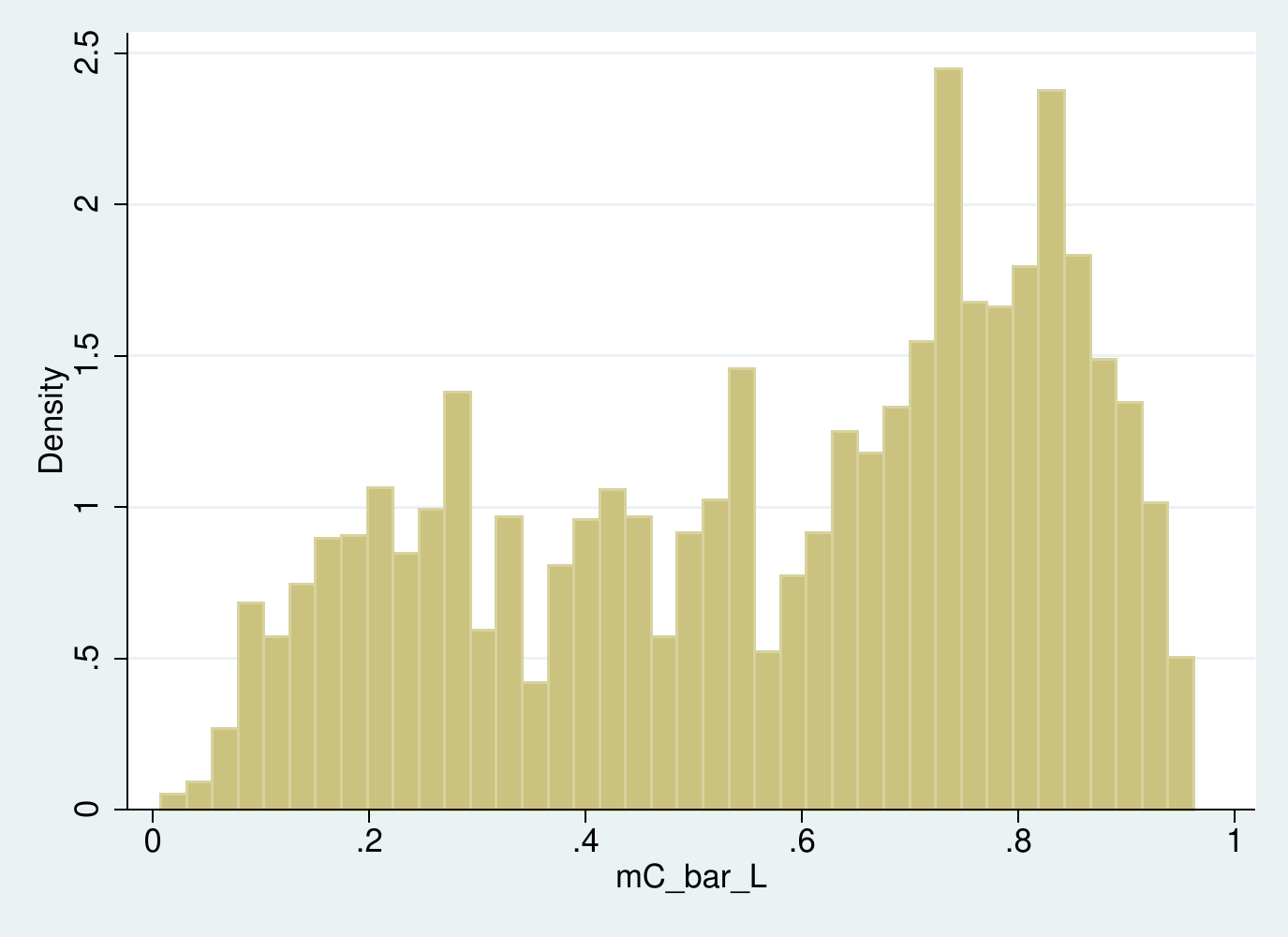}
     \subcaption{4-digit Electric Audio Equipment}\label{fig:hist-audio-4dgt}
     \end{minipage}
\end{figure}

Figures \ref{fig:hist-concrete-2dgt}, \ref{fig:hist-concrete-3dgt}, and \ref{fig:hist-concrete-4dgt} contrast histograms of plant-level averages of intermediate input shares for ceramics and clay (2-digit), cement products (3-digit), and concrete products (4-digit) industries. These figures suggest that while dispersion decreases somewhat from the 2-digit to the 4-digit level, the degree of heterogeneity remains notably high even at the 4-digit industry classification. Likewise, Figures \ref{fig:hist-audio-2dgt}, \ref{fig:hist-audio-3dgt}, and \ref{fig:hist-audio-4dgt} reveal that the dispersion of material shares does not decline considerably when transitioning from electric parts, devices, and circuits (2-digit) to electric devices (3-digit) and subsequently to electric audio equipment (4-digit).

As illustrated in Table \ref{tab:hetero-2-4dgt}, the 90th-10th percentile difference in intermediate input shares decreases from 0.38 (2-digit) to 0.28 (4-digit) for the ceramics and clay industry. In contrast, the 90th-10th percentile difference for the electric parts industry changes only marginally from 2-digit to 4-digit, ranging from 0.61 to 0.62. Similar patterns are observed for the plant-level averages of intermediate cost shares in variable costs. The 90th-10th percentile differences in intermediate cost shares for concrete products (4-digit) and  electric audio equipment (4-digit) are substantial, measuring 0.27 and 0.67, respectively. Therefore, classifying industries at a more refined level does not substantially reduce the heterogeneity in output elasticities with respect to intermediate and labor inputs.


\begin{table}[!htbp]
\caption{The 90th-10th Percentile Difference of Intermediate and Labor Cost Shares for  Concrete Product and Electric Audio} \label{tab:hetero-2-4dgt}
\centering
\begin{tabular}[t]{l|c|c|c }           \hline
 & No. of& 90-10 diff & 90-10 diff   in  \\
Industry Code :  Name &  Obs. & in $ \Big(\frac{PM_{it}}{PY_{it}}\Big)_i$ &  $ \Big(\frac{PM_{it}}{PM_{it}+WL_{it}}\Big)_i$\\ \hline
 22: \ \ \ Ceramics and Clay  	&	53,042	&	0.38 	&	0.38 	\\
 222: \  Cement Product 	&	22,834	&	0.35 	&	0.32 	\\
 2223: Concrete Product 	&	14,463	&	0.28 	&	0.27 	\\	\hline
28: \ Electric Parts/Devise/Circuit 	&	30,814	&	0.61 	&	0.66 	\\
281: \    Electric Device 	&	19,901	&	0.62 	&	0.66 	\\
2814: Electric Audio Equipment	&	11,325	&	0.62 	&	0.67 	\\
\hline
\end{tabular}
\end{table}

\begin{table}[!htbp]
\caption{The 90th-10th Percentile Difference of Intermediate Input Shares}  \label{tab:hetero-all-2-4dgt}
\centering
\begin{tabular}[t]{l|c|c|c |c}           \hline
Industry & No. of &  Ave. 90-10 diff    &  Ave. 90-10 diff & Ave. No. \\
Classifications  & Industries &    in ${\Big(\frac{PM_{it}}{PY_{it}}\Big)_i}$ &    in  ${\Big(\frac{PM_{it}}{PM_{it}+WL_{it}}\Big)_i}$& of   Obs. \\ \hline
2-digit	&	24	&	0.46 	&	0.44 	&	49,512	\\
&& (0.08) & (0.11) & \\\hline
3-digit	&	149	&	0.42 	&	0.39 	&	7,975	\\
&& (0.10) & (0.13) & \\\hline
4-digit	&	479	&	0.38 	&	0.35 	&	2,481	\\
&& (0.11) & (0.14) & \\
\hline
\end{tabular}
\begin{flushleft}
Notes: Standard deviations across industries are shown in parentheses.
\end{flushleft}
\end{table}

\begin{table}[!htbp]
\caption{The 90th-10th percentile ratio of Permanent Component of Intermediate Input Shares}  \label{tab:hetero-all-2-4dgt-perm}
\centering
\begin{tabular}[t]{l|c|c|c |c}           \hline
Industry & No. of &  Ave. 90-10 diff    &  Ave. 90-10 diff & Ave. No. \\
Classifications  & Industries &    in $\hat\xi_i$ for
$\frac{PM_{it}}{PY_{it}}$ &    in   $\hat\xi_i$ for
$\frac{PM_{it}}{PM_{it}+WL_{it}}$& of   Obs. \\ \hline
2-digit	&	24	&	0.30 	&	0.16 	&	49,512	\\
&& (0.05) & (0.02) & \\\hline
3-digit	&	149	&	0.28 	&	0.14 	&	7,975	\\
&& (0.05) & (0.03) & \\\hline
4-digit	&	479	&	0.26 	&	0.13 	&	2,481	\\
&& (0.06) & (0.04) & \\
\hline
\end{tabular}
\begin{flushleft}
Notes: Standard deviations across industries are shown in parentheses.
\end{flushleft}
\end{table}


Analogous patterns are observed across various industries. Table \ref{tab:hetero-all-2-4dgt} displays the average differences between the 90th and 10th percentiles for all industries, classified at the 2-digit, 3-digit, and 4-digit levels, with their corresponding standard deviations presented in parentheses. The findings reveal that dispersion decreases only marginally when refining industry classification from the 2-digit to the 4-digit levels. In general, substantial dispersion persists even at the 4-digit classification level, indicating that output elasticities of variable inputs exhibit variation across plants, even when adopting a more detailed industry classification.

The implications in (\ref{input-share}) are valid exclusively under the Cobb-Douglas production function. For a more generalized production function, the elasticities of output for inputs are dependent on the levels of material, labor, and capital inputs, even without heterogeneity in production technology. Consequently, we investigate whether the intermediate input cost-to-output value ratio remains similar across firms, even after adjusting for variations in capital, labor, and intermediate inputs. To achieve this, we regress $\frac{P_{M,t}M_{it}}{P_{Y,t}Y_{it}}$ or $\frac{P_{M,t}M_{it}}{P_{M,t}M_{it}+W_{t}L_{it}}$ on second-order polynomials of the natural logarithm of materials, the number of workers, and capital, resulting in residuals denoted by $e_{it}$. Subsequently, we compute the plant-level average $\hat\xi_i := (1/25)\sum_{t=1986}^{2010} e_{it}$ to assess production technology heterogeneity, conditional on inputs.

Table \ref{tab:hetero-all-2-4dgt-perm} presents the averages of the 90th-10th percentile differences in $\hat\xi_i$ for $\frac{P_{M,t}M_{it}}{P_{Y,t}Y_{it}}$ or $\frac{P_{M,t}M_{it}}{P_{M,t}M_{it}+W_{t}L_{it}}$ across all industries, classified at the 2-digit, 3-digit, and 4-digit levels. The findings reveal considerable variation in production technology after accounting for observable inputs, even at the 4-digit industry classification. This provides further evidence supporting the existence of heterogeneity in production technology.

\section{The Model}

We consolidate the notation as follows. Let $:=$ stand for "equal by definition". Bold letters denote vectors or matrices. For a continuous random variable $X$, a calligraphic letter $\mathcal{X}$ denotes its support, while its probability density function is represented by $g_{X}(x)$ for $x \in \mathcal{X}$.

Output, capital, intermediate inputs, labor input in effective units of labor, and total wage bills are denoted by $(Y_{it},K_{it},M_{it},L_{it},B_{it}) \in \mathcal{Y} \times \mathcal{K} \times \mathcal{M} \times \mathcal{L} \times \mathcal{B} \subset \mathbb{R}_{++}^5$, respectively, where $\mathcal{Y}$, $\mathcal{K}$, $\mathcal{M}$, $\mathcal{L}$, and $\mathcal{B}$ are the supports of the corresponding variables. We assume that $(Y_{it},K_{it},M_{it},L_{it},B_{it})$ are continuously distributed with strictly positive density on connected supports. We combine capital, intermediate, and labor inputs into a vector as ${\boldsymbol X}_{it} := (K_{it},M_{it},L_{it})' \in \mathcal{X} := \mathcal{K} \times \mathcal{M} \times \mathcal{L}$.

%

We allow firms' production technologies to differ beyond Hick's neutral productivity shocks. Specifically, we use a finite mixture specification to capture the unobserved heterogeneity in firms' production technologies as well as the process through which Hick's neutral productivity shocks evolve. We assume that there are $J$ unobserved types. Define the latent random variable $D_{i} \in\{1,2,...,J\}$ representing the type of firm $i$ such that $D_i=j$ if the production technology of firm $i$ is of the $j$-th type.
In the following, the superscript $j$ indicates that the functions are specific to technology type $j$, while the subscript $t$ indicates that the functions are specific to period $t$.


For the $j$-th type of production technology at time $t$, the output is related to inputs as follows:
\begin{equation}
Y_{it} = e^{\omega_{it} + \epsilon_{it}} F_t^j({\boldsymbol X}_{it}) \quad\text{with}\quad \epsilon_{it} \overset{iid}{\sim} g_{\epsilon_t}^j(\cdot), \label{prod}
\end{equation}
where $\epsilon_{it}$ is an idiosyncratic productivity shock with its density function $g_{\epsilon_t}^j(\cdot)$, and $\omega_{it}$ follows an exogenous first-order stationary Markov process given by:

\begin{align}
\omega_{it} &= h_t^j(\omega_{it-1}) + \eta_{it}, \quad\text{with}\quad \eta_{it} \overset{iid}{\sim} g_{\eta_t}^j(\cdot), \label{omega}
\end{align}
where $\eta_{it}$ is an innovation to the productivity process. As indicated by the subscript $t$ in $F_{t}^j(\cdot)$, $h_t^j(\cdot)$,  $g_{\epsilon_t}^j(\cdot)$, and $g_{\eta_t}^j(\cdot)$, production functions and productivity processes differ not only between latent types but also across periods. For example, this reflects type-specific aggregate shocks or type-specific biased technological changes.

 We assume that labor input in effective units of labor, $L_{it}$, is not directly observable because firms differ in their labor quality and working hours, while we only observe the number of workers. Instead, $L_{it}$ is related to the number of workers, denoted by $\widetilde L_{it}$, as
\begin{equation}\label{labor-input}
L_{it} = e^{\psi_t^j} \widetilde L_{it}.
\end{equation}
Equation (\ref{labor-input}) imposes a specific structure on how labor input in effective units of labor is related to the observed number of workers, where $\psi_t^j$ represents latent worker quality and working hours specific to type $j$. 
With this specification, measurement errors in observed labor input (i.e., the number of workers) are captured by the latent type-specific value of $\psi_t^j$.


The total wage bills, $B_{it}$, are related to the labor input in effective units as follows:
\begin{equation}\label{wb}
B_{it} = e^{v_{it}+\zeta_{it}}P_{L,t} L_{it},
\end{equation}
where $P_{L,t}$ represents the market wage. The random variable $v_{it}$ is a transitory wage shock known to firm $i$ at the time of choosing intermediate and labor inputs, while $\zeta_{it}$ is an idiosyncratic wage shock not included in the information set when firm $i$ selects these two inputs. Thus, $\zeta_{it} \in \boldsymbol{\mathcal{I}}_{it}$, but $v_{it} \notin \boldsymbol{\mathcal{I}}_{it}$.

We assume that firms make flexible choices regarding both $M_{it}$ and $L_{it}$ after observing their serially correlated productivity shock, $\omega_{it}$, but before observing $\epsilon_{it}$. In contrast, $K_{it}$ is predetermined at the end of the previous period, prior to the observation of the serially correlated productivity shock $\omega_{it}$. Denote the information available to a firm for making decisions on $M_{it}$ and $L_{it}$ by $\boldsymbol{\mathcal{I}}_{it}$.

We present the model assumptions. For continuous random variables $Z_{it}$ and $W_{it}$, we denote the probability density function and expectation conditional on $D_i=j$ as $g_{Z_t}^j(z_t) := g_{Z_t|D=j}(z_t|D_i=j)$ and $E^j[Z_{it}] := E[Z_{it}|D_i=j]$, respectively. Additionally, we denote the probability density function of $Z_{it}$ conditional on $D_i=j$ and $W_{it}=w_t$ as $g_{Z_t|W_t}^j(z_t|w_t)$. The unconditional probability density function of $Z_{it}$ is denoted by $g_{Z_t}(z_t)$.

\begin{assumption}\label{A-1}
(a) Each firm belongs to one of the $J$ types, where the population probability of belonging to type $j$ is given by $\pi^j=\text{Pr}(D_i=j)$, and $J$ is known to econometricians. (b) A firm knows its type, i.e., $D_i\in \bs{\mathcal{I}}_{it}$.
\end{assumption}

\begin{assumption}\label{A-2}
(a) $(v_{it},\omega_{it}) \in\bs{\mathcal{I}}_{it}$.
 (b)  $(\epsilon_{it},\zeta_{it})\not\in\bs{\mathcal{I}}_{it}$.
(c) For the $j$-th type,
$\eta_{it}$ and $v_{it}$ are  mean-zero i.i.d. continuous random variables on $\mathbb{R}$ with its probability density functions $g_{\eta_t}^j(\cdot)$ and $g_{v_t}^j(\cdot)$, respectively, while  $(\epsilon_{it},\zeta_{it})$ is a  mean-zero i.i.d. continuous random variable on $\mathbb{R}^2$ with the joint probability density function $g_{\epsilon_t,\zeta_t}^j(\cdot,\cdot)$. (d)
The unconditional mean of $\omega_{it}$ is zero, i.e.,  $E_t^j[\omega_{it}]=0$ for every $t$.
\end{assumption}

\begin{assumption}\label{A-3}
(a)  $K_{it}\in  \bs{\mathcal{I}}_{it}$ but  $K_{it}\not\in \bs{\mathcal{I}}_{it-1}$. (b) the conditional density function of $K_{it}$  given $\bs{\mathcal{I}}_{t-1}$ is type specific and only depends on  $K_{it-1}$ and $\omega_{it-1}$,  i.e., $g_{K_t|\bs{\mathcal{I}}_{t-1},D}(K_{it}|\bs{\mathcal{I}}_{i,t-1},D_i=j) = g_{K_t|K_{it-1},\omega_{t-1}}^j(K_{it}|K_{it-1},\omega_{it-1})$.  
\end{assumption}

\begin{assumption}\label{A-4}
(a)  $M_{it}$  and $L_{it}$ are  chosen at time $t$ by maximizing expected profit conditional on $\bs{\mathcal{I}}_{it}$ as
\begin{align}
(M_{it},L_{it})&=(\mathbb{M}_t^j(K_{it},\omega_{it},v_{it}),\mathbb{L}_t^j(K_{it},\omega_{it},v_{it}))\nonumber\\
&:  = \argmax_{(M,L)\in\mathcal{M}\times\mathcal{L}} P_{Y,t}E^j[e^{\epsilon_{it}}]e^{\omega_{it}}F_t^j(K_{it},L,M)-P_{M,t} M -E^j[ e^{\zeta_{it}}]e^{v_{it}} P_{L,t}  L,\label{max_LM}
\end{align}
where $(\mathbb{M}_t^j(K_{it},\omega_{it},v_{it}),\mathbb{L}_t^j(K_{it},\omega_{it},v_{it}))$ is a type-specific deterministic function of  $(K_{it}, \omega_{it},v_{it})$. (b) For any given $K_{it}\in\mathcal{K}$, $(\mathbb{M}_t^j(K_{it},\omega_{it},v_{it}),\mathbb{L}_t^j(K_{it},\omega_{it},v_{it}))$ is invertible with respect to  $(\omega_{it},v_{it})$ with probability one.  (c) For any given $K_{it}\in\mathcal{K}$, the function $F_{t}^t(K_{it},L_{it},M_{it})$  is continuously differentiable and strictly concave in $(L_{it}, M_{it})$.
\end{assumption}

\begin{assumption} \label{A-5}
(a)  A firm is a price taker. (b) The intermediate input price  $P_{M,t}$,  the output price $P_{Y,t}$, and the market wage $P_{L,t}$ at time $t$ are common across firms.
(c) $(P_{M,t},P_{Y,t},P_{L,t})\in \bs{\mathcal{I}}_{it} $ and $(P_{M,t},P_{Y,t})$ is known to an econometrician.   \end{assumption}

\begin{assumption} \label{A-6} Labor input in effective unit of labour $L_{it}$ is not directly observable but $L_{it}$ is related to the observed number of workers as in (\ref{labor-input}) with $\sum_{j=1}^J \pi^j e^{\psi_t^j}=1$
 \end{assumption}

 Under Assumptions \ref{A-1}-\ref{A-6}, the information set at the time of choosing $M_{it}$ and $L_{it}$  is given by $\bs{\mathcal{I}}_{it} =\{D_i,\omega_{it},v_{it}, K_{it}, P_{M,t},P_{Y,t}, P_{L,t}, \bs V_{it-1},\bs V_{it-2},...\}$ with $\bs V_{it}=\{\zeta_{it},\epsilon_{it},\omega_{it},v_{it}, K_{it}, P_{M,t},P_{Y,t},P_{L,t}\}$.


Assumption \ref{A-1}(a) presumes that the number of types is known. \cite{Kasahara09} and \cite{Kasahara2014} discuss nonparametric identification of a lower bound for the number of types. \cite{Kasahara2019} and \cite{HaoKasahara2022} develop a likelihood-based testing procedure for the number of types in multivariate and panel data normal mixture models. Assumption \ref{A-1}(b) assumes that a firm is aware of its type.

Assumption \ref{A-2}(a)(b) asserts that $(\omega_{it},v_{it})$ is known when $L_{it}$ and $M_{it}$ are chosen, while $(\epsilon_{it},\zeta_{it})$ is not known when $L_{it}$ and $M_{it}$ are chosen. The presence of the wage shock $v_{it}$ in (\ref{wb}) provides an additional source of variation for $L_{it}$ beyond $\omega_{it}$ and $K_{it}$; as a result, $L_{it}$ and $M_{it}$ are not collinear, avoiding the identification problem discussed by \cite{bond05} and \cite{Ackerberg15}. Assumption \ref{A-2}(c) introduces notation for the probability density function of $\eta_{it}$, $v_{it}$, $\epsilon_{it}$, and $\zeta_{it}$, allowing for correlation between $\epsilon_{it}$ and $\zeta_{it}$. Assumption \ref{A-2}(d) is a normalization assumption to identify the location of $F_{t}^j$.

%


Assumption \ref{A-3}(a) presumes that $K_{it}$ is determined at time $t-1$, implying that $(\eta_{it},\omega_{it},v_{it})$ is not known when $K_{it}$ is chosen. Assumption \ref{A-3}(b) can be explicitly derived from a dynamic model of investment decisions with convex/non-convex adjustment costs under the first-order Markov productivity process (\ref{omega}).

Assumption \ref{A-4}(a) introduces the demand function for $M_{it}$ and $L_{it}$, derived from the static profit maximization problem when $M_{it}$ and $L_{it}$ are flexibly chosen in each period. Assumption \ref{A-4}(b) holds when there is a one-to-one relationship between $(M_{it},L_{it})$ and $(\omega_{it},v_{it})$, except for a set of measure zero conditional on the value of $K_{it}$, and is satisfied in the case of a Cobb-Douglas function. Under Assumption \ref{A-4}(c), the first-order condition for the maximization problem (\ref{max_LM}) characterizes the optimal choice for $L_{it}$ and $M_{it}$.

%

Assumption \ref{A-5}(a) posits that the firm has no market power. Under Assumption \ref{A-5}(b), the intermediate input price $P_{M,t}$ cannot be used for instrumenting $M_{it}$. When intermediate prices are exogenous and heterogeneous across firms, the production function could be identified using the intermediate input prices as instruments \citep{doraszelski2018measuring}. In Assumption \ref{A-5}(c), an alternative approach assumes that a firm is subject to an idiosyncratic price shock $\xi_{it}$ such that, for example, $P_{Y,it}=\exp(\xi_{it})P_{Y,t}$ with $\xi_{it}\not\in \bs{\mathcal{I}}_{it}$, then $\xi_{it}$ plays a similar role to $\epsilon_{it}$. We may assume that $(P_{M,t},P_{Y,t})$ is not known to the econometrician by treating $P_{M,t}/P_{Y,t}$ as parameters to be estimated; in such a case, we can identify the production function up to scale.

Appendix \ref{CES} discusses an alternative assumption to Assumption \ref{A-5} when a firm produces differentiated products and faces a demand function with constant price elasticity.

Assumption \ref{A-6} suggests that the quality of workers or the average working hours per worker differ across types and periods, as captured by the parameter $\psi_t^j$, which leads to the systematic difference in the average wage of workers  latent types. The assumption that $\sum_{j=1}^J \pi^j e^{\psi_t^j}=1$ serves as a normalization for identification.

\section{Nonparametric identification}


Assume that we have panel data for firms $i=1,...,n$ over periods $t=1,...,T$ consisting of output, capital, intermediate inputs, the number of workers, and total wage bills, denoted by $(Y_{it},B_{it},K_{it},M_{it},\widetilde L_{it})\in \mathcal{Y}\times\mathcal{B}\times\mathcal{K}\times \mathcal{M}\times\tilde{\mathcal{L}}$, respectively. For brevity, define $\bs{\bs{\widetilde X}}:= (K_{it},M_{it},\widetilde L_{it}) \in \widetilde{ \mathcal{X}}:=\mathcal{K}\times \mathcal{M}\times\tilde{\mathcal{L}}$.
 Each firm's observation $\{Y_{it},B_{it},\widetilde {\bs X}_{it}\}_{t=1}^T$ is randomly sampled from a population distribution with a density function given by  $g_{\{Y_{t},B_{t},\widetilde {\bs X}_{t}\}_{t=1}^T}(\{Y_{t},B_{t},\widetilde {\bs X}_{t}\}_{t=1}^T)$.

Let $g_{\epsilon_t}(\epsilon):= \int_{\mathbb{R}}g_{\epsilon_t,\zeta_t}(\epsilon,\zeta)d\zeta$ and $g_{\zeta_t}(\zeta):= \int_{\mathbb{R}} g_{\epsilon_t,\zeta_t}(\epsilon,\zeta)d\epsilon$ be the probability density functions of $\epsilon$ and $\zeta$, respectively.  Under Assumptions \ref{A-1},  \ref{A-2}, \ref{A-3}(a), \ref{A-4}(a), and \ref{A-5}, the first order conditions with respect to  $M_{it}$ and $L_{it}$ for maximizing the expected static profit in Assumption \ref{A-4}(a) are given by:
\begin{equation}
 P_{Y,t} F_{M,t}^j({\bs X}_{it}) E_t^j[e^{\epsilon}]e^{\omega_{it} }= P_{M,t},\quad P_{Y,t}    F_{L,t}^j({\bs X}_{it}) E_t^j[e^{\epsilon}] e^{\omega_{it}   }  =  E_t^j[e^{\zeta}]e^{v_{it}}P_{L,t},\label{foc}
\end{equation}
where  $F_{M,t}^j(\bs X):=\frac{\partial F_{t}^j(X) }{\partial M}$, $F_{L,t}^j(X):=\frac{\partial F_{t}^j(\bs X) }{\partial L}$,   $E_t^j[e^{\epsilon}] :=\int e^{\epsilon} dg^j_{\epsilon_t}(\epsilon)d\epsilon$,   and  $E_t^j[e^{\zeta}] :=\int e^{\zeta} dg^j_{\zeta_t}(\zeta)d\zeta$.  Rearranging equations (\ref{prod}), (\ref{omega}), (\ref{wb}), and (\ref{foc}) gives a system of equations:
\begin{equation}\label{system}
\begin{aligned}
\ln Y_{it} &= \ln F_t^j({\bs X}_{it}) + \omega_{it}+ \epsilon_{it}\quad\text{with}\quad \omega_{it}=h^j(\omega_{it-1}) + \eta_{it},\\
\ln S_{it}^m & =  \ln \left(\Gamma_{M,t}^j({\bs X}_{it})E_t^j[e^{\epsilon}]\right)-\epsilon_{it},\\
\ln S_{it}^\ell-\ln S_{it}^m  & =  \ln \left(\frac{\Gamma_{L,t}^j({\bs X}_{it})}{\Gamma_{M,t}^j({\bs X}_{it})E_t^j[e^{\zeta}] }\right)   +\zeta_{it}, \\
\ln P_{M,t}M_{it}  &=   \ln  \left(\frac{P_{L,t}\tilde L_{it}\Gamma_{M,t}^j({\bs X}_{it})E_t^j[e^{\zeta}]}{\Gamma_{L,t}^j({\bs X}_{it}) }\right)  +\psi_t^j + v_{it},
\end{aligned}
\end{equation}
where
\begin{align*}
&S_{it}^m : =\frac{P_{M,t}M_{it}}{P_{Y,t} Y_{it}},\  S_{it}^\ell : =\frac{B_{it}}{P_{Y,t} Y_{it}}, \  \Gamma_{M,t}^j({\bs X}_{it}):=\frac{F_{M,t}^j({\bs X}_{it}) M_{it}}{F_{t}^j({\bs X}_{it})},\   \text{and}\ \Gamma_{L,t}^j({\bs X}_{it}):=\frac{F_{L,t}^j({\bs X}_{it})L_{it}}{F_{t}^j({\bs X}_{it})}.
\end{align*}


For notational brevity, we drop the subscript $i$ in the rest of this section. Because $Y_t=\frac{P_{M,t}M_t}{S^m_t P_{Y,t}}$ and $B_t=\frac{S^\ell_t P_{M,t}M_t}{S^m_t}$, there exists a one-to-one relationship between $(Y_t,B_t)$ and $(S_t^m,S_t^\ell)$ given $\widetilde {\bs X}_t$ under Assumption \ref{A-5}. Therefore,  denoting $\bs S_t=(S_t^m,S_t^\ell)\in \mathcal{S}$, we consider $\{\bs S_t,\widetilde {\bs X}_t\}_{t=1}^T$ in place of $\{Y_{t},B_{t},\widetilde {\bs X}_{t}\}_{t=1}^T$ as our data.

Let $\bs Z_t := (\bs S_t,\widetilde {\bs X}_t)\in \mathcal{Z}:= \mathcal{S}\times\widetilde{\mathcal{X}}$. We assume that the population density function, denoted by $g_{\bs Z_1,...,\bs Z_T}(\{\bs z_{t}\}_{t=1}^T)$, is directly identified from the data. We are interested in identifying the model structure
\[
\bs{\theta} :=
\left\{ \pi^j,   \{g_{v_t}^j(\cdot), g_{\epsilon_t,\zeta_t}^j(\cdot), ,\Gamma_{M,t}^j(\cdot),\Gamma_{L,t}^j(\cdot),P_{L,t},\psi_t^j\}_{t=1}^T,\{F_t^j(\cdot)\}_{t=2}^T,\{h_t^j(\cdot),g_{\eta_t}^j(\cdot)\}_{t=3}^T\right\}_{j=1}^J
\]
from the population density function  $\text{g}_{\bs Z_1,...,\bs Z_T}(\{\bs z_{t}\}_{t=1}^T)$ given a set of restrictions in (\ref{system}) under Assumptions \ref{A-1}-\ref{A-6}.  

We first establish the nonparametric identification of model structure $\bs\theta$ when $J=1$ as follows.

\begin{proposition}\label{P-0}
Suppose that  $J=1$ and  Assumption  \ref{A-1}-\ref{A-6} holds with  $T\geq 3$.   Then,  $\bs\theta$ is uniquely determined from the population density function  $\text{g}_{\bs Z_1,...,\bs Z_T}(\{\bs z_{t}\}_{t=1}^T)$.
\end{proposition}

\begin{remark}\label{J1}
Proposition \ref{P-0} extends the identification result of GNR to the setting where $L_{it}$ is contemporaneously determined rather than predetermined.
\end{remark}



When $J\geq 2$, the probability density function of ${\{\bs Z_t\}_{t=1}^T}$ follows an $J$-term mixture distribution
\begin{equation}
\begin{aligned}
g_{\bs Z_1,...,\bs Z_T}(\{\bs z_t\}_{t=1}^T)& = \sum_{j=1}^J \pi^j g_{\bs Z_1}^j(\bs z_1)\prod_{t=2}^Tg_{\bs Z_t|\bs Z_{t-1},...,\bs Z_1}^j(\bs z_t|\bs z_{t-1},...,\bs z_{1}).
\end{aligned} \label{mixture-s}
\end{equation}
The number of type $J$ is defined to be the smallest integer $J$ such
that the density function of ${\{\bs Z_t\}_{t=1}^T}$ admits the representation (\ref{mixture-s}).

\begin{proposition}\label{P-1}
Suppose that Assumptions \ref{A-1}-\ref{A-6} hold. Then, the probability density function of ${\{\bs Z_t\}_{t=1}^T}$ defined in (\ref{mixture-s}) can be written as
\begin{align}
g_{\bs Z_1,...,\bs Z_T}(\{\bs z_t\}_{t=1}^T) & =  \sum_{j=1}^J  \pi^j g_{{\bs Z}_1}^j(\bs z_1) \prod_{t=2}^T g_{{\bs Z}_t|{\bs Z}_{t-1}}^j({\bs z}_t|{\bs z}_{t-1})\label{mixture-2}\\
&=\sum_{j=1}^J \pi^j \left(g_{\bs S_1|\widetilde {\bs X}_1}^j(\bs s_1|\widetilde {\bs x}_1)\prod_{t=2}^Tg_{\bs S_t|\widetilde {\bs X}_t}^j(\bs s_t|\widetilde {\bs x}_{t})\right) \times \left(
g_{\widetilde {\bs X}_1}^j(\widetilde {\bs x}_1)
\prod_{t=2}^Tg_{\widetilde {\bs X}_t|\widetilde {\bs X}_{t-1}}^j(\widetilde {\bs x}_t|\widetilde {\bs x}_{t-1})\right),\label{mixture-s2}
\end{align}
where $g_{\bs S_t|\widetilde {\bs X}_t}^j(\bs s_t|\widetilde {\bs x}_t)$ is the type $j$'s conditional probability density function of $\bs S_t$ given $\widetilde {\bs X}_t=\widetilde {\bs x}_t$, $g_{\widetilde {\bs X}_1}^j(\widetilde {\bs x}_1)$ is the type $j$'s marginal probability density function of $\widetilde {\bs X}_1$, $g_{{\bs Z}_1}^j(\bs z_1) =g_{\widetilde {\bs X}_1}^j(\widetilde {\bs x}_1)g_{\bs S_1|\widetilde {\bs X}_1}^j(\bs s_t|\widetilde {\bs x}_1)$, and $g_{{\bs Z}_t|{\bs Z}_{t-1}}^j({\bs z}_t|{\bs z}_{t-1})=g_{\bs S_t|\widetilde {\bs X}_t}^j(\bs s_t|\widetilde {\bs x}_{t}) g_{\widetilde {\bs X}_t|\widetilde {\bs X}_{t-1}}^j(\widetilde {\bs x}_t|\widetilde {\bs x}_{t-1})$.
\end{proposition}
Therefore,  under the stated model assumption, ${\{\bs Z_t\}_{t=1}^T}$ follows a first order Markov process within subpopulation specified by type.
The result of Proposition \ref{P-1} allows us to establish the nonparametric identification of $\{\pi^{j},g_{\bs Z_1}^{j}(\bs z_1), g_{\bs Z_2|\bs Z_1}^{j}(\bs z_2|\bs z_1), ... , g_{\bs Z_T|\bs Z_{T-1},...,\bs Z_1}^{j}(\bs z_T|\bs z_{T-1}) \}_{j=1}^J$ by extending the argument in \cite{Kasahara09}, \cite{carolletal10jns},  and \cite{Hu12}.  

We now establish identification when $T=4$.  Define 
\begin{equation}
\begin{aligned}
&{\bs L}_{{\bs z}_3}:=\left[
\begin{array}{ccc}
1& \cdots &1\\
\lambda^1_{4}(\bs b_1|{\bs z}_3) & \cdots &\lambda^J_{4}(\bs b_1|{\bs z}_3)\\
\vdots & \ddots &  \vdots \\
\lambda^1_{4}(\bs b_{J-1}|{\bs z}_3) & \cdots &\lambda^J_{4}(\bs b_{J-1}|{\bs z}_3)
\end{array}
\right],\  
\bar {\bs L}_{{\bs z}_2}:=\left[
\begin{array}{ccc}
\bar\lambda^1_{2}(\bs a_1,{\bs z}_2)&\cdots&\bar\lambda^J_{2}(\bs a_1,{\bs z}_2)\\
\vdots & \ddots & \dots  \\
\bar\lambda^1_{2}(\bs a_{J},{\bs z}_2)&\cdots&\bar\lambda^J_{2}(\bs a_{J},{\bs z}_2)\\
\end{array}
\right], \text{ and}\\
&{\bs D}_{{\bs z}_3|{\bs z}_2} := \text{diag}\left(\lambda^1_3({\bs z}_3|{\bs z}_2),...,\lambda^J_3({\bs z}_3|{\bs z}_2)\right),
\end{aligned} \label{LL}
\end{equation}
where $\bar \lambda^j_2(\bs a,{\bs z}_2):=\pi^j g_{{\bs Z}_2|{\bs Z}_1}^j({\bs z}_2|\bs a)g_{{\bs Z}_1}^j(\bs a)$, $\lambda^j_3({\bs z}_3|{\bs z}_2):= g_{{\bs Z}_3|{\bs Z}_2}^j({\bs z}_3|{\bs z}_2)$, and  $\lambda^j_{4}(\bs b|{\bs z}_3):=g_{{\bs Z}_4|{\bs Z}_3}^j(\bs b|{\bs z}_3)$.

\begin{assumption}\label{A-P-2} There exists a value $\bs z_3^*$ that satisfies the following condition: for every $\bs z_3\in\mathcal{Z}_3$, we can find $(\bar {\bs z}_2,\check{\bs z}_2,\bar {\bs z}_3)\in \mathcal{Z}_2\times  \mathcal{Z}_2\times \mathcal{Z}_3$, $(\bs a_1,...,\bs a_{J})\in \mathcal{Z}_1^{J}$ and $(\bs b_1,...,\bs b_{J-1})\in  \mathcal{Z}_4^{J-1}$ such that (a)  $\bs L_{{\bs z}_3^*}$, $\bs L_{{\bs z}_3}$, $\bs L_{\bar {\bs z}_3}$, $\bar {\bs  L}_{\check{\bs z}_2}$, and $\bar {\bs L}_{\bar {\bs z}_2}$ are non-singular, and (b) all the diagonal elements of ${\bs D}_{\bs z_3,\bs z_3}:= {\bs D}_{{\bs z}_3|\check{\bs z}_2} {\bs D}_{\bar {\bs z}_3|\check{\bs z}_2} ^{-1} {\bs D}_{\bar {\bs z}_3|\bar {\bs z}_2}{\bs D}_{{\bs z}_3|\bar {\bs z}_2}^{-1}$   take  distinct values. Furthermore,
(c)  for every $(\bs z_2,\bs z_3)\in\mathcal{Z}_2\times \mathcal{Z}_3$,  $g^j_{{\bs Z}_3|{\bs Z}_2}({\bs z}_3|{\bs z}_2)> 0$ for $j=1,...,J$.
\end{assumption}

\begin{proposition}\label{P-2}
Suppose that    Assumptions \ref{A-1}-\ref{A-P-2} hold and $T\geq 4$.   Then,  \\ $\{\pi^j, g_{{\bs Z}_1}^j({\bs z}_1),g^j_{{\bs Z}_2|{\bs Z}_1}({\bs z}_2|{\bs z}_1),...,  g_{{\bs Z}_T|{\bs Z}_{T-1}}^j({\bs z}_T|{\bs z}_{T-1})\}_{j=1}^J$ is uniquely determined from $g_{{\bs Z}_1,...,{\bs Z}_T}(\{{\bs z}_{t}\}_{t=1}^T)$ up to a common permutation of the latent types.
\end{proposition}

\begin{remark}\label{r-ks}
Under the additional assumption of stationarity, i.e., the conditional density function $g_{{\bs Z}_t|{\bs Z}_{t-1}}^j({\bs z}_{t}|{\bs z}_{t-1})$ does not depend on $t$ for $t=2,...,T$, \cite{Kasahara09} establish the nonparametric identification of the model (\ref{mixture-s2}) when $T=6$ while \cite{Hu12} show that $T=4$ suffices for identification.
\end{remark}
\begin{remark}\label{r-hu-shum}Considering serially correlated continuous unobserved variables $\{X_t^*\}_{t=1}^T$, \cite{Hu12} analyze the nonparametric identification of the model
\[
g_{{\bs Z}_1,...,{\bs Z}_T}(\{{\bs z}_{t}\}_{t=1}^T)= \int g_{{\bs Z}_1|X_1^*}({\bs z}_{1},x_{1}^*)\prod_{t=2}^Tg_{{\bs Z}_t,X_t^*|{\bs Z}_{t-1},X_{t-1}^*}({\bs z}_{t},x_{t}^*|{\bs z}_{t-1},x_{t-1}^*)d(\{x_t^*\}_{t=1}^T).
\]
Given panel data ${\{\bs {\bs Z}_t\}_{t=1}^T}$ with $T=5$, Theorem 1 and Corollary 1 of \cite{Hu12} state  that, under their Assumptions 1-4, $g_{{\bs Z}_3,X_3^*}({\bs z}_{3},x_{3}^*)$, $g_{{\bs Z}_4,X_4^*|{\bs Z}_3,X_3^*}({\bs z}_{4},x_{4}^*|{\bs z}_{3},x_{3}^*)$,  and\\ $g_{{\bs Z}_5,X_5^*|{\bs Z}_4,X_4^*}({\bs z}_{5},x_{5}^*|{\bs z}_{4},x_{4}^*)$ are nonparametrically identified, but the identification of $g_{{\bs Z}_1,X_1^*}({\bs z}_{1},x_{1}^*)$, $g_{{\bs Z}_2,X_2^*|{\bs Z}_1,X_1^*}({\bs z}_{2},x_{2}^*|{\bs z}_{1},x_{1}^*)$,  and $g_{{\bs Z}_3,X_3^*|{\bs Z}_2,X_2^*}({\bs z}_{3},x_{3}^*|{\bs z}_{2},x_{2}^*)$ remains unresolved.
Our Proposition \ref{P-2} shows a new identification result that, for a  model in which unobserved heterogeneity is discrete and finite,  we can nonparametrically identify the type-specific distribution of $\{{\bs Z}_{t}\}_{t=1}^T$, including the first two periods of the data, from  $T=4$ periods of panel data without imposing stationarity.
 \end{remark}

\begin{remark}\label{r-permutation}
In the identification argument of \cite{Kasahara09},  the type-specific distribution is identified  only up to an arbitrary ordering of the latent types that differs across different  evaluation points  in $\{{\bs z}_1,...,{\bs z}_T\}$.
  Our proof of Proposition \ref{P-2} on the identification of the common order of the latent types   is  based on \cite{HigginsJochmans21}.
 \end{remark}

\begin{remark}\label{r-rank}
Assumption \ref{A-P-2}(a) assumes the rank condition of matrices ${\bs L}_{{\bs z}^*_3}$, ${\bs L}_{{\bs z}_3}$, ${\bs L}_{\bar {\bs z}_3}$, $\bar {\bs L}_{\check{\bs z}_2}$,  and $\bar {\bs L}_{\bar {\bs z}_2}$ defined in (\ref{LL}), of which elements are constructed by evaluating  $g^j_{{\bs Z}_4|{\bs Z}_3}({\bs z}_{4}|{\bs z}_3)$ and $\pi^jg^j_{{\bs Z}_2|{\bs Z}_1}({\bs z}_{2}|{\bs z}_1)g^j_{{\bs Z}_1}({\bs z}_1)$ at different points. These conditions are similar to the assumption stated in Proposition 1 of Kasahara and Shimotsu (2009), implying that all columns in these matrices must be linearly independent. For example, because each column of ${\bs L}_{{\bs z}_3}$ represents  the type-specific conditional density function of $\bs z_4$ across different values of $\bs z_4$ given $\bs z_3$, the changes in the value of $\bs z_4$ must induce sufficiently different changes in the values of conditional density function across types.  One needs to find only one set of values $(\check{\bs z}_2,\bar {\bs z}_2,\bar {\bs z}_3)\in \mathcal{Z}_2^2\times \mathcal{Z}_3$ and one set of $J-1$ and $J$ points of ${\bs Z}_1$ and ${\bs Z}_4$ to construct nonsingular $\bs L_{{\bs z}_3^*}$, $\bs L_{{\bs z}_3}$, $\bs L_{\bar {\bs z}_3}$, $\bar {\bs L}_{\check{\bs z}_2}$,  and $\bar {\bs L}_{\bar {\bs z}_2}$ for each ${\bs z}_3\in \mathcal{Z}_3$
and these rank conditions are not stringent when ${\bs Z}_t$ has continuous support. The identification of $g_{{\bs Z}_4|{\bs Z}_3}^j({\bs z}_4|{\bs z}_3)$ and $\pi^jg_{{\bs Z}_2|{\bs Z}_1}^j({\bs z}_2|{\bs z}_1)g_{{\bs Z}_1}^j({\bs z}_1)$ at all other points of ${\bs Z}_4$, $\bs Z_2$, and ${\bs Z}_1$ follows without any further requirement on the rank condition.
\end{remark}

Once the type-specific distribution of $\{{\bs Z}_t\}_{t=1}^T$ is identified, we can use the argument in the proof of Proposition \ref{P-0} to prove nonparametric identification for the model structure of each type.

\begin{proposition}\label{P-3}
Suppose that    Assumptions  \ref{A-1}-\ref{A-P-2} hold and $T\geq 4$.   Then,  $\bs\theta$ is uniquely determined from $g_{{\bs Z}_1,...,{\bs Z}_T}(\{{\bs z}_{t}\}_{t=1}^T)$.
\end{proposition}

Therefore, type-specific production functions, as well as the distribution of unobserved variables, can be identified nonparametrically. In the estimation, we focus on the case where the type-specific production functions are Cobb-Douglas.
\begin{example}[Random Coefficients Model] \label{ex-1}
Consider a Cobb-Douglas production function with  time-varying  random coefficients:
\begin{equation}
\tilde f_t^j(\tilde {\bs X}_t) = \beta_{0,t}^j + \beta_{k,t}^j \ln K_{t}  + \beta_{m,t}^j \ln M_{t}+ \beta_{\ell,t}^j( \psi_t^j + \ln \tilde L_t), \label{rc}
\end{equation}
where $\tilde f_t^j(\tilde {\bs X}_t) := \ln F_t^j(K_t,e^{\psi_t^j} \widetilde L_t,M_t)$ while
 the intermediate and labor cost share equations are given by
\begin{align*} \label{CD-share}
\ln S_t ^m& =  \ln(\beta_{m,t}^j) + \ln E_t^j[e^{\epsilon}] - \epsilon_t,\quad \ln S_t ^\ell-\ln S_t ^m =  \ln(\beta_{\ell,t}^j/\beta_{m,t}^j) - \ln \left(E_t^j[e^{\zeta}]\right)  + \zeta_t,\\
\ln  M_{it}-  \ln \tilde L_{it} & = \ln (P_{L,t}/P_{M,t})+ \ln  \left(\beta_{m,t}^j/\beta_{\ell,t}^j\right) +\ln \left(E_t^j[e^{\zeta}]\right)  + \psi_t^j +v_{it},
\end{align*}
Under Assumptions  \ref{A-1}-\ref{A-P-2}, $\bs\theta=\{\pi^j,   \{g^j_{v_t}(\cdot), \beta_{m,t}^j, \beta_{\ell,t}^j,g_{\epsilon_t,\zeta_t}^j(\cdot),\psi_t^j\}_{t=1}^4,\{\beta_{0,t}^j,  \beta_{k,t}^j \}_{t=2}^4, \{h_t^j(\cdot), g^j_{\eta_t}(\cdot)\}_{t=3}^4 \}$ for $j=1,...,J$ is nonparametrically identified  from the panel data $\{\bs S_t,\widetilde{\bs X}_t\}_{t=1}^4$.
\end{example}
In Appendix \ref{appendix-6}, we discuss the sufficient conditions under which Assumption \ref{A-P-2} holds when the production function is Cobb-Douglas as in Example \ref{ex-1}.


\section{Estimation of production function with finite mixture random coefficients models}

In this section, we present a finite mixture model of the random coefficient Cobb-Douglas production function, based on our nonparametric identification analysis and with consideration for computational efficiency. We develop a penalized maximum likelihood estimator for this model.

Let us denote the logarithmic values of $(Y_{it},K_{it},M_{it},\tilde L_{it},S_{it}^m,S_{it}^\ell,B_{it})$ using the corresponding lowercase letters, such that $(y_{it},k_{it},m_{it},\tilde \ell_{it},s_{it}^m,s_{it}^\ell,b_{it})$, where $y_{it} :=\log Y_{it}$, and so on. Define $\bs s_{it}:=(s_{it}^m,s_{it}^\ell)$ and $\tilde {\bs x}_{it}:=(k_{it},m_{it},\tilde \ell_{it})$. For estimation purposes, we assume that the data generation follows the parametric assumptions outlined below.


\begin{assumption}\label{A-8} (a) $T$ is fixed at $T\geq 4$ and $N\rightarrow\infty$.
 (b) Equation (\ref{prod}) holds with
\begin{equation}
Y_{it}= F_t^j(K_{it},M_{it},e^{\psi_t^j} \widetilde L_{it})e^{\omega_{it}+ \epsilon_{it}}\ \text{with}\ F_t^j(K_{it},M_{it},e^{\psi_t^j} \widetilde L_{it})=\exp((\beta_{0,t}^j + \beta_{\ell}^j  \psi_t^j)+ \beta_{k}^j k_{it} + \beta_m^j m_{it}  + \beta_{\ell}^j  \tilde \ell_{it}  ).\label{production}
\end{equation}
(c)  $(\epsilon_{it},\zeta_{it})^\top |D_i=j \overset{d}{\sim} N\left(\bs 0, \bs\Sigma_{\epsilon\zeta}\right)$ with
$\bs\Sigma_{\epsilon\zeta}=
\begin{pmatrix}
(\sigma_\epsilon^j)^2&\rho_{\epsilon\zeta}^j\sigma_\epsilon^j\sigma_\zeta^j\\
\rho_{\epsilon\zeta}^j\sigma_\epsilon^j\sigma_\zeta^j&(\sigma_\zeta^j)^2
\end{pmatrix}$,
$g_{\eta}^j(\eta)=\phi(\eta/\sigma_{\eta}^j)/\sigma_{\eta}^j$, and
$g_{v}^j(v)=\phi(v/\sigma_{v}^j)/\sigma_{v}^j$, where $\phi(t) = \exp(-t^2/2)/\sqrt{2}\pi$. Furthermore, we assume  $h_t^j(\omega_{it})=\rho_\omega^j \omega_{it}$ in  (\ref{omega}) so that
\begin{equation}
\omega_{it}=\rho_\omega^j \omega_{it-1} + \eta_{it}. \label{omega-ar1}
\end{equation}
  (d) Conditional on being type $j$, $k_{it}$ given $(k_{it-1},\omega_{it-1})$ is normally distributed with mean $\rho_{k0}^j+\rho_{kk}^j k_{it-1}+ \rho_{k\omega}^j\omega_{it-1}$ and variance $(\sigma_k^j)^2$ while the distribution of $(k_{i1},\omega_{i1})$ follows a bivariate normal distribution with mean $\bs{\mu}_1^j$ and variance  $\bs\Sigma_1^j$.

\end{assumption}

Assumption \ref{A-8}(a) posits that the length of panel data is short, while Assumption \ref{A-8}(b) enforces the Cobb-Douglas functional form assumption. Assumptions \ref{A-8}(c) and \ref{A-8}(d) impose Gaussian distribution assumptions under Assumptions  \ref{A-2} and \ref{A-3}, where $\omega_{it}$ follows a first-order autoregressive (AR(1)) process.

In equation (\ref{production}), since $\ln L_{it} = \psi_t^j +\tilde \ell_{it}$, the intercept term encompasses both $\beta_{0,t}^j$ and $\beta_{\ell}^j\psi_t^j$. The latter term captures the variation in worker quality across types. The normality assumption in Assumptions \ref{A-8}(c) and \ref{A-8}(d) could potentially be relaxed; for instance, by employing the maximum smoothed likelihood estimator of finite mixture models proposed by \cite{lhc11}, in which the type-specific distribution of $\epsilon_{it}$ and $\zeta_{it}$ is nonparametrically specified. Additionally, \cite{kasaharashimotsu15jasa} develop a likelihood-based procedure to test the number of components in normal mixture regression models.

%

Suppose we have a random sample of $n$ independent observations $\{\{\bs S_{it},\widetilde {\bs X}_{it}\}_{t=1}^T\}_{i=1}^n$ from the $J$-component mixture model $\sum_{j=1}^J \pi^j g_{\{\bs S_t,\widetilde {\bs X}_t\}_{t=1}^T}^j(\{\bs s_{t},\widetilde {\bs x}_{t}\}_{t=1}^T)$ that satisfies Assumptions \ref{A-1}-\ref{A-8}. We propose a penalized maximum likelihood estimator (PMLE) that directly maximizes the log-likelihood function of a finite mixture model of production functions. The likelihood function is a parametric version of (\ref{mixture-s2}). To address the issue of unbounded likelihood for a normal mixture model  \citep[c.f.,][]{hartigan85book}, we introduce a penalty term to the log-likelihood function. The maximum likelihood estimator, which leverages distributional information, is consistent even when $T$ is small, provided $T \geq 4$. Given the nonparametric identification result established in Proposition \ref{P-2}, if the parametric assumptions are invalid and the parametric model is misspecified, the parametric maximum likelihood estimator converges in probability to the pseudo-true value of the parameter that minimizes the Kullback-Leibler Information Criterion between the density of the parametric model and the true population density \citep{White1982}.

Our estimation procedure is based on the two-stage identification proof from Proposition \ref{P-2}. To address the computational complexity of maximizing the log-likelihood function for the finite mixture model, the EM algorithm is employed.

Under Assumptions \ref{A-3}-\ref{A-6}, \ref{A-8}, the first order conditions for the expected profit maximization imply that
\begin{align}
&s_{it}^m   = \ln \beta_m^j   +  0.5(\sigma_{\epsilon}^j)^2  - \epsilon_{it}, \quad
s_{it}^\ell -s_{it}^m   = \ln(\beta_\ell^j/\beta_m^j)   -   0.5  (\sigma_{\zeta}^j)^2  + \zeta_{it},\label{share} \\
&m_{it}-  \tilde \ell_{it}  =\alpha_t+ \ln   (\beta_{m}^j/\beta_{\ell}^j) +0.5  (\sigma_{\zeta}^j)^2 + \psi^j +v_{it},\label{share2}
\end{align}
where   $\alpha_t:= \ln (P_{L,t}/P_{M,t})$ and (\ref{share2}) follows from (\ref{share}), $v_{it}=b_{it}-(\psi^j +\tilde \ell_t + \ln P_{L,t}+ \zeta_{it})$, and $s_{it}^\ell-s_{it}^m= b_{it}  - (\ln P_{M,t}+ m_{it})$.

Collect the model parameter $\bs\theta$ into $\bs\pi$, $\bs{\theta}_1$ and $\bs{\theta}_2$ as
\[
\bs\theta:=(\bs\pi',\bs\theta_1',\bs\theta_2')'\quad\text{with}\quad\bs{\theta}_1= (\bs{\alpha}',(\bs \theta_1^1)',...,(\bs{\theta}_1^J)')'\ \text{and}\ \bs{\theta}_2 = ((\bs{\theta}_2^1)',...,(\bs{\theta}_2^J)')',
\]
where $\bs\theta_{1}^j =(\beta_m^j,\beta_\ell^j, \psi^j, (\sigma_{\epsilon}^j)^2,(\sigma_{\zeta}^j)^2,(\sigma_v^j)^2)'$ and   $\bs\theta_2^j=(\beta_{2}^j,...,\beta_T^j, \beta_k^j,(\bs\mu_1^j)',\text{vech}(\bs\Sigma_1^j)',\rho_{k0}^j,\rho_{kk}^j,$\\ $\rho_{k\omega}^j,(\sigma_k^j)^2,\rho_{\omega}^j,(\sigma_\eta^j)^2)'$ for $j=1,...,J$, and $\bs{\alpha}=(\alpha_1,....,\alpha_T)'$.

Denote $\bs\theta^j:=((\bs\theta_1^j)',(\bs\theta_2^j)')'$. Then, under Assumption \ref{A-8}, we may write the probability density function  of $\{\bs s_{it},\tilde{\bs x}_{it}\}_{t=1}^T$ for type $j$ as
\begin{equation}
\begin{aligned}
 g^j(\{\bs s_{it},\tilde{\bs x}_{it}\}_{t=1}^T;\bs\theta) &= \underbrace{\prod_{t=1}^T g_t^j(\bs s_{it}  , \tilde \ell_{it}-m_{it} ;\bs \theta_1^j, \alpha_t) }_{:= L_{1i}(\bs\theta_{1}^j,\bs{\alpha})}\times  \underbrace{g_1^j(\tilde{\bs x}_{it}|\tilde \ell_{i1}-m_{i1};\bs \theta^j)\prod_{t=2}^Tg_t^j(\tilde{\bs x}_{it}| \tilde \ell_{it}-m_{it},\tilde{\bs x}_{it-1};\bs \theta^j)}_{:=L_{2i}(\bs \theta_1^j,\bs \theta_{2}^j)},
\end{aligned}
\label{type-j}
\end{equation}
where the exact expression for $L_{1i}(\bs\theta_1^j,\bs{\alpha})$ and $L_{2i}(\bs \theta_2^j,\bs \theta_1^j)$ is derived below.

To deal with the issue of unbounded log-likelihood function of a normal mixture model \citep{hartigan85book,HaoKasahara2022}, we estimate the model parameter $\bs\theta$ by a penalized likelihood method proposed by \cite{chentan09jmva}. Let $(\hat\sigma_{\epsilon,0}^2,\hat\sigma_{\zeta,0}^2,\hat\sigma_{v,0}^2, \hat\sigma_{k,0}^2,\hat\sigma_{\eta,0}^2, \hat{\bs\Sigma}_{1,0})$ be the estimator of $(\sigma_{\epsilon}^2,\sigma_{\zeta}^2,\sigma_{v}^2, \hat\sigma_{k}^2,\hat\sigma_{\eta}^2, \hat{\bs\Sigma}_{1})$ for the one-component model with $J=1$. Then, we consider the following penalized maximum likelihood estimator (PMLE):
\begin{equation}\label{pmle}
\hat{\bs\theta} =\argmax_{\bs\theta\in\Theta}\ \sum_{i=1}^n Q_i(\bs\theta)+
\sum_{j=1}^J \left\{  \sum_{s\in\{v,k,\eta\}} p_n((\sigma_s^j)^2;(\hat\sigma_{s,0})^2)   + p_n(\bs\Sigma_{\epsilon\zeta}^j;\hat{\bs\Sigma}_{\epsilon\zeta,0})   + p_n(\bs\Sigma_1^j;\hat{\bs\Sigma}_{1,0}) \right\},
\end{equation}
where
\[
Q_i(\bs\theta):= \log\left(\sum_{j=1}^J \pi^j L_{i}(\bs \theta^j,\bs\alpha)\right) \quad\text{with}\quad L_{i}(\bs \theta^j,\bs\alpha):=L_{1i}({\bs \theta}_1^j,{\bs{\alpha}})L_{2i}({\bs\theta}_1^j,{\bs\theta}_2^j)
\]
and
\begin{align}
 p_{n}((\sigma_s^j)^2;\hat\sigma_{s,0}^2) &=- n^{-1}  \left\{ \hat\sigma_{s,0}^2/(\sigma_s^j)^2 - \log(\hat\sigma_{s,0}^2/(\sigma_s^j)^2)\right\},\ \text{and} \label{eq:penalty1}\\
 p_n(\bs\Sigma^j;\hat{\bs\Sigma}_{0}) &= - n^{-1} \left\{\text{tr}(\hat{\bs\Sigma}_{0} (\bs\Sigma^j)^{-1})-\log(\text{det}((\hat{\bs\Sigma}_{0} (\bs\Sigma^j)^{-1}))\right\}. \label{eq:penalty2}
\end{align}
The expression for $L_{1i}({\bs \theta}_1^j,{\bs{\alpha}})$ and $L_{2i}({\bs\theta}_1^j,{\bs\theta}_2^j)$ are derived below.

To reduce the computational burden of finding the penalized maximum likelihood estimator, we follow a three-stage procedure.



  In the first stage, from equations (\ref{share})-(\ref{share2}), we can express $\epsilon_{it}$, $\zeta_{it}$, and $v_{it}$  as a function of $\bs s_{it}$, $\tilde \ell_{it}-m_{it}$, $\bs\theta_1^j$, and $\alpha_t$ as
 \begin{align}
&  \epsilon^*(\bs s_{it};\bs \theta_1^j)  := - s_{it}^m+  \ln\beta_m^j+0.5 (\sigma_{\epsilon}^j)^2,\quad  \zeta^*(\bs s_{it};\bs \theta_1^j)   := s_{it}^\ell - s_{it}^m -  \ln(\beta_\ell^j/\beta_m^j)+0.5 (\sigma_{\zeta}^j)^2, \label{eps-star}\\
&v^*(\tilde \ell_{it}-m_{it} ;\bs\theta_1^j,\alpha_t) := - \left( \tilde \ell_{it}-m_{it}  + \alpha_t + \ln (\beta_{m}^j/\beta_{\ell}^j) + 0.5  (\sigma_{\zeta}^j)^2 +\psi^j\right).  \label{v_hat}
  \end{align}
Then, we estimate  $\bs\theta_1$ by maximizing the log likelihood function as
  \begin{align*}
(\tilde{\bs\pi}, \tilde{\bs\theta}_1,\tilde{\bs\alpha})  &=  \argmax_{ \bs\pi,\bs\theta_1,\bs\alpha } \sum_{i=1}^n
 \ln \left(\sum_{j=1}^J \pi^j L_{1i}(\bs\theta_{1}^j,\bs\alpha)\right) + \sum_{j=1}^J \left\{ p_n((\sigma_v^j)^2;\hat\sigma_{v,0}^2)    + p_n(\bs\Sigma_{\epsilon\zeta}^j;\hat{\bs\Sigma}_{\epsilon\zeta,0})\right\},
 \end{align*}
 where, under Assumption \ref{A-8}(b),  the likelihood function $L_{1i}(\bs\theta_1^j,\bs{\alpha})$ is  given by
  \begin{align*}
 L_{1i}(\bs\theta_{1}^j,\bs{\alpha})&:= \prod_{t=1}^T \frac{1}{\sqrt{1-(\rho_{\epsilon\zeta}^j)^2}\sigma_\epsilon^j\sigma_\zeta^j}\phi\left(\frac{\epsilon^*(\bs s_{it};\bs\theta_1^j)}{\sigma_\epsilon^j}\right)\phi\left(\frac{\zeta^*(\bs s_{it};\bs\theta_1^j)-\rho_{\epsilon\zeta}^j(\sigma_{\zeta}^j/\sigma_{\epsilon}^j)\epsilon^*(\bs s_{it};\bs\theta_1^j) }{\sqrt{1-(\rho_{\epsilon\zeta}^j)^2}\sigma_\zeta^j}\right)\\
  &\qquad \times\frac{1}{\sigma_v^j}\phi\left(\frac{v^*(m_{it}- \tilde \ell_{it} ;\bs\theta_1^j,\alpha_t)}{\sigma_v^j}\right)
  \end{align*}
  with $ \epsilon^*(\bs s_{it};\bs \theta_1^j)$, $\zeta^*(\bs s_{it};\bs \theta_1^j) $, and $v^*(\tilde \ell_{it}-m_{it} ;\bs\theta_1^j,\alpha_t)$ defined in (\ref{eps-star})-(\ref{v_hat}).

In the second stage,  from (\ref{production}),  $\epsilon_{it} = E[s_{it}^m| \tilde{\bs x}_{it}]-s_{it}^{m}$, and $y_{it}+s_{it}^m=m_{it}+\ln (P_{M,t}/P_{Y,t})$, we have
\begin{equation}
\omega_{it}=\omega_t^*(m_{it},\tilde \ell_{it}-m_{it},k_{it};\bs\theta^j)  : =  (1-\beta^j_m- \beta_\ell^j) m_{it}  - \beta_\ell^j \psi^j-\beta_t^j-
 \beta_\ell^j (\tilde \ell_{it}-m_{it}) - \beta_k^j k_{it},\label{omega_hat}
 \end{equation}
 where $\beta_t^j:=\beta_{0,t}^j +\ln (P_{M,t}/P_{Y,t})-\ln \beta_m^j - 0.5(\sigma_{\epsilon}^j)^2$.

In view of equation (\ref{omega_hat}), by a change of variables, we may relate  the  density function of $m_{it}$ conditional on  $\tilde \ell_{it}-m_{it}$ and $k_{it}$
to the  density function of $\omega_{it}$, denoted by $g_{\omega,t}$, as $g_t^j(m_{it}|\ell_{it}-m_{it},k_{it})=(1-\beta^j_m-\beta^j_\ell)g_{\omega,t}^j(\omega_t^*(m_{it},\tilde \ell_{it}-m_{it},k_{it};\bs\theta^j))$.
Then, from (\ref{v_hat})-(\ref{omega_hat}) and Assumptions \ref{A-2}-\ref{A-3}, we have
\begin{align}
&g_1^j(m_{i1}|\ell_{i1}-m_{i1},k_{i1};\bs\theta^j)=(1-\beta^j_m-\beta^j_\ell)g_{\omega|k,1}^j(\omega_{i1}^*(\bs\theta^j) |k_{i1}),\label{prob0}\\
&g_t^j(m_{it}|\ell_{it}-m_{it},k_{it},\tilde{\bs x}_{it-1};\bs\theta^j)= (1-\beta^j_m-\beta^j_\ell) g_{\eta}^j(\eta_{it}^*(\bs\theta^j)
)\quad \text{for $t\geq 2$},\label{prob1}\\
&g_t^j(k_{it}|\tilde{\bs x}_{it-1};\bs\theta^j)= g_{k,t}^j(k_{it}|k_{it-1},\omega_{i,t-1}^*(\bs\theta^j))\quad \text{for $t\geq 2$},\label{prob3}
\end{align}
where $g_{\omega|k,1}^j(\omega_{i1} |k_{i1})$ is the density function of $\omega_{i1}$ conditional on $k_{i1}$, $g_{k,t}^j(k_{it}|k_{it-1},\omega_{it-1})$
is the density function of $k_{it}$ given $(k_{it-1},\omega_{it-1})$,
$\omega_{it}^*(\bs\theta^j):=\omega_t^*(m_{it},\ell_{it}-m_{it},k_{it};\bs\theta^j) $,  and
\begin{equation}\label{eta_hat}
\eta_{it}^*(\bs\theta^j) :=
\omega_{it}^*(\bs\theta^j)
-\rho_\omega^j \omega_{i,t-1}^*(\bs\theta^j).
\end{equation}
Therefore, under Assumption \ref{A-8}, it follows from (\ref{type-j}) and (\ref{prob0})-(\ref{prob3}) that
\begin{align*}
&L_{2i}(\bs\theta^j)=g_1^j(m_{i1}|\ell_{i1}-m_{i1},k_{i1};\bs\theta^j) g_1^j(k_{i1};\bs\theta^j) \times \prod_{t=2}^Tg_t^j(m_{it}|\ell_{it}-m_{it},k_{it},x_{it-1};\bs\theta^j) g_t^j(k_{it}|x_{it-1};\bs\theta^j)\\
&=   (1-\beta^j_m-\beta^j_\ell)^Tg_{\omega k,1}^j(\omega_{i1}^*(\bs\theta^j),k_{i1})  \prod_{t=2}^T
g_{\eta}^j(\eta_{it}^*(\bs\theta^j))
g_{k,t}^j(k_{it}|k_{it-1},\omega_{i,t-1}^*(\bs\theta^j)),
\end{align*}
where
\begin{align*}
& g_{\eta}^j(\eta_{it}^*(\bs\theta^j)) := \frac{1}{\sigma_\eta^j}\phi\left(\frac{\eta_{it}^*(\bs\theta^j)}{\sigma_\eta^j}\right),\\
&g_{\omega k,1}^j(\omega_{i1}^*(\bs\theta^j),k_{i1}):=
(2\pi)^{-3/2} |\bs\Sigma_1^j|^{-1/2}\exp\left(-\frac{1}{2}\left(\begin{pmatrix}
k_{i1}\\
\omega_{i1}^*(\bs\theta^j)
\end{pmatrix}-\bs\mu_1^j\right)'(\bs\Sigma_1^j)^{-1}\left(\begin{pmatrix}
k_{i1}\\
\omega_{i1}^*(\bs\theta^j)
\end{pmatrix}-\bs\mu_1^j\right)\right),\\
&g_{k,t}^j(k_{it}|k_{it-1},\omega_{i,t-1}^*(\bs\theta^j)):=\frac{1}{\sigma_k^j}\phi\left(\frac{k_{it}-(\rho_{k0}^j+\rho_{kk}^j k_{it-1}+ \rho_{k\omega}^j\omega_{it-1})}{\sigma_k^j}\right).
\end{align*}


Given the first stage estimate $\tilde{\bs\theta}_1$,  the second stage estimates parameters $\bs{\pi}$ and $\bs\theta_2$  by maximizing the log-likelihood function as
\[
(\tilde{\bs{\pi}}_2,\tilde{\bs\theta}_2) = \argmax_{\bs{\pi},\bs\theta_2} \sum_{i=1}^n \log\left(\sum_{j=1}^J \pi^j L_{1i}(\tilde{\bs\theta}_1^j,\tilde{\bs{\alpha}}) L_{2i}(\tilde{\bs\theta}_1^j,\bs\theta_2^j) \right)+ \sum_{j=1}^J  \left\{ \sum_{s\in\{k,\eta\}} p_n((\sigma_s^j)^2;\hat\sigma_{s,0}^2)   + p_n(\bs\Sigma_1^j;\hat{\bs\Sigma}_{1,0})\right\}.
\]

%

 Finally, using $\tilde{\boldsymbol{\theta}} := (\tilde{\boldsymbol{\pi}}_2', \tilde{\boldsymbol{\theta}}_1', \tilde{\boldsymbol{\theta}}_2')'$ as an initial value, we obtain the PMLE $\hat{\boldsymbol{\theta}}$ by maximizing the full-information penalized log-likelihood function as in Equation (\ref{pmle}) using the EM algorithm.

Let the true value of the model parameter be denoted by $\boldsymbol{\theta}^*$. When the number of types $J$ is correctly specified, the Fisher Information matrix is given by
\[
\boldsymbol{I}(\boldsymbol{\theta}^*) := -\mathbb{E}\left[ \nabla_{\boldsymbol{\theta}\boldsymbol{\theta}'} Q_i(\boldsymbol{\theta}^*) \right] = \mathbb{E}\left[ \nabla_{\boldsymbol{\theta}}Q_i(\boldsymbol{\theta}^*) \nabla_{\boldsymbol{\theta}'} Q_i(\boldsymbol{\theta}^*) \right],
\]
which is positive definite. The following proposition demonstrates that the PMLE is consistent and asymptotically normal.

\begin{proposition}\label{consistency}
Suppose that Assumptions \ref{A-1}-\ref{A-8}  hold. Then, $\hat{\bs\theta}\overset{p}{\rightarrow} \bs \theta^*$ and $\sqrt{n}(\hat{\bs\theta}-\bs\theta^*) \overset{d}{\rightarrow} N(\bs 0, \bs{{I}}(\bs\theta^*))$.
\end{proposition}

\section{Empirical Application}

\subsection{Data}\label{data}

We utilize plant-level panel data from the Census of Manufacture of Japan spanning 1986-2010. This dataset encompasses production information for manufacturing plants in Japan. Our analysis focuses on plants with 30 or more employees, as detailed data are consistently available only for these establishments.\footnote{The survey employs distinct questionnaires based on plant size: 1. Plants with 30 or more employees; 2. Plants with 4-29 employees; 3. Plants with 1-3 employees. The questionnaire for plants with 30 or more employees provides more comprehensive information. For instance, beginning in 2000, the census collects fixed asset data every five years, rather than annually, for plants with fewer than 30 employees.}

At the 4-digit industry classification level accessible in the Census of Manufacture, we identify a total of 276 industries. In our empirical application, we primarily concentrate on concrete products and electric audio equipment for two reasons: 1. Both industries have a substantial number of observations; 2. The former exhibits relatively small variation in intermediate input share, while the latter displays significant variation, as demonstrated in Section \ref{sec:evidence-hetero}. Thus, examining these two industries proves useful for assessing the significance of unobserved heterogeneity.

Output ($Y$) is defined as the sum of shipments, revenue from repair and maintenance services, and revenue from performing subcontracted work. Initial capital value ($K$) is determined as the fixed asset value minus land, and subsequent capital values are constructed using the perpetual inventory method. The observed labor input ($\tilde{L}$) is represented by the number of employees. The intermediate input ($M$) is defined as the sum of material input, energy input, and subcontracting expenses for consigned production.

Flow data, such as shipments and various production costs, pertain to the calendar year. The number of employees refers to the value at the end of the year, while the stock of fixed assets corresponds to the beginning of the period. Table \ref{tab:sumstat} presents summary statistics for the variables employed in our empirical analysis.

\begin{table}[tb]
\centering
  \caption{Summary Statistics}
  \label{tab:sumstat}
  \smallskip
  \begin{tabular}{l|ccccc|ccccc}
  \hline
  & \multicolumn{5}{c|}{Concrete Products} & \multicolumn{5}{c}{Electric Audio Equipment}\\
 & Obs & Mean & Std. Dev & Min & Max & Obs & Mean & Std. Dev & Min & Max\\
  \hline
  $y_{it}$ 	&	13892	&	11.37	&	0.68	&	7.60	&	14.39	&	10913	&	11.24	&	1.78	&	5.51	&	17.35	\\
  $\tilde \ell_{it}$ 	&	13892	&	3.97	&	0.41	&	3.40	&	6.81	&	10913	&	4.51	&	0.90	&	3.40	&	8.53	\\
  $k_{it}$ 	&	13892	&	10.91	&	0.86	&	5.22	&	14.06	&	10913	&	10.04	&	1.99	&	1.67	&	16.22	\\
  $m_{it}$ 	&	13892	&	10.34	&	0.83	&	-0.13	&	13.89	&	10913	&	10.37	&	2.31	&	3.27	&	16.89	\\
$s^m_{it}$	&	13892	&	-0.98	&	0.35	&	-1.83	&	-0.33	&	10913	&	-0.99	&	0.79	&	-3.13	&	-0.08	\\
$s^\ell_{it}$	&	13892	&	-1.47	&	0.43	&	-2.39	&	-0.50	&	10913	&	-1.39	&	0.78	&	-3.11	&	-0.16	\\
  $I_{it}/K_{it}$ 	&	13892	&	0.10	&	1.06	&	-1.04	&	115.02	&	10913	&	0.34	&	11.49	&	-1.28	&	1030.83	\\
   \hline
\end{tabular}
\end{table}

\subsection{Estimation of Production Function}

This section presents estimation results for a random-coefficient Cobb-Douglas production function featuring three technology types and two unobserved labor types within each technology type. 

Table \ref{tab:est-concrete} and Table \ref{tab:est-audio} present the parameter estimates for the concrete products and electric audio equipment industries, considering both the unobserved heterogeneity case ($J = 3 \times 2 = 6$) and the homogeneous case ($J = 1$). The estimated coefficients in both industries demonstrate economically significant differences in output elasticities associated with labor, capital, and intermediate inputs across various firm types.

Comparing the two industries, the variation in $\hat \beta_m^j$ across types is more substantial for electric audio equipment than for concrete products, which aligns with the dispersion of intermediate input shares discussed in Section \ref{sec:evidence-hetero}. As $\hat \beta_\ell^j$ and $\hat\beta_k^j$ also exhibit variation across types, the ratio of output elasticities between capital and labor, $\hat\beta_k^j/\hat\beta_\ell^j$, varies as well. For electric audio equipment, the value of $\hat\beta_k^j/\hat\beta_\ell^j$ ranges from 0.29 (Type 1 and 2) to 1.14 (Type 3 and 4), while for concrete products, it spans from 0.56 to 0.78. As demonstrated below, the value of $\hat\beta_k^j/\hat\beta_\ell^j$ serves as a crucial determinant of the capital investment response to productivity $\hat\omega_{it}$.

Furthermore, the productivity growth processes display variations across latent types, as indicated by the estimated AR(1) coefficient $\rho_\omega^j$ and standard deviation $\sigma_\eta^j$.  In both industries, the estimated AR(1) coefficient for the homogeneous case ($J = 1$) is considerably larger than those for the unobserved heterogeneity case ($J = 6$). This suggests that neglecting unobserved heterogeneity may result in an upward bias in the estimates of the AR(1) coefficient for productivity processes.

In various latent types, the returns to scale for concrete products, denoted by $(\hat{\beta}_m^j + \hat{\beta}_\ell^j + \hat{\beta}_k^j)$, is approximately 0.7, whereas for electrical audio equipment, it ranges between 0.78 and 0.91. When considering the homogeneous case, the returns to scale are comparatively lower at 0.63 and 0.65 for these respective industries. The estimates of $(\hat{\psi}^j)$ indicate the existence of considerable unobserved heterogeneity in labor quality or working hours among manufacturing plants.


\begin{table}[tb]
\centering
\caption{Estimates of Production Function (Concrete Products)}
\label{tab:est-concrete}
\begin{tabular}{c|c|cc|cc|cc}
  \hline
  & J = 1 & \multicolumn{6}{c}{J = 6} \\\cline{3-8}
 & & Type 1& Type 2& Type 3& Type 4& Type 5& Type 6 \\
 \hline
$\beta_m^j$ & 0.332 & \multicolumn{2}{c|}{0.287} & \multicolumn{2}{c|}{0.326}  & \multicolumn{2}{c}{0.394} \\
   & (0.004) &\multicolumn{2}{c|}{(0.005)}&\multicolumn{2}{c|}{(0.012)}&\multicolumn{2}{c}{(0.005)} \\
$\beta_{\ell}^j$ & 0.224 & \multicolumn{2}{c|}{0.263}  & \multicolumn{2}{c|}{0.220}  & \multicolumn{2}{c}{0.186} \\
   & (0.003) &\multicolumn{2}{c|}{(0.005)}&\multicolumn{2}{c|}{(0.009)}&\multicolumn{2}{c}{(0.003)} \\
$\beta_k^j$ & 0.075 & \multicolumn{2}{c|}{0.147}  & \multicolumn{2}{c|}{0.172} & \multicolumn{2}{c}{0.103}  \\
   & (0.015) &\multicolumn{2}{c|}{(0.036)}&\multicolumn{2}{c|}{(0.034)}&\multicolumn{2}{c}{(0.014)} \\ \hline
$\rho_\omega^j$ &0.935 &\multicolumn{2}{c|}{ 0.834}&\multicolumn{2}{c|}{	0.869}&\multicolumn{2}{c}{	0.877}\\
&(0.007)&\multicolumn{2}{c|}{(0.014)}	&\multicolumn{2}{c|}{(0.024)}	&\multicolumn{2}{c}{(0.012)}\\
$\sigma_\eta^j$&0.155&\multicolumn{2}{c|}{0.137}&\multicolumn{2}{c|}{	0.248}&\multicolumn{2}{c}{	0.114}\\
&(0.004)&\multicolumn{2}{c|}{(0.004)}&\multicolumn{2}{c|}{(0.014)}&\multicolumn{2}{c}{(0.003)}\\
 \hline
$\beta_m^j + \beta_{\ell}^j + \beta_k^j$ & 0.632 & \multicolumn{2}{c|}{0.696} & \multicolumn{2}{c|}{0.719} & \multicolumn{2}{c}{0.683} \\
  $\beta_k^j/\beta_{\ell}^j$ & 0.336 & \multicolumn{2}{c|}{0.558} & \multicolumn{2}{c|}{0.780} & \multicolumn{2}{c}{0.556} \\
  $\Sigma_{t=1}^T\beta_{0,t}^j/T$ & 6.218 & \multicolumn{2}{c|}{5.604} & \multicolumn{2}{c|}{5.303} & \multicolumn{2}{c}{5.449} \\
   \hline
$\psi^j$ & 0.000 & -0.108 & 0.344 & -1.454 & 0.534 & 0.056 & 0.627 \\
   & (NA) &(0.086)&(0.090)&(0.339)&(0.081)&(0.085)&(0.088) \\
 \hline
$\pi$ & 1.000 & 0.174 & 0.240 & 0.051 & 0.119 & 0.263 & 0.152 \\
   & (NA) &(0.014)&(0.019)&(0.015)&(0.021)&(0.018)&(0.015) \\\hline
 Obs & \multicolumn{7}{c}{13892} \\
 No. Plants & \multicolumn{7}{c}{914} \\
 \hline
\end{tabular}
\begin{flushleft}
Notes: Standard errors are reported in parentheses.
\end{flushleft}
\end{table}

\begin{table}[tb]
\centering
\caption{Estimates of Production Function (Electric Audio Equipment)}
\label{tab:est-audio}
\begin{tabular}{c|c|cc|cc|cc}
  \hline
  & J = 1 & \multicolumn{6}{c}{J = 6} \\
 & & Type 1& Type 2& Type 3& Type 4& Type 5& Type 6 \\
 \hline
$\beta_m^j$ & 0.281 & \multicolumn{2}{c|}{0.135} & \multicolumn{2}{c|}{0.430} & \multicolumn{2}{c}{0.536} \\
   & (0.008) &\multicolumn{2}{c|}{(0.022)}&\multicolumn{2}{c|}{(0.029)}&\multicolumn{2}{c}{(0.054)} \\
$\beta_{\ell}^j$ & 0.296 & \multicolumn{2}{c|}{0.496} & \multicolumn{2}{c|}{0.163} & \multicolumn{2}{c}{0.195} \\
   & (0.006) &\multicolumn{2}{c|}{(0.016)}&\multicolumn{2}{c|}{(0.010)}&\multicolumn{2}{c}{(0.041)} \\
$\beta_k^j$ & 0.076 & \multicolumn{2}{c|}{0.146} & \multicolumn{2}{c|}{0.186} & \multicolumn{2}{c}{0.175} \\
   & (0.016) &\multicolumn{2}{c|}{(0.031)}&\multicolumn{2}{c|}{(0.027)}&\multicolumn{2}{c}{(0.049)} \\
 \hline
$\rho_\omega^j$ &0.960&\multicolumn{2}{c|}{0.769} & \multicolumn{2}{c|}{0.845}&\multicolumn{2}{c}{0.895} \\
&(0.003)&\multicolumn{2}{c|}{(0.017)}&\multicolumn{2}{c|}{(0.018)}&\multicolumn{2}{c}{(0.032)} \\
$\sigma_\eta^j$&0.378&\multicolumn{2}{c|}{0.457} & \multicolumn{2}{c|}{0.376}&\multicolumn{2}{c}{0.135} \\
&(0.011)&\multicolumn{2}{c|}{(0.040)}&\multicolumn{2}{c|}{(0.037)}&\multicolumn{2}{c}{(0.014)} \\
 \hline
$\beta_m^j + \beta_{\ell}^j + \beta_k^j$ & 0.652 & \multicolumn{2}{c|}{0.776}  & \multicolumn{2}{c|}{0.778} & \multicolumn{2}{c}{0.906} \\
  $\beta_k^j/\beta_{\ell}^j$ & 0.256 & \multicolumn{2}{c|}{0.294} &\multicolumn{2}{c|}{1.140} & \multicolumn{2}{c}{0.893} \\
  $\Sigma_{t=1}^T\beta_{0,t}^j/T$ & 6.125 & \multicolumn{2}{c|}{5.747} & \multicolumn{2}{c|}{4.560} & \multicolumn{2}{c}{2.992} \\
   \hline
$\psi^j$ & 0.000 & -1.084 & 0.452 & -0.551 & 1.178 & -0.534 & 0.540 \\
   & (NA) &(0.249)&(0.080)&(0.062)&(0.109)&(0.390)&(0.146) \\
 \hline
$\pi$ & 1.000 & 0.278 & 0.168 & 0.163 & 0.098 & 0.109 & 0.184 \\
   & (NA) &(0.033)&(0.046)&(0.015)&(0.044)&(0.027)&(0.022) \\\hline
 Obs & \multicolumn{7}{c}{10913} \\
 No. Plants & \multicolumn{7}{c}{907} \\
 \hline
\end{tabular}
\begin{flushleft}
Notes: Standard errors are reported in parentheses.
\end{flushleft}
\end{table}

\begin{figure}[tb]
    \caption{Posterior Probabilities (Concrete Products)}
	\centering
    \includegraphics[width=.7\linewidth]{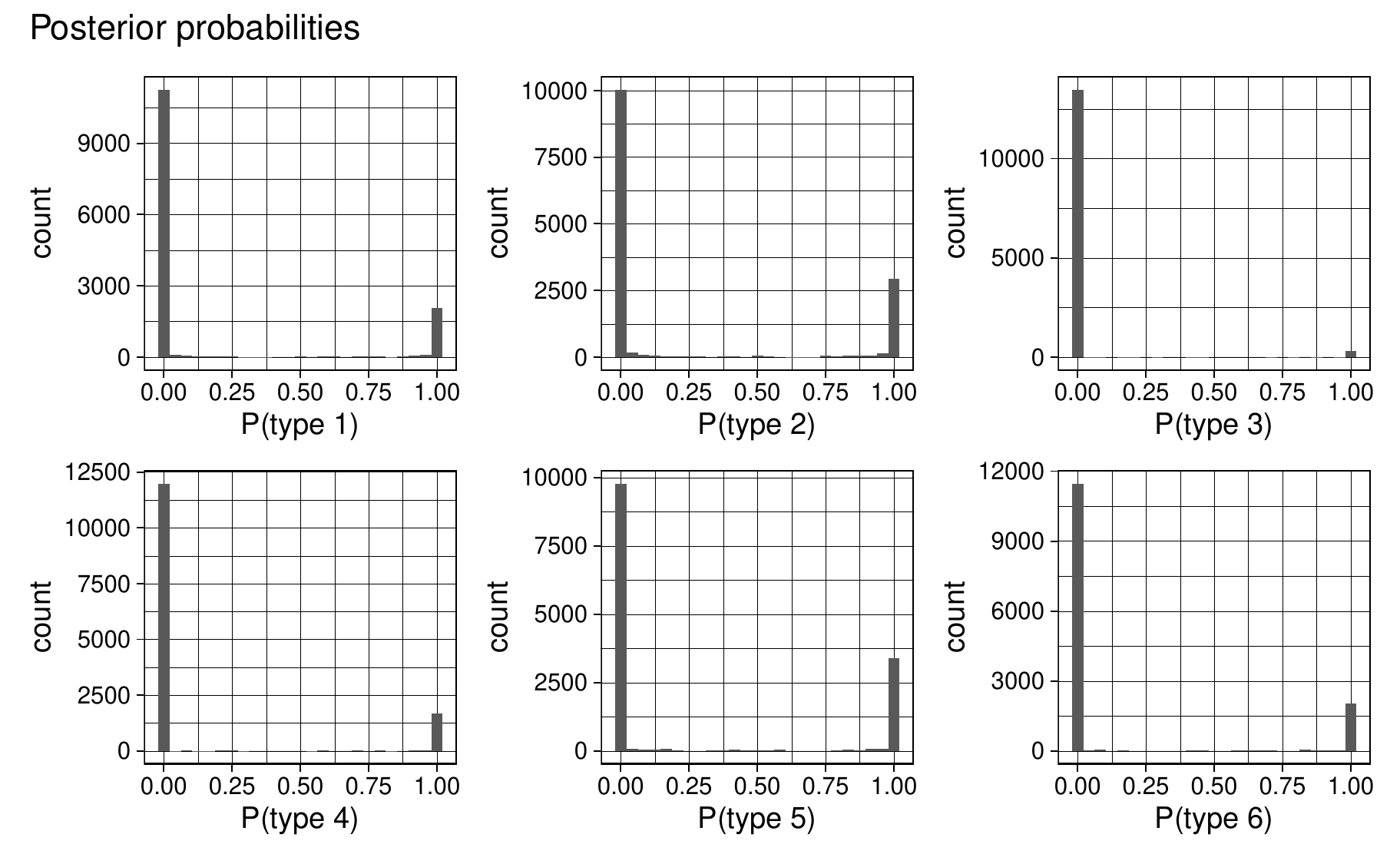}
	\label{fig:posterior_2223}
\end{figure}
\begin{figure}[tb]
    \caption{Posterior Probabilities (Electric Audio Equipment)}
	\centering
    \includegraphics[width=.7\linewidth]{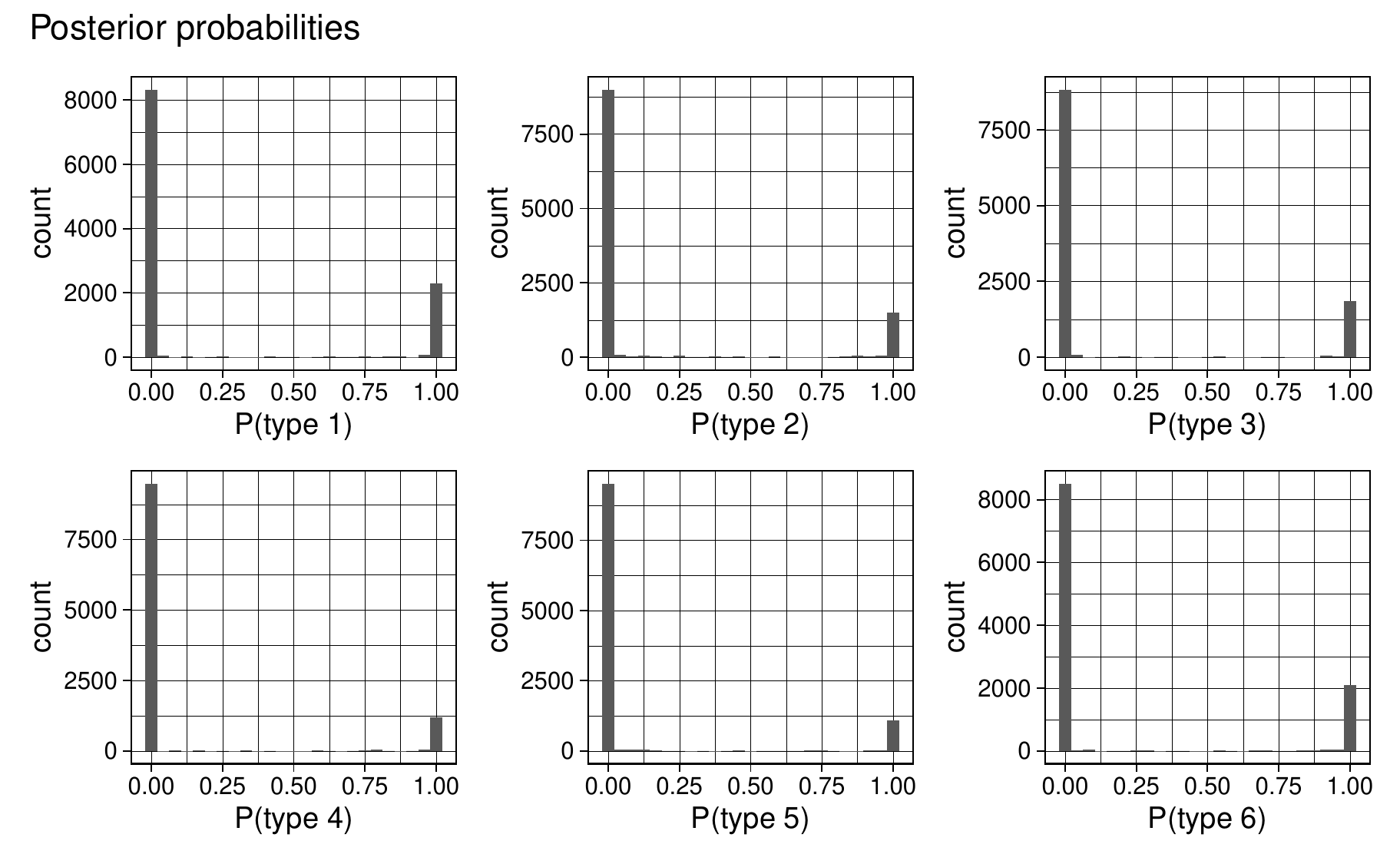}
	\label{fig:posterior_2814}
\end{figure}

Figures \ref{fig:posterior_2223} and \ref{fig:posterior_2814} display the distribution of posterior type probabilities, defined by $\hat\pi^j_i := \frac{\hat\pi^j L_{i}(\hat\theta^j)}{\sum_{k=1}^J \hat\pi^{k} L_{i}(\hat\theta^{k})}$ for $j=1,...,J$, across plants for the model with $J=6$. The posterior probabilities for each type are concentrated around 0 or 1. In the subsequent analysis, we assign one of the $J$ types to each plant based on its posterior type probability that achieves the highest value across the $J$ types.


Ignoring unobserved heterogeneity may lead to significant biases in measuring productivity growth. To examine this issue, we consider a specification with $J=6$ as the true model and compute the bias in measuring productivity growth when using a misspecified model with $J=1$. Specifically, let $\Delta\omega_{it}:= \Delta y_{it}-(\hat\beta_{t}^j+\hat\beta_{m}^j  \Delta m_{it}+ \hat\beta_{\ell}^j  \Delta \tilde\ell_{it} + \hat\beta_{k}^j \Delta  k_{it}  + \Delta \hat\epsilon_{it}^j)$ for $j=1,2,...,6$ be the estimated productivity growth when $J=6$ and let $\Delta \tilde\omega_{it}:= \Delta y_{it}-(\bar\beta_t+  \bar\beta_{m} \Delta m_{it}+  \bar\beta_{\ell}\Delta  \tilde\ell_{it} + \bar\beta_{k}  \Delta k_{it}  + \Delta \bar\epsilon_{it})$ be the estimated productivity growth when $J=1$, where $\{\hat\beta_{t}^j,\hat\beta_{m}^j,\hat\beta_{\ell}^j, \hat\beta_{k}^j\}_{j=1}^6$ and  $\{\bar\beta_{t},\bar\beta_{m}^j,\bar\beta_{\ell}^j, \bar\beta_{k}^j\}$ denote estimated coefficients when $J=6$ and $J=1$, respectively. Then, we compute the bias as
\[
\Delta\tilde \omega_{it}=\Delta\omega_{it}
+ \underbrace{ (\bar \beta_{m}-\hat\beta_{m}^j)  \Delta m_{it}+ (\bar\beta_{\ell} - \hat\beta_{\ell}^j) \Delta \ell_{it} + (\bar\beta_{k}^j- \hat\beta_{k}^j)  \Delta k_{it}  +(\Delta \bar\epsilon_{it}- \Delta \hat\epsilon_{it}^j)}_{:=\text{Bias}_{it}}.
\]

The first row of Table \ref{tab:bias}, labeled as $\frac{\text{Mean of } |\text{Bias}_{it}|}{\text{Mean of }|\Delta\tilde\omega_{it}|}$, presents the ratio of the average absolute value of bias to the average productivity growth within each of three subsamples, which are classified by technology types. The magnitude of the bias is around 0.10 for concrete products, while it ranges from 0.23 to 0.35 for electric audio equipment.

The second row of Table \ref{tab:bias}, denoted by $\frac{\text{Mean of } \text{Bias}_{it}}{\text{Mean of } |\Delta\tilde\omega_{it}|}\Big|_{\Delta\omega_{it}>0}$, reports the ratio of the average value of bias to the average productivity growth conditional on positive productivity growth measured by the model with $J=6$. Note that $\text{Bias}_{it}\approx (\hat\beta_{m}^j-\bar \beta_{m}) \Delta m_{it}$ and $Corr(\Delta \omega_{it},\Delta m_{it})>0$. Consequently, the average bias conditional on $\Delta\omega_{it}>0$ tends to be positive when $\hat\beta_m^j > \bar\beta_m^j$. The empirical results confirm this pattern: for both concrete products and electric audio equipment, Types 5-6 have $\hat\beta_m^j$ higher than $\bar\beta_m$, and thus the estimated bias $\frac{\text{Mean of } \text{Bias}_{it}}{\text{Mean of } |\Delta\tilde\omega_{it}|}\Big|_{\Delta\omega_{it}>0}$ is positive for these types, while the estimated bias  when $\Delta\omega_{it}>0$ is negative for Types 1-2 that have lower values of $\hat\beta_m^j$.

These findings imply that neglecting unobserved heterogeneity could lead to significant bias in estimating productivity growth, and the bias is likely to exhibit a systematic pattern depending on the values of $\hat\beta_m^j$.

\begin{table}[tb]
\centering
\caption{Bias in $\Delta\tilde\omega$}
\smallskip
\label{tab:bias}
\begin{tabular}{l|ccc|ccc}
  \hline
  & \multicolumn{3}{c|}{Concrete Products} &\multicolumn{3}{c}{Electric Audio Equipment} \\\cline{2-7}
  & \multicolumn{3}{c|}{J = 6} &\multicolumn{3}{c}{J = 6}\\\cline{2-7}
 & Type 1 - 2& Type 3 - 4& Type 5 - 6 & Type 1 - 2& Type 3 - 4& Type 5 - 6\\
 \hline
$\frac{\text{Mean of } |\text{Bias}_{it}|}{\text{Mean of }
                     |\Delta\tilde\omega_{it}|}$ & 0.098 & 0.104 & 0.099 & 0.230 & 0.229 & 0.350 \\

 \hline
$\frac{\text{Mean of } \text{Bias}_{it}}
                  {\text{Mean of } |\Delta\tilde\omega_{it}|}
                  \Big|_{\Delta\omega_{it}>0}$ & -0.080 & -0.025 & 0.089 & -0.201 & 0.156 & 0.203 \\
   \hline
$\beta_m^j$ & 0.287 & 0.326 & 0.394 & 0.135 & 0.430 & 0.536\\
   \hline
\end{tabular}
\end{table}

As an application of using the estimated productivity growth in empirical analysis, we now investigate whether unobserved heterogeneity, as captured by type-specific production function parameters, is significant for investment decisions. Specifically, for each subsample classified by type, we estimate the following linear investment model:
\[
\frac{I_{it}}{K_{it}} = \alpha_0 + {\alpha^j_\omega} \hat\omega_{it} + \text{quadratic of $k_{it}$} + \zeta_{it},
\]
where ${I_{it}}/{K_{it}}$ represents the ratio of investment to capital stock.

Table \ref{tab:investment-concrete} displays the OLS estimates of $\alpha^j_\omega$ in the first row as well as the quantile regression estimates of $\alpha^j_\omega$ at the 10th, 25th, 50th, 75th, and 90th percentiles across different types for $J=1$ and $6$ for concrete products. Table \ref{tab:investment-audio} presents the same estimates for the electric audio equipment industry. When $J=1$, the OLS coefficient of $\omega_{it}$ is estimated significantly at $0.06$ for concrete products and at $0.5$ for electric audio equipment.

For the model with $J=6$, the estimated coefficients of $\omega_{it}$ differ considerably across different types of plants, indicating that the investment response to a productivity shock varies across plants. Both OLS and quantile regression results show that the estimated coefficients tend to be higher for the types with higher $\hat\beta_m^j$ and $\hat\beta_k^j/\hat\beta_\ell^j$, suggesting that firms invest more given a positive productivity shock if their production technology features high material shares and high capital-labor ratios. In the case of quantile regressions, this pattern is particularly pronounced for firms with high investment ratios.

Overall, these results highlight the importance of accounting for unobserved heterogeneity in the production function when estimating plant-level productivity and its impact on investment.

\begin{table}[tb]
\caption{The Effect of $\omega_{it}$ on Investment (Concrete Products)}
\label{tab:investment-concrete}
\begin{center}
\begin{tabular}{l |c| c c c c c c }
\hline
 & J = 1 & \multicolumn{6}{c}{J=6}\\\cline{3-8}
 & & Type 1 & Type 2 & Type 3 & Type 4 & Type 5 & Type 6 \\
\hline
$\alpha^j_{\omega}$          & $0.059$   & $0.059$   & $0.061$   & $0.048$   & $-0.163$  & $0.076$   & $0.081$   \\
                             & $(0.024)$ & $(0.013)$ & $(0.014)$ & $(0.037)$ & $(0.140)$ & $(0.011)$ & $(0.011)$ \\ \hline
$\alpha^j_{\omega}(0.10)$    & $0.002$   & $0.004$   & $0.001$   & $-0.000$  & $0.014$   & $0.016$   & $0.018$   \\
                             & $(0.000)$ & $(0.003)$ & $(0.000)$ & $(0.004)$ & $(0.004)$ & $(0.003)$ & $(0.004)$ \\
$\alpha^j_{\omega}(0.25)$    & $0.007$   & $0.016$   & $0.011$   & $0.001$   & $0.010$   & $0.023$   & $0.029$   \\
                             & $(0.001)$ & $(0.003)$ & $(0.003)$ & $(0.003)$ & $(0.002)$ & $(0.003)$ & $(0.004)$ \\
$\alpha^j_{\omega}(0.50)$    & $0.017$   & $0.031$   & $0.031$   & $0.011$   & $0.022$   & $0.044$   & $0.040$   \\
                             & $(0.001)$ & $(0.005)$ & $(0.005)$ & $(0.006)$ & $(0.004)$ & $(0.005)$ & $(0.005)$ \\
$\alpha^j_{\omega}(0.75)$    & $0.037$   & $0.049$   & $0.065$   & $0.050$   & $0.030$   & $0.073$   & $0.080$   \\
                             & $(0.003)$ & $(0.009)$ & $(0.008)$ & $(0.016)$ & $(0.007)$ & $(0.008)$ & $(0.009)$ \\
$\alpha^j_{\omega}(0.90)$    & $0.072$   & $0.077$   & $0.125$   & $0.091$   & $0.025$   & $0.124$   & $0.110$   \\
                             & $(0.007)$ & $(0.025)$ & $(0.024)$ & $(0.021)$ & $(0.018)$ & $(0.025)$ & $(0.023)$ \\
\hline
$\beta_m^j$                  & $0.332$   & $0.287$   & $0.287$   & $0.326$   & $0.326$   & $0.394$   & $0.394$   \\
$\beta_k^j/\beta_{\ell}^j$   & $0.336$   & $0.558$   & $0.558$   & $0.780$   & $0.780$   & $0.556$   & $0.556$   \\
$\overline{(I_{it}/K_{it})}$ & $0.104$   & $0.080$   & $0.085$   & $0.141$   & $0.264$   & $0.075$   & $0.069$   \\
\hline
\end{tabular}
\label{table:coefficients}
\end{center}
\begin{flushleft}
Notes: Standard errors are reported in parentheses. The first row presents the OLS estimate, while the second to the sixth rows present the quantile regression estimates at the $\tau$-th quantile  for $\tau=0.10, 0.25, 0.50, 0.75, 0.90$.
\end{flushleft}
\end{table}

\begin{table}[tb]
\caption{The Effect of $\omega_{it}$ on Investment (Electric Audio Equipment)}
\label{tab:investment-audio}
\begin{center}
\begin{tabular}{l |c| c c c c c c }
\hline
 & J = 1 & \multicolumn{6}{c}{J=6}\\\cline{3-8}
 & & Type 1 & Type 2 & Type 3 & Type 4 & Type 5 & Type 6 \\
\hline
$\alpha^j_{\omega}$          & $0.500$   & $-0.215$  & $-0.113$  & $0.188$   & $0.171$   & $0.080$   & $0.079$   \\
                             & $(0.110)$ & $(0.138)$ & $(0.974)$ & $(0.401)$ & $(0.079)$ & $(0.020)$ & $(0.014)$ \\ \hline
$\alpha^j_{\omega}(0.10)$    & $0.002$   & $0.002$   & $-0.000$  & $0.002$   & $0.014$   & $0.004$   & $0.004$   \\
                             & $(0.000)$ & $(0.001)$ & $(0.000)$ & $(0.001)$ & $(0.006)$ & $(0.002)$ & $(0.003)$ \\
$\alpha^j_{\omega}(0.25)$    & $0.001$   & $0.000$   & $0.000$   & $0.000$   & $0.013$   & $0.006$   & $0.012$   \\
                             & $(0.000)$ & $(0.002)$ & $(0.003)$ & $(0.000)$ & $(0.004)$ & $(0.003)$ & $(0.003)$ \\
$\alpha^j_{\omega}(0.50)$    & $0.010$   & $0.003$   & $0.010$   & $0.010$   & $0.030$   & $0.030$   & $0.050$   \\
                             & $(0.001)$ & $(0.001)$ & $(0.003)$ & $(0.002)$ & $(0.006)$ & $(0.009)$ & $(0.005)$ \\
$\alpha^j_{\omega}(0.75)$    & $0.028$   & $0.016$   & $0.031$   & $0.039$   & $0.048$   & $0.071$   & $0.074$   \\
                             & $(0.002)$ & $(0.004)$ & $(0.006)$ & $(0.007)$ & $(0.011)$ & $(0.017)$ & $(0.011)$ \\
$\alpha^j_{\omega}(0.90)$    & $0.060$   & $0.011$   & $0.032$   & $0.097$   & $0.060$   & $0.115$   & $0.106$   \\
                             & $(0.004)$ & $(0.011)$ & $(0.014)$ & $(0.018)$ & $(0.015)$ & $(0.024)$ & $(0.028)$ \\
\hline
$\beta_m^j$                  & $0.281$   & $0.135$   & $0.135$   & $0.430$   & $0.430$   & $0.536$   & $0.536$   \\
$\beta_k^j/\beta_{\ell}^j$   & $0.256$   & $0.294$   & $0.294$   & $1.140$   & $1.140$   & $0.893$   & $0.893$   \\
$\overline{(I_{it}/K_{it})}$ & $0.335$   & $0.350$   & $0.828$   & $0.471$   & $0.165$   & $0.073$   & $0.079$   \\
\hline
\end{tabular}
\label{table:coefficients}
\end{center}
\begin{flushleft}
Notes: Standard errors are reported in parentheses. The first row presents the OLS estimate while the second to the sixt rows present the estimates of quantile regressions at the $\tau$-th quantile for $\tau=0.10, 0.25, 0.50, 0.75, 0.90$.
\end{flushleft}
\end{table}

\section{Conclusion}
This paper establishes the nonparametric identifiability of production functions when unobserved heterogeneity exists across firms in the form of latent technology groups. Building on our nonparametric identification analysis and considering computational simplicity, we propose an estimation procedure for the production function with random coefficients employing a finite mixture specification.

Our analysis of Japanese plant-level panel data reveals a substantial degree of variation in estimated input elasticities and productivity growth processes across latent types within narrowly defined industries. We demonstrate that neglecting unobserved heterogeneity in input elasticities can lead to significant and systematic bias in estimated productivity growth. Moreover, we highlight the critical role played by unobserved disparities in input elasticities in plant-level investment decisions. Specifically, we find that the correlation between estimated productivity and investment is notably stronger among high capital-intensive latent type firms than among low capital-intensive type firms.

As future research topics, we may extend our framework in several directions.  First, the framework may be extended to explicitly account for plant- or firm-specific biased technological change, which may be continuously distributed. This can be achieved by integrating the structural assumption discussed in \cite{doraszelski2018measuring}, \cite{ZHANG2019}, \cite{Demirer20}, and \cite{Raval2023} into our framework. The identification approach of this paper---based on panel data with a Markov structure---is different from, but complementary to, the structural approaches of these existing papers. Adopting both identification approaches simultaneously can provide an empirical framework for estimating production functions that incorporates a broader range of unobserved heterogeneity.

Second, although we establish nonparametric identification, we employ a parametric finite mixture model for estimation due to computational complexity. An important research direction would be to relax the parametric assumption and develop a nonparametric estimator for a finite mixture model of the production function. This could be achieved, for instance, by extending the maximum smooth likelihood estimator proposed by \cite{lhc11}  or the estimator presented by  \cite{Bonhomme16as}.

Finally, the assumption of perfect competition or monopolistic competition with constant price elasticity may not be realistic. If data on quantities and prices are available separately, our framework can be employed to estimate the production function using the quantity of output, instead of relying on sales as a proxy for output. However, it is frequently the case that firm- or plant-level datasets do not contain output quantities and prices separately. Therefore, developing a framework for concurrently identifying the production function and demand structure from revenue data, as demonstrated in the work by \cite{kasahara_sugita2020}, while incorporating unobserved heterogeneity, is an important future research topic.
\clearpage

\bibliographystyle{apalike}

\bibliography{references}

\appendix
\section{Appendix}

\subsection{Proof of Proposition \ref{P-0}}

We partition  $\bs\theta$ as $\theta=(\theta_1',\theta_2')$ with $\theta_1:=    \{g_{v_t}(\cdot), g_{\epsilon_t\zeta_t}(\cdot), \Gamma_{M,t}(\cdot), \Gamma_{L,t}(\cdot),P_{L,t}\}_{t=1}^T$ and $\theta_2:= \{\{F_t(\cdot)\}_{t=2}^T,\{h_t(\cdot),g_{\eta,t}(\cdot)\}_{t=3}^T\}$. We drop the superscript $j$ and we have $L_{t}=\widetilde L_{t}$ and $X_{t}=\bs{\widetilde X}_{t}$ because $J=1$ and $\psi_t^1=0$ under Assumption \ref{A-6}.

We first prove the identification of $\theta_1$. Let $(s^\ell_t,s^m_t)= (\ln S_t^\ell,\ln S_{t}^m)$  and let $\Delta s_t := s_t^\ell-s_t^m$. Because $E[s_t^m | X_{t}]=\ln \left(\Gamma_{M,t}(X_{t})E_t[e^{\epsilon}]\right)$ and  $E[\Delta s_t | X_{t}]=\ln \left(\Gamma_{L,t}(X_{t})/\Gamma_{M,t}(X_{t})E_t[e^{\zeta}]\right)$, we may identify the value of $(\epsilon_{t},\zeta_{t})$ across all observations as $\epsilon_{t}=E[s_t^m | X_{t}]-s_t^{m}$ and    $\zeta_{t}=\Delta s_t-E[\Delta s_t | X_{t}]$  from the second and third equations in (\ref{system}), and $g_{\epsilon\zeta,t}(\cdot)$ is identified from the identified values of  $(\epsilon_{t},\zeta_{t})$. The marginal density functions of $\epsilon_t$ and $\zeta_t$ are also identified as $g_{\epsilon_t}(\epsilon)=\int g_{\epsilon\zeta,t}(\epsilon,\zeta)d\zeta$ and $g_{\zeta_t}(\zeta)=\int g_{\epsilon\zeta,t}(\epsilon,\zeta)d\epsilon$.
Furthermore,  because $E[s_t^m | X_t]=\ln \Gamma_{M,t}(X_t)+ \ln \int e^{\epsilon} g_{\epsilon_t}(\epsilon)d\epsilon$, we may identify $\Gamma_{M,t}(X_t)$ as $\Gamma_{M,t}(X_t)=\exp\left(E[s_t^m | X_t]-\ln \int e^{\epsilon} g_{\epsilon_t}(\epsilon)d\epsilon \right)$ and, similarly, $\Gamma_{L,t}(X_t)=\exp\left(E[\Delta s_t | X_t] +\ln \int e^{\zeta} g_{\zeta_t}(\zeta)d\zeta\right)$. Given  the identification of $\Gamma_{M,t}(X)$, $\Gamma_{L,t}(X)$, and $g_{\zeta_t}(\cdot)$, we may identify  the value of $v_t$ for all observations as $v_t =   \ln P_{M,t}M_{t}  -  \ln  \left({L_{it}\Gamma_{L,t}(X_{t}) \int e^{\zeta}g_{\zeta_t}(\zeta)d\zeta}/{\Gamma_{M,t}(X_{t}) }\right)$, and the identification of $g_{v,t}(\cdot)$ follows.  $P_{L,t}$ is identified as $\ln P_{L,t} = E_t[\ln \left({S^\ell_t P_{M,t}M_t}/{S^m_t  L_t}\right)]$ given that $v_t$ and $\zeta_t$ are mean zero random variables. This proves that  $\theta_1$ is identified.

We proceed to show that $\theta_2$ is identified. Fix  $(L_0,M_0)\in \mathcal{L}\times \mathcal{M}$ such that $L_0<L_t$ and $M_0<M_t$. Because $\frac{\Gamma_{L,t}(X_t)}{L_t}=\frac{\partial \ln F_t(X_t)}{\partial L_t}$ and $\frac{\Gamma_{M,t}(X_t)}{M_t}=\frac{\partial \ln F_t(X_t)}{\partial M_t}$,
we have
\begin{equation} \label{G-m}
\ln F_t(K_t,L_t,M_t) = \int_{L_0}^{L_t} \frac{\Gamma_{L,t}(K_t,L,M_t)}{L}dL +  \int_{M_0}^{M_t} \frac{ \Gamma_{M,t}(K_t,L_0,M)}{M}dM + \ln F_t(K_t,L_0,M_0).
\end{equation}
It follows from (\ref{prod}), (\ref{G-m}), $\epsilon_t = E[s_t^m| X_{t}]-s_t^{m}$, and $\ln Y_t+s^m_t=\ln M_t+\ln (P_{M,t}/P_{Y,t})$ that
\begin{equation} \label{omega-2}
\omega_t = \tilde y_t(X_t;\theta_1) - \ln F_t(K_t,L_0,M_0),
\end{equation}
where
\[
\tilde y_t(X_t;\theta_1) :=   \ln M_t+\ln (P_{M,t}/P_{Y,t}) -  \left\{\int_{L_0}^{L_t} \frac{ \Gamma_{L,t}(K_t,L,M_t)}{L}dL +  \int_{M_0}^{M_t} \frac{ \Gamma_{M,t}(K_t,L_0,M)}{M}dM + E[s_t^m| X_{t}]\right\}.
\]
Note that, given the identification of $\theta_1$, we may identify $\tilde y_t(X_t;\theta_1)$ for each value of $X_t$.

Substituting the right-hand side of (\ref{omega-2}) to $\omega_t = h(\omega_{t-1}) +\eta_t$ and rearranging terms give
\begin{equation}\label{reg2}
\tilde y_t(X_t;\theta_1) =   \ln F_t(K_t,L_0,M_0) + h\left(\tilde y_{t-1}(X_{t-1};\theta_1)- \ln F_{t-1}(K_{t-1},L_{0},M_0)\right)  +\eta_t,
\end{equation}
where the second term on the right hand side only depends on $X_{t-1}$.
Fix $K_0\in \mathcal{K}$ and let $C_t := \ln F_t(K_0,L_0,M_0)$. Then, by taking the conditional expectation given $K_t$ and $X_{t-1}$ in  (\ref{reg2}) and noting that $E[\eta_t|K_t,X_{t-1}]=0$, $\ln F_t(K_t,L_0,M_0)$ is identified up to constant $C_{t}$ as
\begin{equation}\label{c0}
\begin{aligned}
\ln F_t(K_t,L_0,M_0) &= C_t+E[\tilde y_t(X_t;\theta_1)|K_t,X_{t-1}]  - E[\tilde y_t(X_t;\theta_1)|K_t=K_0,X_{t-1}].
\end{aligned}
\end{equation}
It follows from the  moment  restriction $E[\omega_t]=0$ with (\ref{omega-2}) and (\ref{c0}) that we may identify $C_t$ as
\begin{align*}
C_t &= E\left\{\tilde y_t(X_t;\theta_1)-E[\tilde y_t(X_t;\theta_1)|K_t,X_{t-1}] +E[\tilde y_t(X_t;\theta_1)|K_t=K_0,X_{t-1}]\right\}.
\end{align*}
Therefore, $\ln F_t(K_t,L_0,M_0)$ is identified from (\ref{c0}), and the identification of $\ln F_t(L_t,K_t,M_t)$ for $t\geq 2$ follows from (\ref{G-m})  given that   the first two terms on the right hand side of (\ref{G-m}) is identified from $\Gamma_{L,t}(X_t)$ and $\Gamma_{M,t}(X_t)$.

Finally, we prove the identification of $g_{\eta,t}(\cdot)$ and $h_t(\cdot)$. Note that the value of $\omega_t$ for all observations is identified as $\omega_t = \tilde y_t(X_t;\theta_1) - \ln F_t(K_t,L_0,M_0)$ for $t\geq 2$. Thus, we may identify the conditional probability density function of $\omega_t$ given $\omega_{t-1}$, denoted by  $g_{\omega,t}(\omega_{t}|\omega_{t-1})$,  from the joint distribution of $\omega_t$ and $\omega_{t-1}$ for $t\geq 3$. Then, $h_t(\omega_{t-1})$ is identified as $h_t(\omega_{t-1})=E_t[\omega_t|\omega_{t-1}]=\int \omega_t g_{\omega,t}(\omega_t|\omega_{t-1})d\omega_{t}$. Given the identification of $\omega_t$, $\omega_{t-1}$, and $h_t(\cdot)$, the value of $\eta_t$ is identified as $\eta_t=\omega_t-h_t(\omega_{t-1})$ for all observations and hence   the probability density function of $\eta_t$ is identified. This proves the identification of $\theta_2$.  $\qedsymbol$

\subsection{Proof of Proposition \ref{P-1}}

For notational brevity, we drop the subscript from the probability density function by writing, for example, $g_{\{\bs S_{t},\bs{\widetilde X}_{t}\}_{t=1}^T}^j(\{\bs s_{t},\bs{\widetilde x}_{t}\}_{t=1}^T)  $ as $g^j(\{\bs s_{t},\bs{\widetilde x}_{t}\}_{t=1}^T)$. The probability density function of $\{\bs s_{t},\bs{\widetilde x}_t\}_{t=1}^T$ for type $j$ can be written as
\begin{equation}
\begin{aligned}
&g^j(\{\bs s_{t},\bs{\widetilde x}_{t}\}_{t=1}^T)  =g^j(\bs s_{1},\bs{\widetilde x}_{1}) \prod_{t=2}^Tg^j(\bs s_{t},\bs{\widetilde x}_{t}|\{\bs s_{t-s},\bs{\widetilde x}_{t-s}\}_{s=1}^{t-1})\\
&=g^j(\bs s_{1}|\bs{\widetilde x}_{1})g^j(\bs{\widetilde x}_1)  \prod_{t=2}^T  g^j(\bs s_{t}|\bs{\widetilde x}_{t},\{\bs s_{t-s},\bs{\widetilde x}_{t-s}\}_{s=1}^{t-1}) g^j(\bs{\widetilde x}_{t}|\{\bs s_{t-s},\bs{\widetilde x}_{t-s}\}_{s=1}^{t-1}).
\end{aligned} \label{mix}
\end{equation}
Define $\widetilde F_t^j(K_t,M_t,\widetilde L_t):= F_t^j(K_t,M_t,e^{\psi_t^j} \widetilde L_t)$. Similarly, define $\widetilde \Gamma_{L,t}^j(K_t,M_t,\widetilde L_t):=  \Gamma_{L,t}^j(K_t,M_t,e^{\psi_t^j} \widetilde L_t)$ and $\widetilde \Gamma_{M,t}^j(K_t,M_t,\widetilde L_t):=  \Gamma_{M,t}^j(K_t,M_t,e^{\psi_t^j} \widetilde L_t)$. Then, we may write a system of equations (\ref{system}) as
\begin{equation}\label{system2}
\begin{aligned}
\ln Y_t &= \ln \widetilde F_t^j(\bs{\widetilde X}_t) + \omega_t+ \epsilon_t,\quad
\ln S_t^m  =  \ln \left(\widetilde \Gamma_{M,t}^j(\bs{\widetilde X}_t)E_t^j[e^{\epsilon}]\right)-\epsilon_t,\\
\ln S_t^\ell-\ln S_t^m  & =  \ln \left(\frac{\widetilde \Gamma_{L,t}^j(\bs{\widetilde X}_t)}{\widetilde \Gamma_{M,t}^j(\bs{\widetilde X}_t)E_t^j[e^{\zeta}] }\right)   +\zeta_t, \quad
\ln P_{M,t}M_t  =   \ln  \left(\frac{P_{L,t}\widetilde L_t\widetilde \Gamma_{M,t}^j(\bs{\widetilde X}_t)E_t^j[e^{\zeta}]}{\widetilde \Gamma_{L,t}^j(\bs{\widetilde X}_t) }\right)  +\psi_t^j + v_t.
\end{aligned}
\end{equation}

In view of the second and the third equations of (\ref{system2}), because $\epsilon_t$ and $\zeta_t$ are i.i.d. under Assumption \ref{A-2}(c), we have
\begin{equation}
g^j(\bs s_{t}|\bs{\widetilde x}_{t},\{\bs s_{t-s},\bs{\widetilde x}_{t-s}\}_{s=1}^{t-1})=g^j(\bs s_{t}|\bs{\widetilde x}_{t}).  \label{eq-p-s}
\end{equation}
Furthermore,
\begin{equation}
\begin{aligned}
&g^j(\bs{\widetilde x}_{t}|\{\bs s_{t-s},\bs{\widetilde x}_{t-s}\}_{s=1}^{t-1}) =g^j(K_t,\omega_t,v_t|\{\bs s_{t-s}, K_{t-s},\omega_{t-s},v_{t-s}\}_{s=1}^{t-1})\\
&=g^j(\omega_t,v_t|K_t,\{\bs s_{t-s}, K_{t-s},\omega_{t-s},v_{t-s}\}_{s=1}^{t-1})g^j(K_t|\{\bs s_{t-s}, K_{t-s},\omega_{t-s},v_{t-s}\}_{s=1}^{t-1})\\
&=g^j(\omega_t|\omega_{t-1})g^j(v_t)g_t^j(K_t|K_{t-1},\omega_{t-1})\\
&=g^j(K_t,\omega_t,v_t|K_{t-1},\omega_{t-1})\\
&=g^j(K_t,\omega_t,v_t|K_{t-1},\omega_{t-1},v_{t-1})\\
&=g^j(\bs{\widetilde x}_t|\bs{\widetilde x}_{t-1}),
\end{aligned}\label{eq-p-x}
\end{equation}
where the first  equality and the last equality hold because there is a one-to-one mapping between $\bs{\widetilde X}_t$ and $(K_t,\omega_t,v_t)$ given $\psi_t^j$  in view of Assumption \ref{A-4}(b); the third equality follows from  Assumptions \ref{A-2}(a) and \ref{A-3}(b);  the fifth equality holds because $v_{t-1}$ is i.i.d. and, thus, independent of $(K_t,\omega_t,v_t)$. Therefore, the stated result follows from   (\ref{mix}), (\ref{eq-p-s}), and (\ref{eq-p-x}).
 $\qedsymbol$

\subsection{Proof of Proposition  \ref{P-2}}

We apply the argument of \cite{Kasahara09}, \cite{carolletal10jns},  and \cite{Hu12}
under the assumption that unobserved heterogeneity is permanent and discrete.   The proof is constructive.

Consider the case that $T=4$. For each value of ${\bs z}_3\in \mathcal{Z}_3$, choose $(\check{\bs z}_2,\bar {\bs z}_2,\bar {\bs z}_3)\in \mathcal{Z}_2\times  \mathcal{Z}_2\times \mathcal{Z}_3$, $(\bs a_1,...,\bs a_{J})\in \mathcal{Z}_1^{J}$, and $(\bs b_1,...,\bs b_{J-1})\in  \mathcal{Z}_4^{J-1}$ that satisfy Assumption \ref{A-P-2}. Evaluating (\ref{mixture-2}) for $T=4$ at $({\bs Z}_1,{\bs Z}_2,{\bs Z}_3,{\bs Z}_4)=(\bs a,{\bs z}_2,{\bs z}_3,\bs b)$ gives
\begin{equation}
\begin{aligned}
g_{{\bs Z}_1,{\bs Z}_2,{\bs Z}_3,{\bs Z}_4}(\bs a,{\bs z}_2,{\bs z}_3,\bs b) &= \sum_{j=1}^J  \pi^j g_{{\bs Z}_1}^j(\bs a) g_{{\bs Z}_2|{\bs Z}_1}^j({\bs z}_2|\bs a)  g_{{\bs Z}_3|{\bs Z}_2}^j({\bs z}_3|{\bs z}_2)  g_{{\bs Z}_4|{\bs Z}_3}^j(\bs b|\bs z_3)  \\
&=\sum_{z=1}^J\bar \lambda^j_2(\bs a,{\bs z}_2)\lambda^j_3({\bs z}_3|{\bs z}_2) \lambda^j_{4}(\bs b|{\bs z}_3),
\end{aligned} \label{p-w}
\end{equation}
where $\bar \lambda^j_2(\bs a,{\bs z}_2):=\pi^j g_{{\bs Z}_1}^j(\bs a)g_{{\bs Z}_2|{\bs Z}_1}^j({\bs z}_2|\bs a)$, $\lambda^j_3({\bs z}_3|{\bs z}_2):= g_{{\bs Z}_3|{\bs Z}_2}^j({\bs z}_3|{\bs z}_2)$, and  $\lambda^j_{4}(\bs b|{\bs z}_3):=g_{{\bs Z}_4|{\bs Z}_3}^j(\bs b|{\bs z}_3)$. Similarly, evaluating (\ref{mixture-s2}) for $T=3$ at $({\bs Z}_1,{\bs Z}_2,{\bs Z}_3)=(\bs a,{\bs z}_2,{\bs z}_3)$ gives
\begin{equation}
g_{{\bs Z}_1,{\bs Z}_2,{\bs Z}_3}(\bs a,{\bs z}_2,{\bs z}_3)   =\sum_{j=1}^J \bar \lambda^j_2(\bs a,{\bs z}_2) \lambda^j_3({\bs z}_3|{\bs z}_2). \label{p-w-2}
\end{equation}
Denote
$q_{{\bs z}_2,{\bs z}_3}(\bs a,\bs b) :=g_{{\bs Z}_1,{\bs Z}_2,{\bs Z}_3,{\bs Z}_4}(\bs a,{\bs z}_2,{\bs z}_3,\bs b)$ and  $\bar q_{{\bs z}_2,{\bs z}_3}(\bs a) :=g_{{\bs Z}_1,{\bs Z}_2,{\bs Z}_3}(\bs a,{\bs z}_2,{\bs z}_3)$. Evaluating (\ref{p-w}) at $\bs a = \bs a_1,...,\bs a_{J}$ and $\bs b= \bs b_1,...,\bs b_{J-1}$ gives $J(J-1)$ equations while evaluating (\ref{p-w-2}) at $\bs a = \bs a_1,...,\bs a_{J}$ gives $J$ equations.

Using matrix notation, we collect  these $J(J-1)+J=J^2$ equations as
\begin{equation}
\bs Q_{{\bs z}_2,{\bs z}_3} = {\bs L}_{{\bs z}_3} {\bs D}_{{\bs z}_3|{\bs z}_2} \bar {\bs L}_{{\bs z}_2}^\top, \label{decomp}
\end{equation}
where ${\bs L}_{{\bs z}_3}$, $\bar {\bs L}_{{\bs z}_2}$, and ${\bs D}_{{\bs z}_3|{\bs z}_2} $ are defined in (\ref{LL}) while
\begin{equation}
\begin{aligned}
&\bs Q_{{\bs z}_2,{\bs z}_3}:= \left[
\begin{array}{cccc}
\bar q_{{\bs z}_2,{\bs z}_3}(\bs a_1) &\bar q_{{\bs z}_2,{\bs z}_3}(\bs a_2) &\cdots &\bar q_{{\bs z}_2,{\bs z}_3}(\bs a_J) \\
q_{{\bs z}_2,{\bs z}_3}(\bs a_1,\bs b_1) &q_{{\bs z}_2,{\bs z}_3}(\bs a_2,\bs b_1) &\cdots &q_{{\bs z}_2,{\bs z}_3}(\bs a_J,\bs b_1) \\
\vdots & \vdots & \ddots & \vdots \\
q_{{\bs z}_2,{\bs z}_3}(\bs a_1,\bs b_{J-1}) &q_{{\bs z}_2,{\bs z}_3}(\bs a_2,\bs b_{J-1}) &\cdots &q_{{\bs z}_2,{\bs z}_3}(\bs a_J,\bs b_{J-1}) \\
\end{array}
\right].
\end{aligned} \label{L}
\end{equation}
Let $\bs z_3^*$ be the value of $\bs z_3$ as defined in Assumption Assumption \ref{A-P-2}. For each $\bs z_3$, choose $\check{\bs z}_2$, $\bar {\bs z}_2$, and $\bar {\bs z}_3$ that satisfy Assumption \ref{A-P-2}(a)(b).
Evaluating (\ref{decomp}) at four different points, $(\check{\bs z}_2,{\bs z}_3^*)$, $(\bar {\bs z}_2,{\bs z}_3)$, $(\check{\bs z}_2,\bar {\bs z}_3)$, and $(\bar {\bs z}_2,\bar {\bs z}_3)$ gives
\[
\begin{aligned}
\bs Q_{\check{\bs z}_2,{\bs z}_3} &= {\bs L}_{{\bs z}_3} {\bs D}_{{\bs z}_3|\check{\bs z}_2} \bar {\bs L}_{\check{\bs z}_2}^\top,\quad
\bs Q_{\bar {\bs z}_2,\bar {\bs z}_3}  = {\bs L}_{\bar {\bs z}_3} {\bs D}_{\bar {\bs z}_3|\bar {\bs z}_2} \bar {\bs L}_{\bar {\bs z}_2}^\top, \\
\bs Q_{\check{\bs z}_2,\bar {\bs z}_3} &= {\bs L}_{\bar {\bs z}_3} {\bs D}_{\bar {\bs z}_3|\check{\bs z}_2} \bar {\bs L}_{\check{\bs z}_2}^\top,\quad \bs Q_{\bar {\bs z}_2,{\bs z}_3^*}  = {\bs L}_{{\bs z}_3^*}{\bs D}_{{\bs z}_3^*|\bar {\bs z}_2} \bar {\bs L}_{\bar {\bs z}_2}^\top.
\end{aligned}
\]
Then, following the identification argument in  \cite{carolletal10jns}, under Assumption \ref{A-P-2}(a)(c), we have
\begin{equation}\label{A_z}
\bs A_{\bs z_3^*,\bs z_3}   := \bs Q_{\check{\bs z}_2,{\bs z}_3} \bs Q_{\check{\bs z}_2,\bar {\bs z}_3}^{-1} \bs Q_{\bar {\bs z}_2,\bar {\bs z}_3}  \bs Q_{\bar {\bs z}_2,{\bs z}_3^*}^{-1}   = {\bs L}_{{\bs z}_3}  {\bs D}_{ {\bs z}_3^*,  {\bs z}_3} {\bs L}_{{\bs z}_3^*} ^{-1},
\end{equation}
where
\begin{equation}
 {\bs D}_{ {\bs z}_3^*,  {\bs z}_3} : = {\bs D}_{{\bs z}_3|\check{\bs z}_2} {\bs D}_{\bar {\bs z}_3|\check{\bs z}_2} ^{-1} {\bs D}_{\bar {\bs z}_3|\bar {\bs z}_2}{\bs D}_{{\bs z}_3^*|\bar {\bs z}_2}^{-1}. \label{D-z}
\end{equation}

We  first identify ${\bs L}_{{\bs z}_3}$ for all $\bs z_3\in \mathcal{Z}_3$ up to an unknown permutation matrix. Evaluating (\ref{A_z}) at $\bs z_3^*=\bs z_3$, we have
\[
\bs A_{\bs z_3,\bs z_3}  {\bs L}_{{\bs z}_3}= {\bs L}_{{\bs z}_3} {\bs D}_{{\bs z}_3, {\bs z}_3}.
\]
Because $\bs A_{\bs z_3,\bs z_3}$ has $J$ distinct eigenvalues under Assumption \ref{A-P-2}(b),  the eignvalues of $\bs A_{\bs z_3,\bs z_3}$ determine the diagonal elements of $ {\bs D}_{ {\bs z}_3,  {\bs z}_3} $ while the right eigenvectors of $\bs A_{\bs z_3,\bs z_3} $ determine the columns of ${\bs L}_{{\bs z}_3} $ up to multiplicative constant and the ordering of its columns.
Namely, collecting the right eigenvectors of $\bs A_{\bs z_3,\bs z_3}$  into a matrix in descending order of their eigenvalues, we identify
\[
\bs B   :=  {\bs L}_{{\bs z}_3} \Delta_{{\bs z}_3}  \bs C,
\]
where $\bs B$ satisfies $\bs A_{\bs z_3,\bs z_3} \bs B =\bs B {\bs D}_{\bs z_3,\bs z_3}$,
 $\Delta_{{\bs z}_3}$ is an unknown permutation matrix, and $\bs C$ is some diagonal matrix with non-zero diagonal elements. 

We can determine the diagonal matrix $\bs C{\bs D}_{\bs z_3,\bs z_3}$ from the first row of $\bs A_{\bs z_3,\bs z_3} \bs B  =\bs B {\bs D}_{\bs z_3,\bs z_3} =  {\bs L}_{{\bs z}_3} \Delta_{{\bs z}_3} \bs C {\bs D}_{\bs z_3,\bs z_3} $  because the first row of ${\bs L}_{{\bs z}_3}  \Delta_{{\bs z}_3}  $ is a vector of ones. Then, $ {\bs L}_{{\bs z}_3}  \Delta_{{\bs z}_3}$ is determined from $\bs A_{\bs z_3,\bs z_3}  \bs B $ and $\bs C{\bs D}_{\bs z_3,\bs z_3} $ as $  {\bs L}_{{\bs z}_3}\Delta_{{\bs z}_3} = \bs A_{\bs z_3,\bs z_3}  \bs B  (\bs C{\bs D}_{\bs z_3,\bs z_3})^{-1}$ in view of $\bs A_{\bs z_3,\bs z_3} \bs B  = {\bs L}_{{\bs z}_3} \Delta_{{\bs z}_3} \bs C {\bs D}_{\bs z_3,\bs z_3} $.   Repeating the above argument for all values of ${\bs z}_3\in\mathcal{Z}_3$, the eigenvalue decomposition algorithm identifies the matrices
\begin{equation}\label{L-tilde}
\tilde  {\bs L}_{{\bs z}_3}:= {\bs L}_{{\bs z}_3}  \Delta_{{\bs z}_3}\quad\text{for all ${\bs z}_3\in\mathcal{Z}_3$},
\end{equation}
where $\Delta_{{\bs z}_3}$ is an unknown permutation matrix that depends on $\bs z_3$.

Next, we identify permutation matrices that re-arrange ${\bs L}_{{\bs z}_3} \Delta_{{\bs z}_3} $ in a common order of latent types across different values of $\bs z_3$ using the identification argument in  \cite{HigginsJochmans21}. Pre- and post- multiplying (\ref{A_z})  by $\tilde  {\bs L}_{{\bs z}_3}^{-1}$ and  $\tilde  {\bs L}_{\bs z_3^*}$, respectively, we have
\[
\tilde {\bs D}_{ \bs z_3^*,  {\bs z}_3} := \tilde  {\bs L}_{{\bs z}_3}^{-1} \bs A_{\bs z_3^*,\bs z_3}  \tilde  {\bs L}_{\bs z_3^*} =  \Delta_{{\bs z}_3}^{-1} {\bs D}_{ \bs z_3^*,  {\bs z}_3} \Delta_{\bs z_3^*}= \Delta_{{\bs z}_3}^{-1} \Delta_{\bs z_3^*}  \left(\Delta_{\bs z_3^*}^{-1}  {\bs D}_{ \bs z_3^*,  {\bs z}_3} \Delta_{\bs z_3^*} \right),
\]
where the last equality uses the fact that $\Delta_{\bs z_3^*}   \Delta_{\bs z_3^*}^{-1}$ is an identity matrix. Because $\Delta_{{\bs z}_3}^{-1} \Delta_{\bs z_3^*}$ is a permutation matrix, $\tilde {\bs D}_{ \bs z_3^*,  {\bs z}_3}$ is a matrix obtained by permutating the rows of the diagonal matrix $\Delta_{\bs z_3^*}^{-1}  {\bs D}_{ \bs z_3^*,  {\bs z}_3} \Delta_{\bs z_3^*} $. Therefore, each diagonal element of  $\Delta_{\bs z_3^*}^{-1}  {\bs D}_{ \bs z_3^*,  {\bs z}_3} \Delta_{\bs z_3^*}$ is identified with the sum of elements in the corresponding column of $\tilde {\bs D}_{ \bs z_3^*,  {\bs z}_3} $, and the identification of $\Delta_{\bs z_3^*}^{-1}  {\bs D}_{ \bs z_3^*,  {\bs z}_3} \Delta_{\bs z_3^*}$ follows.
Then, we may identify  $ \Delta_{{\bs z}_3}^{-1} \Delta_{\bs z_3^*} $ as $  \Delta_{{\bs z}_3}^{-1} \Delta_{\bs z_3^*} = \tilde {\bs D}_{ \bs z_3^*,  {\bs z}_3}\left(\Delta_{\bs z_3^*}^{-1}  {\bs D}_{ \bs z_3^*,  {\bs z}_3} \Delta_{\bs z_3^*} \right)^{-1}$. Therefore, $ {\bs L}_{{\bs z}_3} $ is identified up to a common permutation matrix $ \Delta_{\bs z_3^*} $ that does not depend on $\bs z_3$ from (\ref{L-tilde}) as
\begin{equation}\label{L-star}
 {\bs L}_{{\bs z}_3}^*:=  {\bs L}_{{\bs z}_3}  \Delta_{\bs z_3^*} = \tilde  {\bs L}_{{\bs z}_3}  \Delta_{{\bs z}_3}^{-1} \Delta_{\bs z_3^*} =\tilde  {\bs L}_{{\bs z}_3} \tilde {\bs D}_{ \bs z_3^*,  {\bs z}_3}\left(\Delta_{\bs z_3^*}^{-1}  {\bs D}_{ \bs z_3^*,  {\bs z}_3} \Delta_{\bs z_3^*} \right)^{-1}.
 \end{equation}

In the next step, we identify $\{\pi^j, g_{{\bs Z}_1}^j({\bs z}_1),g^j_{{\bs Z}_2|{\bs Z}_1}({\bs z}_2|{\bs z}_1),g^j_{{\bs Z}_3|{\bs Z}_2}({\bs z}_3|{\bs z}_2),g^j_{{\bs Z}_4|{\bs Z}_3}({\bs z}_4|{\bs z}_3)\}_{j=1}^J$ up to a permutation matrix $\Delta_{{\bs z}_3^*}$. For this purpose,
we evaluate $g_{{\bs Z}_3,{\bs Z}_4|{\bs Z}_2}({\bs Z}_3,{\bs Z}_4|{\bs Z}_2)=\sum_{j=1}^J \pi^j g^j_{{\bs Z}_2,{\bs Z}_3,{\bs Z}_4}({\bs Z}_2,{\bs Z}_3,{\bs Z}_4)/\sum_{k=1}^J \pi^k g^k_{{\bs Z}_2}({\bs Z}_2)$ at $({\bs Z}_2,{\bs Z}_3,{\bs Z}_4)=({\bs z}_2,{\bs z}_3,\bs b)$ as
\begin{equation}
\begin{aligned}
g_{{\bs Z}_3,{\bs Z}_4|{\bs Z}_2}({\bs z}_3,\bs b|{\bs z}_2) &=\frac{ \sum_{j=1}^J  \pi^j g^j_{{\bs Z}_2}({\bs z}_2)g_{{\bs Z}_3|{\bs Z}_2}^j({\bs z}_3|{\bs z}_2) g_{{\bs Z}_4|{\bs Z}_3}^j(\bs b|{\bs z}_3) }{\sum_{j=1}^J \pi^j g_{{\bs Z}_2}^j({\bs z}_2)}\\
&= \sum_{j=1}^J  \tilde\pi_{{\bs z}_2}^j g_{{\bs Z}_3|{\bs Z}_2}^j({\bs z}_3|{\bs z}_2) \underbrace{g_{{\bs Z}_4|{\bs Z}_3}^j(\bs b|{\bs z}_3)}_{= \lambda^j_{4}(\bs b|{\bs z}_3)}, 
\end{aligned} \label{p-w-3}
\end{equation}
where $  \tilde\pi_{{\bs z}_2}^j  :=  \pi^j g^j_{{\bs Z}_2}({\bs z}_2)/\sum_{k=1}^J  \pi^k g^k_{{\bs Z}_2}({\bs z}_2)$. Then, evaluating (\ref{p-w-3})  at  $\bs b= \bs b_1,...,\bs b_{J-1}$ and  collecting them into a vector together with $g_{{\bs Z}_3|{\bs Z}_2}({\bs z}_3|{\bs z}_2)   = \sum_{j=1}^J \tilde\pi_{{\bs z}_2}^j g_{{\bs Z}_3|{\bs Z}_2}^j({\bs z}_3|{\bs z}_2)$ gives
\begin{equation}\label{r}
\bs r_{{\bs z}_3|{\bs z}_2}  = {\bs L}_{{\bs z}_3}   \bs d_{{\bs z}_3|{\bs z}_2}, = {\bs L}_{{\bs z}_3}^* \Delta_{\bs z_3^*}^{-1}   \bs d_{{\bs z}_3|{\bs z}_2}
\end{equation}
with
\[
\bs r_{{\bs z}_3|{\bs z}_2} =
\begin{pmatrix}
g_{{\bs Z}_3|{\bs Z}_2}({\bs z}_3|{\bs z}_2)\\
g_{{\bs Z}_3,{\bs Z}_4|{\bs Z}_2}({\bs z}_3,\bs b_1|{\bs z}_2)\\
\vdots\\
g_{{\bs Z}_3,{\bs Z}_4|{\bs Z}_2}({\bs z}_3,\bs b_{J-1}|{\bs z}_2)
\end{pmatrix}\quad\text{and}\quad
\bs d_{{\bs z}_3|{\bs z}_2}=
\begin{pmatrix}
d_{{\bs z}_3|{\bs z}_2}^1\\
\vdots\\
d_{{\bs z}_3|{\bs z}_2}^J
\end{pmatrix}
:=
\begin{pmatrix}
\tilde\pi_{{\bs z}_2}^1g_{{\bs Z}_3|{\bs Z}_2}^1({\bs z}_3|{\bs z}_2)\\
\vdots\\
\tilde\pi_{{\bs z}_2}^Jg_{{\bs Z}_3|{\bs Z}_2}^J({\bs z}_3|{\bs z}_2)
\end{pmatrix},
\]
where the last equality in (\ref{r}) follows from (\ref{L-star}).
  Therefore,  from (\ref{L-star}) and (\ref{r}), we identify $\tilde\pi_{{\bs z}_2}^j g^j_{{\bs Z}_3|{\bs Z}_2}({\bs z}_3|{\bs z}_2)$ for all values of $(\bs z_2,{\bs z}_3) \in  \mathcal{Z}_3\times\mathcal{Z}_3$ up to $\Delta_{\bs z_3^*}$  as
  \begin{equation}\label{d-3}
 \Delta_{\bs z_3^*}^{-1} \bs d_{{\bs z}_3|{\bs z}_2} :=\begin{pmatrix}
d_{{\bs z}_3|{\bs z}_2}^{\alpha(1)} \\
\vdots\\
d_{{\bs z}_3|{\bs z}_2}^{\alpha(J)}
\end{pmatrix}=\left({\bs L}_{{\bs z}_3}^* \right)^{-1}\bs r_{{\bs z}_3|{\bs z}_2},
  \end{equation}
  where
  \[
  \alpha: \{1,2,...,J\} \rightarrow \{1,2,...,J\}
  \]
  is a permutation implied by $ \Delta_{\bs z_3^*}^{-1}$.
  Furthermore, because  $\tilde \pi_{{\bs z}_2}^j = \int_{\mathcal{Z}_3} \tilde \pi_{{\bs z}_2}^j g^j_{{\bs Z}_3|{\bs Z}_2}({\bs z}_3|{\bs z}_2)d{\bs z}_3$ and $ g^j_{{\bs Z}_3|{\bs Z}_2}({\bs z}_3|{\bs z}_2)=[\tilde \pi_{{\bs z}_2}^j  g^j_{{\bs Z}_3|{\bs Z}_2}({\bs z}_3|{\bs z}_2)]/\tilde \pi_{{\bs z}_2}^j$, we may identify  $g^{\alpha(j)}_{{\bs Z}_3|{\bs Z}_2}({\bs z}_3|{\bs z}_2)$  from $ d_{{\bs z}_3|{\bs z}_2}^{\alpha(j)}$ as
  \begin{align}
  g^{\alpha(j)}_{{\bs Z}_3|{\bs Z}_2}({\bs z}_3|{\bs z}_2) & := \frac{d_{{\bs z}_3|{\bs z}_2}^{\alpha(j)}}{  \int_{\mathcal{Z}_3} d_{{\bs z}_3'|{\bs z}_2}^{\alpha(j)}  d{\bs z}_3'}. \label{g3}
  \end{align}
  Then, we may identify   ${\bs D}_{{\bs z}_3|{\bs z}_2}$ up to   $\Delta_{\bs z_3^*}$  as
  \begin{equation}\label{D23}
 \Delta_{\bs z_3^*}^{-1} {\bs D}_{{\bs z}_3|{\bs z}_2} \Delta_{\bs z_3^*} = \text{diag}\left( g^{\alpha(1)}_{{\bs Z}_3|{\bs Z}_2}({\bs z}_3|{\bs z}_2),...,  g^{\alpha(J)}_{{\bs Z}_3|{\bs Z}_2}({\bs z}_3|{\bs z}_2)\right),
  \end{equation}
and $\bar {\bs L}_{{\bs z}_2}^\top$ is identified from (\ref{decomp}), (\ref{L-star}), and (\ref{D23}) up to $ \Delta_{\bs z_3^*}$  as
   \begin{equation}\label{Lz2}
  \Delta_{\bs z_3^*}^{-1}  \bar {\bs L}_{{\bs z}_2}^\top = (\Delta_{\bs z_3^*}^{-1}  {\bs D}_{{\bs z}_3|{\bs z}_2}  \Delta_{{\bs z}_3^*})^{-1}  \left({\bs L}_{{\bs z}_3}^*\right)^{-1}\bs Q_{{\bs z}_2,{\bs z}_3},
 \end{equation}
 where the invertibility of $\bs D_{{\bs z}_3|{\bs z}_2}$  follows from Assumption \ref{A-P-2}(c).

Once ${\bs D}_{{\bs z}_3|{\bs z}_2}$ and $\bar {\bs L}_{{\bs z}_2}$ are identified up to $ \Delta_{{\bs z}_3^*}$ as in (\ref{D23})-(\ref{Lz2}), we determine $\bs \ell_{{\bs z}_3}(\bs z_4):=
  (\lambda_4^1(\bs z_4|{\bs z}_3),...,\lambda_4^J(\bs z_4|{\bs z}_3))=(g_{{\bs Z}_4|{\bs Z}_3}^1(\bs z_4|{\bs z}_3),...,g_{{\bs Z}_4|{\bs Z}_3}^J(\bs z_4|{\bs z}_3))$ for any $(\bs z_3,\bs z_4)\in \mathcal{Z}_3\times \mathcal{Z}_4$  up to $ \Delta_{{\bs z}_3^*}$  by constructing
\[
{\bs p}_{{\bs z}_2,{\bs z}_3}(\bs z_4):= (q_{{\bs z}_2,{\bs z}_3}(\bs a_1,\bs z_4) ,q_{{\bs z}_2,{\bs z}_3}(\bs a_2,\bs z_4),...,q_{{\bs z}_2,{\bs z}_3}(\bs a_J,\bs z_4))
\] from the observed data, and using the relationship
\begin{equation}\label{g43}
 \bs  \ell_{{\bs z}_3}(\bs z_4)  \Delta_{\bs z_3^*} = \left(g_{{\bs Z}_4|{\bs Z}_3}^{\alpha(1)}(\bs z_4|{\bs z}_3),...,g_{{\bs Z}_4|{\bs Z}_3}^{\alpha(J)}(\bs z_4|{\bs z}_3)\right)=   {\bs p}_{{\bs z}_2,{\bs z}_3}(\bs z_4) ( \Delta_{\bs z_3^*}^{-1}  \bar {\bs L}_{{\bs z}_2}^\top)^{-1} (\Delta_{\bs z_3^*}^{-1}  {\bs D}_{{\bs z}_3|{\bs z}_2}  \Delta_{{\bs z}_3^*})^{-1}
\end{equation}
for all values of $(\bs z_3,\bs z_4)\in\mathcal{Z}_3\times \mathcal{Z}_4$. Therefore, $\{g_{{\bs Z}_4|{\bs Z}_3}^j(\bs z_4|{\bs z}_3)\}_{j=1}^J$ is identified up to $\Delta_{{\bs z}_3^*}$.

 Similarly, we determine $\bar {\bs \ell}_{{\bs z}_2}(\bs z_1):=(\bar \lambda_2^1(\bs z_1,{\bs z}_2),...,\bar \lambda_2^J(\bs z_1,{\bs z}_2))^\top= (\pi^1 g_{{\bs Z}_2|{\bs Z}_1}^1({\bs z}_2|\bs a)g_{{\bs Z}_1}^1(\bs z_1),...,\\ \pi^J g_{{\bs Z}_2|{\bs Z}_1}^J({\bs z}_2|\bs a)g_{{\bs Z}_1}^J(\bs z_1))^\top$
up to $\Delta_{\bs z_3^*}$  for any $(\bs z_1,\bs z_2)\in \mathcal{Z}_1\times\mathcal{Z}_2$ from (\ref{L-star}) and (\ref{D23}) by constructing
\[
\bar {\bs p}_{{\bs z}_2,{\bs z}_3}(\bs z_1):= (\bar q_{{\bs z}_2,{\bs z}_3}(\bs z_1),q_{{\bs z}_2,{\bs z}_3}(\bs z_1,\bs b_1) ,q_{{\bs z}_2,{\bs z}_3}(\bs z_1,\bs b_2),...,q_{{\bs z}_2,{\bs z}_3}(\bs z_1,\bs b_{J-1}))
\] and using the relationship
\begin{equation}\label{lz2}
  \Delta_{\bs z_3^*}^{-1}\bar {\bs \ell}_{{\bs z}_2}(\bs z_1)
=
\begin{pmatrix}
\bar \lambda_2^{\alpha(1)}(\bs z_1,{\bs z}_2)\\
\vdots\\
\bar \lambda_2^{\alpha(J)}(\bs z_1,{\bs z}_2)
\end{pmatrix} =  (\Delta_{\bs z_3^*}^{-1}  {\bs D}_{{\bs z}_3|{\bs z}_2}  \Delta_{{\bs z}_3^*})^{-1}  \left({\bs L}_{{\bs z}_3}^*\right)^{-1} \bar {\bs p}_{{\bs z}_2,{\bs z}_3}(\bs z_1)^\top.
\end{equation}
Then, $\{\pi^j, g_{{\bs Z}_1}^j({\bs z}_1),g_{{\bs Z}_2|{\bs Z}_1}^j({\bs z}_2|{\bs z}_1)\}_{j=1}^J$ is identified up to $\Delta_{{\bs z}_3^*}$  from $\{\bar \lambda_2^{\alpha(j)}(\bs z_1,{\bs z}_2)\}_{j=1}^J$ in (\ref{lz2}) given $\bar \lambda_2^j(\bs z_1,{\bs z}_2)= \pi^j g_{{\bs Z}_1}^j({\bs z}_1)g_{{\bs Z}_2}^j({\bs z}_2|{\bs z}_1)$ as
\begin{align}
&\pi^j := \int_{\mathcal{Z}_1}\int_{\mathcal{Z}_2} \bar \lambda_2^j(\bs z_1,{\bs z}_2)d{\bs z}_2d{\bs z}_1,\quad
g_{{\bs Z}_1}^j({\bs z}_1):=\frac{\int_{\mathcal{Z}_2} \bar \lambda_2^j(\bs z_1,{\bs z}_2)d{\bs z}_2}{\pi^j  },\nonumber\\
 &\text{and}\quad g_{{\bs Z}_2|{\bs Z}_1}^j({\bs z}_2|{\bs z}_1):=\frac{ \bar \lambda_2^j(\bs z_1,{\bs z}_2)}{\pi^j \times g_{{\bs Z}_1}^j({\bs z}_1)}\quad \text{for $j=1,...,J$.} \label{g12}
\end{align}
Therefore, we identify $\{\pi^j, g_{{\bs Z}_1}^j({\bs z}_1),g^j_{{\bs Z}_2|{\bs Z}_1}({\bs z}_2|{\bs z}_1),g^j_{{\bs Z}_3|{\bs Z}_2}({\bs z}_3|{\bs z}_2),g^j_{{\bs Z}_4|{\bs Z}_3}({\bs z}_4|{\bs z}_3)\}_{j=1}^J$ up to a permutation matrix $\Delta_{{\bs z}_3^*}$. $\qedsymbol$

\subsection{Proof of Proposition  \ref{P-3}}

We first show that $P_{L,t}$ and $\{\psi_t^j\}_{j=1}^J$ are identified from $\{\pi^j,g_{B_t}^j(B_t)\}_{j=1}^J$. Because $E^j_t[\ln B_{t}]= \ln (P_{L,t} e^{\psi_t^j})$, we may have $\psi_t^j = E^j_t[\ln B_{t}] - \ln P_{L,t}$ for $j=1,...,J$, where $E^j_t[\ln B_{t}]$ is identified from $g_{Z_t}^j(Z_t)$. Then, $P_{L,t}$ is identified from $\sum_{j=1}^J \pi^j e^{E^j_t[\ln B_{t}] - \ln P_{L,t}}=1$ as $P_{L,t}= \sum_{j=1}^J \pi^j e^{E^j_t[\ln B_{t}] }$.
Once $P_{L,t}$ and $\{\psi_t^j\}_{j=1}^J$ are identified,  then repeating the argument in the proof of Proposition \ref{P-0} for each type  proves the stated result.
 $\qedsymbol$

\subsection{Proof of Proposition \ref{consistency}}
Our panel data model belongs to a class of multivariate normal mixture models analyzed in \cite{chentan09jmva}.  \cite{chentan09jmva} provides the consistency proof under their conditions C1-C3  but \cite{Alexandrovich2014} identifies a soft spot in the proof of  \cite{chentan09jmva} and provides an alternative consistency proof by strengthening the condition C3 of  \cite{chentan09jmva}.

Let $\bs\Omega_j$ be the variance matrix for the vector of random variables $(\bs\epsilon_i',\bs\zeta_i',\bs v',k_{i1},\omega_{i1},\bs\xi_i')'$ for type $j$, where $\bs\epsilon_i=(\epsilon_{i1},...,\epsilon_{iT})'$, $\bs\zeta_i = (\zeta_{i1},...,\zeta_{iT})'$, $\bs v_i=(v_{i1},...,v_{iT})'$, and $\bs\xi_i = (k_{i2}-(\rho_{k0}^j+\rho_{kk}^jk_{i1}+\rho_{k\omega}\omega_{i1}),...,k_{iT}-(\rho_{k0}^j+\rho_{kk}^jk_{iT-1}+\rho_{k\omega}\omega_{iT-1})'$.
Using our notations, the condition C1-C2  in \cite{chentan09jmva} and the condition C3 strengthned by \cite{Alexandrovich2014} are stated as follows:
\begin{description}
\item[C1.]
 The penalty function is written as $\tilde p_n(\bs\theta)=\sum_{j=1}^J p_n(\bs{\Omega}_j)$.
 \item[C2.] For any  fixed $\bs\theta$ with $\text{det}(\bs\Omega_j)>0$ for $j=1,2,...,J$, we have $\tilde p_n(\bs\theta)=o(n)$ and $\sup_{\bs\theta\in\bs\Theta} \max\{0,\tilde p_n(\bs\theta)\}=o(n)$. In addition, $\tilde p_n(\bs\theta)$ is differentiable with respect to $\bs\theta$ and as $n\rightarrow \infty$, $\nabla_{\bs\theta} \tilde p_n(\bs\theta)=o(n^{1/2})$ at any fixed $\bs\theta$ such that $\text{det}(\bs\Omega_j)>0$ for $j=1,2,...,J.$
\item[C3 by \cite{Alexandrovich2014}.] For large enough $n$, $p_n(\bs\Omega_j) \leq \left(\frac{3}{4}\sqrt{n\log\log n}\right)\log(\text{det}(\bs\Omega_j))$, when  $\text{det}(\bs\Omega_j)<c n^{-2}$ for some $c>0$.
\end{description}
The consistency and the asymptotic normality of the PMLE, $\hat {\bs\theta}$,  follows from Theorems 1 and 2 of \cite{chentan09jmva} and Corollary 3 of \cite{Alexandrovich2014} if we can show that the above three conditions hold for our penalty function  defined in (\ref{pmle}).   C1 trivially holds with  $p_n(\bs{\Omega}_j)= \sum_{s\in\{\epsilon,\zeta,v,k,\eta\}} p_n((\sigma_s^j)^2;\hat\sigma_{s,0}^2)   + p_n(\bs\Sigma_1^j;\hat{\bs\Sigma}_{1,0}^j)$.    C2 also holds because $\nabla_{\sigma_s^j} p_{n}((\sigma_s^j)^2;\hat\sigma_{s,0}^2) $ for $s\in\{\epsilon,\zeta,v,k,\eta\}$ and $\nabla_{\text{vech}(\bs\Sigma_1^j)} p_n(\bs\Sigma_1^j;\hat{\bs\Sigma}_{1,0}^j)$ are $O(n^{-1})$ for $\sigma_s^j>0$ and $\text{det}(\bs\Sigma_1^j)>0$. For C3, suppose that $(\sigma_{s}^j)^2< n^{-2}$  for $s\in\{\epsilon,\zeta,v,k,\eta\}$.  Then,  $p_n((\sigma_s^j)^2;\hat\sigma_{s,0}^2) =- n^{-1}  \left\{ \hat\sigma_{s,0}^2/(\sigma_s^j)^2 - \log(\hat\sigma_{s,0}^2/(\sigma_s^j)^2)\right\}<-n\hat\sigma_{s,0}^j +n^{-1}\log(\hat\sigma_{s,0}^j) + n^{-1} 2\log(n)
< - \frac{3}{4} \sqrt{n\log\log n}\times 2\log n$ for large $n$. Similarly, we may show that, if $\text{det}(\bs\Sigma_1^j)< n^{-2}$, then $p_n(\bs\Sigma_1^j;\hat{\bs\Sigma}_{1,0}^j) < - \frac{3}{4} \sqrt{n\log\log n}\times 4\log n$ for large $n$. Therefore, C3 holds.  Consequently,  $\tilde p_n(\bs\theta_M)$ satisfies the above three conditions, and  the stated result follows follows from Theorems 1 and 2 of \cite{chentan09jmva} and Corollary 3 of \cite{Alexandrovich2014}.

\newpage

\section{Online Appendix}

\subsection{Assumption \ref{A-P-2} under Cobb-Douglas production function }\label{appendix-6}

In the following, we discuss the conditions under which Assumption \ref{A-P-2} holds when the production function is Cobb-Douglas.
\setcounter{example}{0}
\begin{example}[continued]   To simplify our identification analysis, we also assume the followings.  First,
we fix the value of $\{\tilde {\bs X}_t\}_{t=1}^T$ at, say, $\{\tilde {\bs x}_t\}_{t=1}^4$ so that the variation in the values of $\bs a_j$'s and $\bs b_j$'s are due the variation in the values of $\bs S_1$ and $\bs S_4$. Let $\bs a_j= (\bar{S}^m_{1,j},\bar{S}^\ell_{1,j}, \tilde{\bs x}_1)'$ for $j=1,...,J$ and let $\bs b_j =  (\bar{S}^m_{4,j},\bar{S}^\ell_{4,j},  \tilde{\bs x}_4)'$ for $j=1,...,J-1$.  Second,   $\ln \bar S^\ell_{4,j}-\ln \bar S^m_{4,j}$ takes the same value at $\overline{\Delta \ln S}_{4}$ for $j=1,...,J-1$  in the values of $\bs b_j$'s and that $\ln \bar S^\ell_{1,j}-\ln \bar S^m_{1,j}=\overline{\Delta \ln S}_{1}$   for $j=1,...,J$  in the values of $\bs a_j$'s. Third, we assume that the probability density function of $\epsilon_{t}$ does not vary across types, i.e., $g_{\epsilon_t}^j=g_{\epsilon_t}$ for all $j=1,..., J$. These assumptions impose restrictions that make it more difficult to satisfy Assumption \ref{A-P-2} but help the identification argument to be transparent. Then, in view of Proposition \ref{P-1},  given $\bs z_3=(\bs S_3, \tilde {\bs x}_3)$,
\[
\begin{aligned}
\lambda_{\bs z_3}^j(\bs b_k)  = g_{{\bs Z}_4|{\bs Z}_3}^j(\bs b_k|{\bs z}_3) =c_{4}^j g_{\epsilon_4}(\ln(\beta_{m,4}^j E_4[e^{\epsilon}])-\ln  \bar   S^m_{4,k})
\end{aligned}
\]
with
\[
c_4^j := g^j_{\tilde {\bs X}_4|\tilde {\bs X}_3} (\tilde {\bs x}_4|\tilde {\bs x}_3)
 g_{\zeta_4}^j\left(\overline{\Delta \ln S}_{4}  -    \ln(\beta_{\ell,4}^j/\beta_{m,4}^jE_4^j[e^{\zeta}]) \right)
 \]
for $j=1,...,J$ and $k =1,...,J-1$, where the dependence of $c_4^j$ on the value of $\tilde {\bs x}_3$, $\tilde {\bs x}_4$, and $\overline{\Delta \ln S}_{t}$ is implicit.

Therefore, we have
\begin{equation}\label{L3-CD}
L_{\bs z_3}=\text{diag}\{c_4^1,....,c_4^J\}
\left[
\begin{array}{cccc}
1&g_{\epsilon_4}(\ln(\beta_{m,4}^1 E_4[e^{\epsilon}])-\ln  \bar   S^m_{4,1})&\cdots&g_{\epsilon_4}(\ln(\beta_{m,4}^1 E_4[e^{\epsilon}])-\ln  \bar   S^m_{4,J-1})\\
\vdots & \vdots & \dots & \vdots \\
1&g_{\epsilon_4}(\ln(\beta_{m,4}^J E_4[e^{\epsilon}])-\ln  \bar   S^m_{4,1})&\cdots&g_{\epsilon_4}(\ln(\beta_{m,4}^J E_4[e^{\epsilon}])-\ln  \bar   S^m_{4,J-1})
\end{array}
\right].
\end{equation}

Similarly,  given $\bs z_2=(\bs S_2, \tilde {\bs x}_2)=(S_2^m,S_2^\ell,\tilde{\bs x}_2)$, we have
\[
\begin{aligned}
\bar \lambda^j_2(\bs a,{\bs z}_2)&=\pi^j g_{{\bs Z}_2|{\bs Z}_1}^j({\bs z}_2|\bs a)g_{{\bs Z}_1}^j(\bs a)=c_{2}^j g_{\epsilon_1}^j(\ln(\beta_{m,1}^j E_1[e^{\epsilon}])-\ln  \bar   S^m_{1,k})
\end{aligned}
\]
with
\begin{align*}
c_2^j& := \pi^j g^j_{\tilde {\bs X}_2|\tilde {\bs X}_1} (\tilde {\bs x}_2|\tilde {\bs x}_1)    g^j_{\tilde {\bs X}_1} (\tilde {\bs x}_1) g_{\zeta_1}^j\left(\overline{\Delta \ln S}_{1}  -    \ln(\beta_{\ell,1}^j/\beta_{m,1}^jE_1^j[e^{\zeta}]) \right) \\
&\qquad
 \times
g_{\epsilon_2}(\ln(\beta_{m,2}^j E_2[e^{\epsilon}])-\ln  S^m_{2})g_{\zeta_2}^j\left(\ln(S_2^\ell/ S_2^m) -    \ln(\beta_{\ell,2}^j/\beta_{m,2}^jE_2^j[e^{\zeta}]) \right)
 \end{align*}
for $j,k=1,...,J$. Therefore,
\[
\bar L_{\bs z_2}=\text{diag}\{c_2^1,....,c_2^J\}
\left[
\begin{array}{ccc}
g_{\epsilon_1}(\ln(\beta_{m,1}^1 E_1[e^{\epsilon}])-\ln  \bar   S^m_{1,1})&\cdots&g_{\epsilon_1}(\ln(\beta_{m,4}^1 E_1[e^{\epsilon}])-\ln  \bar   S^m_{1,J})\\
 \vdots & \dots & \vdots \\
g_{\epsilon_1}(\ln(\beta_{m,1}^J E_1[e^{\epsilon}])-\ln  \bar   S^m_{1,1})&\cdots&g_{\epsilon_1}(\ln(\beta_{m,4}^J E_1[e^{\epsilon}])-\ln  \bar   S^m_{1,J})
\end{array}
\right].
\]

For Assumption \ref{A-P-2}(a), we have  $c_2^j \neq 0$ and $c_4^j\neq 0$ for any $j$ when $g^j_{\tilde {\bs X}_4|\tilde {\bs X}_3} (\tilde {\bs x}_4|\tilde {\bs x}_3)>0$, $g^j_{\tilde {\bs X}_2|\tilde {\bs X}_1} (\tilde {\bs x}_2|\tilde {\bs x}_1) >0$,  $g^j_{\tilde {\bs X}_1} (\tilde {\bs x}_1)>0$,  $g^j_{\epsilon_t}(\epsilon)>0$, and $g^j_{\zeta_t}(\zeta)>0$ for $\epsilon,\zeta\in \mathbb{R}$. Note that the value of $g_{\epsilon_4}(\ln(\beta_{m,4}^j E_4[e^{\epsilon}])-\ln  \bar   S^m_{4,k} $ for $j=1,...,J$ and $k=1,...,J-1$ in the element of $L_{\bs z_3}$ in (\ref{L3-CD}) represents the value of the probability density function of $\ln S^m_{4}$ for the $j$-th type evaluated at $\ln S^m_{4}=\ln  \bar   S^m_{4,k}$. Therefore, the full rank condition of $L_{\bs z_3}$ holds if the value of probability density function of $\ln S^m_4$ changes heterogenously across types when we change the value of $\ln S_4^m$. Similarly,  the full rank condition of $\bar L_{\bs z_2}$ holds if the value of probability density function of $\ln S^m_1$ changes heterogenously across types when we change the value of $\ln S_4^m$.  

Assumption \ref{A-P-2}(b) holds if $g_{\tilde {\bs X}_3|\tilde {\bs X}_2}^j(\tilde{\bs x}_3|\bar{\tilde{\bs x}}_2)\neq 0$ and $g_{\tilde {\bs X}_3|\tilde {\bs X}_2}^j(\bar{\tilde{\bs x}}_3|\tilde{\bs x}_2)\neq 0$ for all $j$. Then, we have
\begin{align*}
&D_{\bs z_3|\bs z_2} (D_{\bar {\bs z}_3|\bs z_2})^{-1} D_{\bar {\bs z}_3|\bar {\bs z}_2}(D_{{\bs z}_3|\bar {\bs z}_2})^{-1} \\
&\quad =
 \text{diag}\left\{ \frac{g_{\tilde {\bs X}_3|\tilde {\bs X}_2}^1(\tilde{\bs x}_3|\tilde{\bs x}_2)}{g_{\tilde {\bs X}_3|\tilde {\bs X}_2}^1(\bar{\tilde{\bs x}}_3|{\tilde{\bs x}}_2)}\frac{g_{\tilde {\bs X}_3|\tilde {\bs X}_2}^1(\bar{\tilde{\bs x}}_3|{\bar{\tilde{\bs x}}}_2)}{g_{\tilde {\bs X}_3|\tilde {\bs X}_2}^1(\tilde{\bs x}_3|\bar{\tilde{\bs x}}_2)},
 ..., \frac{g_{\tilde {\bs X}_3|\tilde {\bs X}_2}^J(\tilde{\bs x}_3|\tilde{\bs x}_2)}{g_{\tilde {\bs X}_3|\tilde {\bs X}_2}^J(\bar{\tilde{\bs x}}_3|{\tilde{\bs x}}_2)}\frac{g_{\tilde {\bs X}_3|\tilde {\bs X}_2}^J(\bar{\tilde{\bs x}}_3|{\bar{\tilde{\bs x}}}_2)}{g_{\tilde {\bs X}_3|\tilde {\bs X}_2}^J(\tilde{\bs x}_3|\bar{\tilde{\bs x}}_2)}\right\}.
 \end{align*}

Therefore, Assumption \ref{A-P-2}(c) requires that $\frac{g_{\tilde {\bs X}_3|\tilde {\bs X}_2}^j(\tilde{\bs x}_3|\tilde{\bs x}_2)}{g_{\tilde {\bs X}_3|\tilde {\bs X}_2}^j(\bar{\tilde{\bs x}}_3|{\tilde{\bs x}}_2)}\frac{g_{\tilde {\bs X}_3|\tilde {\bs X}_2}^j(\bar{\tilde{\bs x}}_3|{\bar{\tilde{\bs x}}}_2)}{g_{\tilde {\bs X}_3|\tilde {\bs X}_2}^j(\tilde{\bs x}_3|\bar{\tilde{\bs x}}_2)}$ takes different values across different $j$'s. 

\end{example}

\subsection{Constant Price Elasticity of Demand} \label{CES}


In place of Assumption \ref{A-5}, we may alternatively consider the case where firms produce differentiated products and face a demand function with constant price elasticity as follows.

\begin{assumption}[Constant Demand Elasticity]\label{A-CES}
(a) A firm faces an inverse demand function with constant elasticity given by $P_{Y,it} = Y_{it}^{-1/\sigma_Y^j} e^{\epsilon_{d,it}^j}$, where $\epsilon_{d,it}\notin \bs{\mathcal{I}}_{it}$  is an i.i.d. ex-post shock that is not known when $M_{it}$ is chosen at time $t$.
(b)  A firm is a price taker for intermediate and labour inputs and the intermediate price and the market wage at time $t$,  $P_{M,t}$ and $P_{L,t}$, are common across firms.
(c)  $P_{Y,it}$ and $Y_{it}$ are not separately observed in the data.
 \end{assumption}
 Under Assumption \ref{A-CES}, the ``revenue'' production function is given by $P_{Y,it}Y_{it} = \overline F_t^j(\bs X_{it}) e^{\overline \omega_{it} +\overline\epsilon_{it}}$, where $\overline F_t^j(\bs X_{it}):=  [F_t^j(\bs X_{it})]^{\frac{\sigma_{Y}^j-1}{\sigma_{Y}^j}}$,  $\overline \omega_{it}:= \frac{\sigma_{Y}^j-1}{\sigma_{Y}^j}\omega_{it}$,  $\overline\zeta_{it}:= \frac{\sigma_{Y}^j-1}{\sigma_{Y}^j}\zeta_{it}$, and  $\overline\epsilon_{it}:= \epsilon_{it}^{d} +  \frac{\sigma_{Y}^j-1}{\sigma_{Y}^j}\epsilon_{it}$. Then,  in place of (\ref{system}), we have
\begin{equation}\label{system-2}
\begin{aligned}
\ln P_{Y,it}Y_{it} &=  \ln \overline F_t^j(\bs X_{it}) + \overline\omega_{it}+\overline\epsilon_{it},\quad
\ln S_{it}^m   =  \ln \left(\overline \Gamma_{M,t}^j(\bs X_{it}) \right)+  \ln \left(E_t^j[e^{\overline\epsilon}]\right)-\overline\epsilon_{it},\\
\ln S_{it}^\ell-\ln S_{it}^m  & =  \ln \left(\frac{\overline \Gamma_{L,t}^j(\bs X_{it})}{\overline \Gamma_{M,t}^j(X_{it})E_t^j[e^{\overline\zeta}]} \right)   +\overline\zeta_{it},  \quad
\ln P_{M,t}M_{it} =   \ln \left(\frac{P_{L,t}L_{it}\overline \Gamma_{M,t}^j(\bs X_{it}) E_t^j[e^{\overline \zeta}]}{\overline \Gamma_{L,t}^j(X_{it})}\right)  +v_{it},
\end{aligned}
\end{equation}
where $\overline \Gamma_{M,t}^j(\bs X_{it}):=\frac{\overline F_{M,t}^j(\bs X_{it}) M_{it}}{\overline F_{t}^j(\bs X_{it})}$   and $\overline \Gamma_{L,t}^j(\bs X_{it}):=\frac{\overline F_{L,t}^j(\bs X_{it}) L_{it}}{\overline F_{t}^j(\bs X_{it})}$.
When $P_{Y,it}$ and $Y_{it}$ are not separately observed in the data,  the observable implication of (\ref{system-2}) are the same as that of (\ref{system}). In particular, we cannot separately identify the parameter $\sigma_{Y}^j$ and the production function $F_t^j$. Therefore, we focus on the identification analysis under Assumption \ref{A-5} although we should be careful in interpreting the empirical result because the unobserved heterogeneity in revenue production function could partly reflect in difference in demand elasticity.

 \end{document}